\documentclass[11pt]{article}
\usepackage{amsmath,mathtools}
\usepackage{amsfonts}
\usepackage{amssymb}
\usepackage{mathrsfs}
\usepackage{scalerel}
\usepackage[dvipsnames]{xcolor}
\usepackage{graphicx}
 \usepackage{caption}
 \usepackage{subcaption}
 \usepackage{eso-pic}% http://ctan.org/pkg/eso-pic
\usepackage{multirow,cite}

\topmargin - 0.7 in

\def\a{\alpha}
\def\b{\beta}
\def\d{\delta}
\def\e{\epsilon}
\def\s{\sigma}
\def\l{\lambda}
\def\p{\partial}
\def\r{\rightarrow}
\def\O{\mathcal{O}}
\def\N{\mathcal{N}}
\def\H{\mathcal{H}}
\def\X{\mathcal{X}}
\def\T{\mathcal{T}}
\def\V{\mathcal{V}}
\def\D{\Delta}

\newcommand{\bi}{\begin{itemize}}
\newcommand{\ei}{\end{itemize}}

\newcommand{\be}{\begin{equation}}
\newcommand{\ee}{\end{equation}}

\newcommand{\bea}{\begin{eqnarray}}
\newcommand{\eea}{\end{eqnarray}}

\numberwithin{equation}{section}
\makeatletter
\renewcommand{\@seccntformat}[1]{%
  \csname the#1\endcsname.\ }
\makeatother

\title{States, symmetries  and correlators of $T\bar T$ and $ J\bar T $  symmetric  orbifolds\vskip3mm}

\author{Soumangsu Chakraborty$^{\S,\P}$, Silvia Georgescu$^{\S, \dag}$ and
Monica Guica$^{\S,\ddag,\sharp }$ \vspace{0mm} \\
\\\vspace{1mm}
${}^\S$\emph{\small Universit\'e Paris-Saclay, CNRS, CEA,} 
\emph{\small Institut de Physique Th\'eorique, 91191 Gif-sur-Yvette, France} \\ %\vspace{1mm}
${}^\P$\emph{\small Institute for Theoretical Physics},
\emph{\small University of Amsterdam, PO Box 94485, 1090GL, Amsterdam, The Netherlands} \\ \vspace{1mm}
${}^\dag$\emph{\small 
CPHT, CNRS, \'Ecole polytechnique, Institut Polytechnique de Paris, 91120 Palaiseau, France} \\  \vspace{1mm}
${}^\ddag$\emph{\small Institute of Physics, Ecole Polytechnique Fed\'erale de Lausanne, CH-1015 Lausanne, Switzerland} \\ 
${}^\sharp$\emph{\small Theoretical Physics Department, CERN, CH-1211 Geneva 23, Switzerland}}

\date{}
\begin{document}

%\AddToShipoutPictureBG*{%
 % \AtPageUpperLeft{%
  %  \hspace{\paperwidth}%
   % \raisebox{-\baselineskip}{%
    %  \makebox[0pt][r]{\today ~~ \currenttime %~~~~~}
%}}}%

\maketitle

\abstract{
\vskip 2mm

\noindent We derive various properties of symmetric product orbifolds of $T\bar T$ and $J\bar T$ - deformed CFTs from a field-theoretical perspective. First, we  generalise the known formula for the torus partition function of a symmetric orbifold theory in terms of the one of the seed  to  non-conformal two-dimensional QFTs;
%
%First, we show that the spectrum of any symmetric product orbifold of two-dimensional QFTs whose partition function can be rendered modular invariant via an appropriate transformation of the couplings can be determined using the group-theoretical approach of Bantay.
 specialising this to seed   $T\bar T$ and $J\bar T$ - deformed CFTs  reproduces previous results in the literature. Second, we show that the single-trace $T\bar T$ and $J\bar T$ deformations preserve 
  the Virasoro and Kac-Moody symmetries of the undeformed symmetric product orbifold CFT, including their fractional counterparts, as well as the KdV charges.     %; the higher-spin symmetries  of the orbifold CFT  do, however, appear to be broken. \emph{Are we sure about this?}  %case are likely broken by the deformation, as indicated by the smaller degree of degeneracy of $T\bar T$ and $J\bar T$ SPO as compared to their CFT counterparts. 
   Finally, we discuss correlation functions in these theories. By extending a previously-proposed basis of operators for $J\bar T$ - deformed CFTs to the single-trace case, we
       explicitly compute the   correlation functions of both 
   untwisted and twisted-sector operators and compare them % show that the resuts for the two-point functions match 
   to an appropriate set of holographic correlators. %in single-trace $J\bar T$ deformed CFTs. 
   %   discuss correlation functions in $J\bar T$ and propose an expression for correlation functions both in the twisted and untwisted sector.
    Our derivations are based mainly on Hilbert space techniques and completely avoid the use of conformal invariance, which is not present in these models.  
 %   
%     completely avoids the use of conformal maps to the covering space, as conformal symmetry is broken in these theories.

}

\tableofcontents

\section{Introduction}

The study of symmetric product orbifolds of $T\bar T$ and $J\bar T$ - deformed CFTs is interesting for a number of reasons. First, symmetric product orbifolds of two-dimensional QFTs play an important role in holography, as  their large $N$ behaviour is compatible with that of a gravitational dual where quantum-gravitational corrections are supressed \cite{El-Showk:2011yvt}.  When the seed theory is a CFT, they enter concrete realisations of the AdS$_3$/CFT$_2$ correspondence \cite{Maldacena:1997re,Dijkgraaf:1998gf,Giveon:1998ns,Seiberg:1999xz,Argurio:2000tb,
David:2002wn}.  According to the proposals of \cite{kutasov,Apolo:2018qpq, Chakraborty:2018vja},  symmetric product orbifolds of $T\bar T$ \cite{Smirnov:2016lqw,Cavaglia:2016oda} and $J\bar T$ - deformed CFTs \cite{Guica:2017lia}  - a set of non-local, yet  UV-complete and solvable two-dimensional QFTs -   should provide tractable models of three-dimensional non-AdS holography. %[While the dual of an exact SPO is always highly stringy, they do provide a very useful concrete handle over the correspondence. 
More precisely, the $T\bar T$ symmetric orbifold should  be related to a spacetime that is asymptotically flat with a linear dilaton, whereas the $J\bar T$ one should correspond to a warped AdS$_3$ background,  which is relevant to understanding  the Kerr/CFT correspondence \cite{Guica:2008mu,El-Showk:2011euy}.

The study of symmetric product orbifolds of $T\bar T$ and $J\bar T$ - deformed CFTs
is also interesting  from the point of view of the original motivation of \cite{Smirnov:2016lqw,Cavaglia:2016oda} - namely, to understand the space of integrable two-dimensional QFTs. The existence of exactly solvable irrelevant deformations of two-dimensional QFTs whose UV behaviour is not governed by a standard UV CFT fixed point, yet is entirely under control \cite{%Dubovsky:2017cnj,
Dubovsky:2013ira}, is quite remarkable. The orbifold construction provides a simple way to enlarge the set of tractable examples of such QFTs. 
 The properties of the resulting theories are similar - though not exactly the same - as those of the seed QFTs. It is a useful exercise to work them out explicitly from first principles, which is the main  goal of this article.

% The standard integrable $T\bar T$ and other SZ deformations provide one way of exploring this space, leading to theories that are exactly solvable. While it is a simple observation that the single-trace $T\bar T$ deformation of an SPO will lead to an SPO of $T\bar T$ - deformed CFTs, and the data in the latter is in principle determined by that of the seed, it is still useful to perform the excercise of extracting it, which is the main goal of this article.  

Another motivation  for studying this problem is that neither $T\bar T$, nor $J\bar T$ - deformed CFTs possess (full) conformal invariance, which is nevertheless  omnipresent in the symmetric product orbifold literature. We would therefore like to use these examples to illustrate the fact that  many observables in symmetric orbifold QFTs \emph{can} be obtained without the conformality asssumption. %In this article, for example, we  use instead the language of states and symmetries to describe the effect of the symmetric product orbifold on the theories of interest. 
Depending on the specifics of the system under study, these observables can even include
  twisted-sector correlation functions, as we show explicitly for the case of single-trace $J\bar T$ - deformed CFTs.
 % can be computed without it, but  using instead the language of states and symmetries.

% to understand how many results on SPOs depend on conformal invariance. We show, as a proof of principle, that many can be obtained without it. However, the techinques we use are more or less heavily adapted to the type of deformations we study.  

%In this article, we will be interested in symmetric product orbifolds that are relevant to non-AdS holography, which apppears to require QFTs with a non-standard UV behaviour. In certain cases, these theories can be obtained as irrelevant flows of a CFT that yield a UV-complete theory, yet non-local. The discovery of the exactly solvable $T\bar T$ and $J\bar T$ flows of two-dimensional QFTs has opened up a new window into studying the UV behaviour of such theories.  Moreover, the symmetric product orbifolds of such theories have been argued to be related to non-AdS holography, for ALD backgrounds in the $T\bar T$ case, and for warped AdS$_3$ in the $J\bar T$ one, which makes a connection to the Kerr/CFT correspondence.    

The analysis presented in this article is purely field-theoretical, and the QFTs under study are \emph{exact} symmetric product orbifolds of  $T\bar T$ or $J\bar T$ - deformed CFTs, obtained via a single-trace $T\bar T$/$J\bar T$ deformation of an \emph{exact} symmetric orbifold of two-dimensional CFTs. As a result, the large $N$ holographic duals of these theories are highly stringy. 
Our setup is thus different from that used in the holographic proposals \cite{kutasov,Apolo:2018qpq, Chakraborty:2018vja}%\cite{Argurio:2000tb} 
, who deformed an \emph{approximate} symmetric product orbifold of CFTs - namely, the CFT dual to the near horizon of several NS5 branes and a large number of F1 strings\footnote{See    \cite{Eberhardt:2021vsx} for a proposed concrete realisation of this CFT.} - 
% and checked in \cite{Dei:2022pkr,Bufalini:2022toj} -
  by an operator whose action resembles that of the single-trace $T\bar T$ or $J\bar T$ operator\footnote{Throughout this article, the single-trace $T\bar T$/$J\bar T$ operator will simply denote the sum over copies, in a symmetric orbifold QFT, of the corresponding Smirnov-Zamolodchikov operator. By contrast, in  \cite{kutasov,Apolo:2018qpq,Chakraborty:2018vja}   ``single-trace $T\bar T$/$J\bar T$'' is  a nickname given to a certain operator of dimension $(2,2) / (1,2)$  that is single-trace (in the sense of corresponding to a single-particle bulk excitation) and some of whose properties resemble those of $T\bar T$/$J\bar T$.   }
  %\footnote{{\color{ForestGreen}{We would also like to stress the fact that, what we refer to as single trace $T\bar T$/$J\bar T$ deformation in this article, is different from the ``single trace $T\bar T$/$J\bar T$" deformations of holographic CFTs studied in \cite{kutasov,Apolo:2018qpq,Chakraborty:2018vja} where the boundary CFT$_2$ (dual to string theory in $AdS_3$ with pure NS-NS flux) is deformed by certain  single trace dimension $(2,2) / (1,2)$ quasi-primary operators $D(x,\bar{x})/A(x,\bar{x})$ constructed in \cite{Kutasov:1999xu}.}}}
  .
   These deformations were argued to correspond to  exactly marginal deformations of the worldsheet string theory, which can be studied with a variety of techniques \cite{Giveon:2017myj,Azeyanagi:2012zd,Apolo:2019zai,Chakraborty:2019mdf,Asrat:2017tzd,Giribet:2017imm,soum,demisethesis,Cui:2023jrb,Hashimoto:2019wct,Hashimoto:2019hqo,Benjamin:2023nts}. Most of the results obtained so far in the ``single-trace $T\bar T$'' and $J\bar T$ literature were in fact derived using worldsheet methods that, given the only approximate identification of the boundary deformation with single-trace $T\bar T/J\bar T$,  may or may not agree with the exact symmetric product orbifold calculations. Thus, yet another motivation for this work is  to provide an \emph{independent} derivation of various properties of these theories that were previously predicted via holography.

The first observable we study is the finite-size spectrum of the orbifolded theories. %, which can be extracted from the torus partition function. 
This has been first computed  using worldsheet methods, by studying the effect of the  exactly marginal deformations on the spectrum of long strings in the massless BTZ background \cite{Giveon:2017myj,Apolo:2018qpq, Chakraborty:2018vja}.  
%, for the specific case of  
 %
%Most of the previous work on the observables   of SPOs of $T\bar T$ and $J\bar T$ deformed CFTs consists of dual worldsheet string computations (restricted to the long string sector $\r$ only spectrum,!!!) that should reproduce the SPO answer provided the holographic dictionary proposed in \cite{} is correct. (given match in $ w=1$ sector, match for arbitrary $ w$ is obvious) These calculations include the spectrum of strings 
More precisely, it was shown that the spectrum of singly-wound long strings in the deformed backgrounds precisely coincides with the $T\bar T$  and, respectively,  $J\bar T$ - deformed spectrum, which provided  a  non-trivial check of the proposed duality; the string theory prediction for the spectrum of multiply-wound strings was then naturally conjectured to represent the  contribution of the twisted sectors of the symmetric product orbifold in this specific example. The $T\bar T$ result has been recently confirmed by the field-theoretical analysis of \cite{Apolo:2023aho},  who fixed the partition function by requiring it to be modular invariant in a generalised $T\bar T$ sense \cite{Aharony:2018bad}.
%
%on deformed massless BTZ, which was therefore conjectured to coincide with that in an SPO of $T\bar T$ in \cite{} and $J\bar T$ and arbitrary combinations in \cite{}. 

As already noted in \cite{Aharony:2018bad}, this modular invariance is an automatic  property of 
the partition function of any (UV - complete) QFT with a single dimensionful scale, assuming its path integral on the torus is  well-defined; the generalization to several parameters, including non Lorentz-invariant ones, is straightforward \cite{Aharony:2018ics}.  In this article, we
%
%the $T\bar T$ , which simply follows from dimensional analysis and the fact one is now obliged to include the radius dependence. The same is true of general non-conformal two-dimensional QFTs, including non-Lorentz invariant ones under certain assumptions.  
%We then
 provide a general expression for the partition function of the symmetric  orbifold of such theories,
 %
%  derivation of the spectrum of SPOs of two-dimensional arbitrary QFTs, 
  based on a slight generalisation of Bantay's formula \cite{
Bantay:1998fy,Bantay:1999us,Bantay:2000eq} for the case of CFTs; %Our only assumption is that the seed partition function can be made modular invariant by taking into account the transformation of the coupling; this property follows almost trivially from dimensional analysis/covariance and, in the case  of $T\bar T$ and $J\bar T$ - deformed CFTs, demystifies the observed modular properties of the partition function \cite{}. 
   its modular invariance follows  automatically from that of the seed QFT. When applied  to the case of $T\bar T$ and $J\bar T$ - deformed CFTs, this partition function precisely reproduces or generalises previous results in the literature.

Given the partition function, one may analyse the thermodynamic properties of the  symmetric product orbifold of $T\bar T$/$J\bar T$ - deformed CFTs. The  $T\bar T$ case has been  analysed in detail in \cite{Apolo:2023aho}. We use these results to compare  the entropy of a single-trace to that of a double-trace $T\bar T$ deformation \cite{Chakraborty:2022xmz}  of a symmetric  orbifold CFT and note that while they agree - as they should - in the universal high-energy regime discussed in \cite{Apolo:2023aho}, they disagree outside it. 
%: while the first exhibits Hagedorn behaviour, the second, super-Hagedorn {\color{red}[in the intermediate regime]}.
% Thus,  non-universal observables can behave  differently under the two deformations.
 We also discuss the entropy of single/double-trace $J\bar T$ - deformed CFTs, showing  there  exists a  regime of real high energies where the behaviour of the entropy is either Cardy-like or Hagedorn, depending on the chirality properties of the $U(1)$ current.

% , this was recently studied in detail in \cite{}. We thus simply remark that the behaviour of a $T\bar T$ SPO is different from that of a double-trace $T\bar T$ of SPO, as the latter exhibits and intermediate super-Hagedorn growth \cite{} that is absent in the former. It also obeys the bound. 

Next, we  study the extended symmetries of single-trace $T\bar T$ and $J\bar T$ - deformed CFTs. A non-trivial property of the standard  $T\bar T$ and $J\bar T$ deformations is that they preserve the Virasoro and, if present, the Kac-Moody symmetries of the undeformed CFT \cite{Guica:2020uhm,Guica:2021pzy,Guica:2022gts}. That the same is true of the single-trace deformation is strongly suggested by the results of the 
 asymptotic symmetry group  analysis of  the  linear dilaton spacetime \cite{Georgescu:2022iyx}, which uncovered an infinite  set of symmetries, whose algebra closely resembles the $T\bar T$ symmetry algebra.

  In this article, we provide a purely field-theoretical proof 
  that  these symmetries are indeed preserved, 
  closely following the argument  used in the double-trace case \cite{Guica:2021pzy,Guica:2022gts}. 
  %
  % This involves finding the operator that flows the energy eigenstates in the SPO, and using it to defined flowed Virasoro and Kac-Moody generators in the deformed theory. The fact these are symmetries then follows from the conservation of these operators.  
%
% Finally, a highly non-trivial analysis of the asymptotic symmetries of the ALD spacetime \cite{} an infinite  set of symmetries, which should be those  of single-trace $T\bar T$, if the duality is correct. 
%
%In order to check this proposed duality, it is therefore important to compute the given observables independently in the dual symmetric product orbifold. To date, only  the single-trace $T\bar T$ spectrum  has been studied from a field-theoretical perspective in \cite{}, They found the same answer as the string computation, an encouraging evidence for the duality.  However, our motivation goes beyond this particular setup, and we would like to derive the properties of these field theories irrespective of the proposed duality and, more generally, investigate how far one can get in the study of SPOs of theories that are not CFTs. 
%
This argument requires understanding the operator that drives the flow of the energy eigenstates under the single-trace $T\bar T$/$J\bar T$ deformation, which is technically more complicated than the corresponding double-trace flow in that many of the initial CFT degeneracies are broken when the deformation is first turned on. We also discuss other bases of symmetry generators, which are non-linearly related to the Virasoro one, and argue that they may be  preferred at a global level in the single-trace $T\bar T$ and $J\bar T$ - deformed CFT. Working out the corresponding non-linear symmetry algebra in single-trace $T\bar T$ - deformed CFTs, we show the result agrees precisely with the holographic calculation  \cite{Georgescu:2022iyx}. 
% analysis does present several new features, such as dealing with the fact that (unlike its double-trace counterpart), the single-trace deformations do break part of the degeneracies of the CFT symmetric orbifold, and the existence of fractional Virasoro modes, which can be similarly shown to be preserved. 
In addition, we  show that the KdV charges and the fractional Virasoro and Kac-Moody modes are preserved by the deformation;
%
%The study of the flow operator is more interesting than in double-trace $T\bar T$ because the single-trace deformation breaks many of the degeneracies that were present in the CFT SPO, which needs to be taken into account when discussing the flow operator. 
%
%The result in hand, we prove that all the Virasoro and Kac-Moody symmetries  of the CFT SPO, including their fractional counterparts, do survive the deformation; however, 
the  fate of the higher spin symmetries such as those discussed in \cite{Apolo:2022fya} is less clear. %could be broken, though a more thorough analysis is  necessary to fully ascertain this. %[Thus, even though the Virasoro and KdV charges  survive the standard  $T\bar T$ deformation, the $T\bar T$ SPOs appear to have (much) less symmetry than their CFT counterparts.] %It would be interesting in the future to understand the precise symmetry structure.

Finally, we turn our attention to correlation functions. For standard $J\bar T$ - deformed CFTs, these have  been  understood  in \cite{Guica:2021fkv} (see also \cite{Guica:2019vnb}), and recently have  also been computed in  $T\bar T$ - deformed CFTs \cite{Aharony:2023dod} (see also \cite{Cardy:2019qao%,Aharony:2018vux
}), using rather different methods. In addition, several holographic calculations of two-point functions -  using either worldsheet or supergravity techniques -  were performed in \cite{Asrat:2017tzd,Giribet:2017imm,soum,demisethesis,Cui:2023jrb,Azeyanagi:2012zd}.  We provide explicit expressions for the correlation functions of a proposed set of  both untwisted and twisted-sector operators in single-trace $J\bar T$ - deformed CFTs, which we
%
% We exemplify the computation of correlation functions from both the untwisted and the twisted sector. To achieve the latter without using conformal maps, we use the special property that $J\bar T$ correlator are obtained from those of the undeformed CFT by spectral flow by the right-moving Hamiltonian, a Hilbert space relation that can be generalised to the twisted case without the use of conformal maps. 
 then compare  with  a  holographic computation of the two-point functions of long string vertex operators - the only worldsheet operators that are described by a symmetric product orbifold - performed using the methods of \cite{Azeyanagi:2012zd,Cui:2023jrb}. The two results are found to  slightly differ, and we comment on possible reasons for this. %and show the result    precisely agrees with the field-theoretical expression  when it is expected to, namely for operators corresponding to the long string sector of the theory, which are the only ones
 %
% , which is due to the fact that the operators whose correlation functions were computed holographically  are not 
% described by a symmetric product orbifold. 

%indicating this observable is sensitive to/which can be understood as being due to the approximate nature of the dictionary proposed in {\color{red}\cite{kutasov}}.   \emph{Careful!}

This article is organised as follows. In section \ref{specent}, we study the torus partition function of symmetric product orbifolds of general two-dimensional QFTs and show that it can be obtained via a slight generalisation of Bantay's formula; %the derivation assumes only that the seed torus partition function is well-defined, namely that the seed is modular invariant in a generalised sense. The modular invariance of the SPO follows as a consequence. W
we work out the  $T\bar T$ and $J\bar T$ case as an example. We also comment on the thermodynamics of
% with comments on the density of stats in 
 single-trace $T\bar{T}$  and $J\bar T$ - deformed CFTs. In section \ref{symflow}, we study the flow of the states and of the Virasoro (- Kac-Moody) generators, including their fractional counterparts,   in single-trace $T\bar{T}$ and $J\bar{T}$ deformed CFTs %. Due to the structure of the flow operator, we can define a set of Virasoro (and KM) generators
  and show that they are still conserved, as are the KdV charges. We also  discuss other possible bases of symmetry generators.  Finally, in  section \ref{correlation} we compute  correlation functions in single-trace $J\bar T$ - deformed CFTs and compare them with an appropriate holographic result. %, and show the field-theoretical results precisely match the holographic ones we briefly derive.
    We end with a summary in section \ref{discussion}. For completeness, each section contains an introductory subsection that summarizes  the relevant results from the double-trace case.

%
%Results:
%\bi
%\item only single-trace $T\bar T$ deformation respects the symmetric orbifold structure. The flow of the states implies that the flow operator is also untwisted and single-trace. \emph{Still need to prove this for flow of twisted states! } Note it acts as $T\bar T$ on a circle of radius $R w$.
%\item this form of the flow operator immediately implies Virasoro symmetry
%
%\item if we understand how the spectrum is organised in terms of fractional Virasoro primaries, then $T\bar T$ flow should respect that. \emph{Is it true that for each seed primary we get fractional primaries $L_{-k/w} |h_w\rangle$?} 
%\item is it true that the states can be characterised as primaries vs decendants of the fractional Virasoro? How do we establish this (e.g. in the CFT) ?
% \ei
% 

\section{The spectrum and the entropy} \label{specent}
In this section, we explain in a simple fashion how  to obtain the finite-size spectrum of a symmetric product orbifold of any  two-dimensional QFT whose partition function is modular invariant  in an appropriately generalised sense. Our results are exemplified by symmetric  orbifolds of $T\bar T$ - deformed CFTs in the Lorentz-invariant case,  and  symmetric  orbifolds of $J\bar T$ - deformed CFTs in the non-Lorentz-invariant one. For completeness, we start this section with a brief review of the spectrum and partition function of standard (double-trace) $T\bar T$ and $J\bar T$ - deformed QFTs. 

% is obtained in terms of the spectrum of the seed QFT. \emph{Rewrite!} We work on a cylinder of radius $R$ throughout. 

\subsection{Review of the $T\bar T$ and $J\bar T$-deformed  spectrum and partition function} \label{seedTTJT}

One remarkable feature of $T\bar T$, $J\bar T$ deformations and their generalisations is that the  spectrum of the deformed QFT on a cylinder of circumference $R$ is entirely determined by the finite-size spectrum of the undeformed QFT, as we now review. 

\subsubsection*{$T\bar T$ - deformed QFTs}  

The $T\bar T$ deformation is a universal irrelevant deformation of a two-dimensional QFT by an operator constructed from the components of the stress tensor 

\be
\p_\mu S = \frac{1}{2} \int d^2 x \, \O_{T\bar T}^{[\mu]} \;, \;\;\;\;\;\; \O_{T\bar T} = \e^{\a\b} \e^{\gamma\delta} T_{\a\gamma} T_{\b \delta}
\ee
which enjoys nice factorization properties in energy eigenstates \cite{Zamolodchikov:2004ce,Smirnov:2016lqw}. These properties imply that
%
%This can be most easily shown for the $T\bar T$ deformation where, 
%
%Let us first review the $T\bar T$ deformation. 
%thanks to 
%the factorization properties of the $T\bar T$ operator in energy eigenstates, 
the energies $E_n^{[\mu]}(R)$ of the eigenstates of the deformed theory on a cylinder of circumference $R$  obey  Burger's equation

\be
\p_\mu E^{[\mu]}_n(R) = E^{[\mu]}_n(R)\frac{\partial E^{[\mu]}_n(R)}{\partial R} + \frac{P_n^2(R)}{R} 
\ee
%where $E,P$ are generic notations for the different energy and momentum eigenvalues $E_n,P_n$ and $R$ is the circumference of the $S^1$ with the identification $\s\sim\s + R$.
%
This equation can be solved via the method of characteristics. For $P=0$, the solution is simply given  by $E^{[\mu]}_n(R) = E^{[0]}_n(R+\mu E_n^{[\mu]})$, where $E_n^{[0]}$ are the undeformed energies; the solution for $P\neq 0$ is a slight generalisation of this result \cite{Cavaglia:2016oda}. % Thus, if the  spectrum of the undeformed QFT is known as a function of $R$, then so is that of the $T\bar T$ - deformed theory. A well-studied example where this is the case is that of 
%
%
% entirely in terms of the undeformed spectrum of the seed QFT (\emph{Notation!})
%
%
%\be
%E(\mu,R) = \cosh \theta_0 \, E^{(0)} (R_0) - \sinh \theta_0 \, P(R_0) 
%\ee
%where $\theta_0, R_0$ are given implicitly by
%%
%\be
%R_0 \cosh \theta_0 = R+\mu E \;, \;\;\;\;\;\; R_0 \sinh \theta_0  = \mu P
%\ee
Thus, if the spectrum of the undeformed QFT is known explicitly as  a function of $R$, then so is the spectrum of the corresponding $T\bar T$ - deformed QFT.
 A well-studied example where this is the case is that of $T\bar T$ - deformed CFTs, where the undeformed energies are inversely proportional to $R$, %i$\propto 1/R%=\frac{1}{R}(h+\bar{h}-\frac{c}{12})
 %$, where the proportionality factor is related to the conformal dimension, 
 and the solution for the deformed  spectrum is 
\begin{align} \label{sqroot}
E_n^{[\mu]}(R)&=\frac{R}{2\mu}\bigg(-1+\sqrt{1+\frac{4\mu E_n^{[0]}(R)}{R}+\frac{4\mu^2 P_n^2}{R^2}}\bigg)%= \frac{R}{2\mu}\bigg(-1+\sqrt{1+\frac{4\mu (h+\bar{h}-\frac{c}{12})}{R^2}+\frac{4\mu^2 (h-\bar{h})^2}{R^4}}\bigg)
\end{align}
This solution can also be written in terms of the conformal dimension $\Delta$ and spin $s$ of the corresponding operator by plugging in the  expressions %\eqref{cftenergy}
 for $E_n^{[0]}, P_n$ as a function of $\Delta$ and $ s$.

The torus partition function of the deformed QFT is defined as usual via the Hilbert space trace
\begin{align}
Z^{[\mu]}(\tau,\bar{\tau},R)&=\sum_{n}  e^{- \tau_2 R \, E_n^{[\mu]}(R)+ i \, \tau_1 R \, P_n } \label{defttbpf}
\end{align}
where $\tau=\tau_1+ i \tau_2$ is the complex structure modular parameter 
and $R$ is the length of the $a$-cycle of the torus, here designated as the spatial one. For a  $T\bar T$ - deformed CFT,  $Z^{[\mu]}$ only depends on $R$ via the dimensionless combination $\mu/R^2$, since $\mu$ is the only dimensionful parameter in the theory.  

Let us now discuss the modular transformation properties of this partition function. The flat metric on the torus can be written as
\begin{align}
ds^2&= R^2 |dx+\tau dy|^2=R^2 dz d\bar{z}
\end{align}
where  $x,y$ are real coordinates of unit periodicity  and  the complex coordinates $z,\bar z$ are defined as $z=x+\tau y,\,\bar{z}=x+\bar{\tau}y$.  This metric is invariant under  large   $PSL(2,\mathbb{Z})$ diffeomorphisms of the torus 
\be
\begin{pmatrix}x \\y \end{pmatrix}\mapsto\begin{pmatrix}\; a & - b \\- c & \;d \end{pmatrix}\begin{pmatrix}x \\y \end{pmatrix} \;, \;\;\;\;\;\;\; 
\tau\mapsto\frac{a\tau+b}{c\tau+d}\;, \;\;\;\;\;\; ad-bc=1 \label{psl2z}
\ee
%
%
%\be
%\begin{pmatrix}x \\y \end{pmatrix}=\begin{pmatrix}d & b \\ c & a \end{pmatrix}\begin{pmatrix}x' \\y' \end{pmatrix} \;, \;\;\;\;\;\;\; 
%\tau\rightarrow\frac{a\tau+b}{c\tau+d}
%\ee
which leave the coordinate periodicities  intact, provided we also transform

\be
%\tau\rightarrow\frac{a\tau+b}{c\tau+d} \;, \;\;\;\;\;\; 
R\r |c\tau+d|R
\ee
%
%where $a,b,c,d\in \mathbb{Z}$ with $ad-bc=1$, 
Note this ensures that the area of the torus, $R^2 \tau_2$, is invariant. Under \eqref{psl2z}, the complex coordinates change as
\be
z\mapsto \frac{z}{c\tau+d} \;, \;\;\;\;\;\; 
\bar z\mapsto \frac{\bar{z}}{c \bar \tau+d} \label{transfzzb}
\ee
%together with an $SL(2,\mathbb{Z})$ transformation of the real coordinates $(x,y)$:
%\begin{align}
%x+\frac{a\tau+b}{c\tau+d}y=\frac{(xd+by)+\tau(ay+xc)}{c\tau+d}=\frac{x' + \tau y'}{c\tau+d}, \hspace{0.4cm} \begin{pmatrix}x' \\y' \end{pmatrix}=\begin{pmatrix}d & b \\ c & a \end{pmatrix}\begin{pmatrix}x \\y \end{pmatrix}
%\end{align}
%Since the area of the torus remains the same, it follows that at a modular transformation, the characteristic length of the torus changes:
%\begin{align}
%R\mapsto |c\tau+d|R
%\end{align}
Assuming the partition function \eqref{defttbpf} can also be computed via an Euclidean path integral over the torus, which is naturally invariant under the diffeomorphisms  discussed above, we conclude that

\be \label{modularttbar}
Z^{[\mu]} \bigg(\frac{a\tau + b}{c\tau+d},\frac{a\bar{\tau}+b}{c\bar{\tau}+d},R |c\tau+d|\bigg) = Z^{[\mu]}(\tau,\bar{\tau},R)
\ee 
While we wrote this relation with $T\bar T$ - deformed CFTs in mind, whose partition function depends on a single dimensionful parameter $\mu$, it 
 should hold in any UV-complete two-dimensional QFT with dimensionful scalar couplings - collectively denoted as `$[\mu]$' - whose partition function can be computed via a path integral over the euclidean torus. %{\color{ForestGreen}(mention here that this is the "generalised modular invariance" we refer to later)}%, and it simply corresponds to the freedom to assign different Hilbert space interpretations to it/diff invariance. 
 In a CFT, the radius dependence drops out by scale invariance, resulting in the usual modular invariance requirement;  \eqref{modularttbar} may then be referred to as ``generalised modular invariance''. It simply states the invariance of the partition function under a relabeling of the torus coordinates, and as such it is natural that these  transformations relate theories defined on tori  with different sizes of the $a$-cycle, where the scalar couplings (which may be dimensionful) are held fixed. 
 
  In a $T\bar T$ - deformed CFT, the partition function $Z^{[\mu]} (\tau,\bar \tau, R) =  Z_{T\bar{T}}(\tau,\bar \tau, \mu/R^2)$, and so the above relation reads %\emph{Notation!}

\be \label{TTbarmodinv}
Z_{T\bar T} \bigg(\frac{a\tau + b}{c\tau+d},\frac{a\bar{\tau}+b}{c\bar{\tau}+d}, \frac{\mu}{R^2 |c\tau+d|^2}\bigg) = Z_{T\bar T}\bigg(\tau,\bar{\tau},\frac{\mu}{R^2}\bigg)
\ee
Thus, in this case one may reinterpret \eqref{modularttbar} as relating theories on a circle of the  same radius, but with different dimensionless couplings. The above relation was checked explicitly in \cite{Datta:2018thy}. 

%[This was derived in \cite{} by expanding the $T\bar T$ partition function \eqref{} in $\mu$, using the explicit solution \eqref{} for the spectrum, and showing that each term in the expansion is a modular form of weight equal to twice the power of $\mu$. Now, we see this had to work this way by the mere existence of a path integral formulation of the torus partition function. ]

The density of states of a $T\bar T$ - deformed CFT follows from the adiabaticity of the deformation, which implies that the number of states is unchanged along the flow. We thus have
%For $T\bar{T}$, the entropy is given by
\be \label{entropytt}
S_{T\bar T} (E) =S_{Cardy} (E^{[0]}(E)) = % 2\pi \sqrt{\frac{c\, E_L^{(0)}R}{12 \pi}} + 2\pi \sqrt{\frac{c \, E_R^{(0)}R}{12\pi}} 
%\ee
%but we need to express it in terms of the physical energies in the deformed theory. Using \eqref{}, we obtain% The relation is given by:
%\be
%E_L^{(0)}  = E_L \left( 1 +   \frac{2\mu  E_R}{ R}\right) \;, \;\;\;\;\;\; E_R^{(0)}  = E_R \left(1 +  \frac{2\mu E_L}{ R}\right) 
%\ee
%Plugging this into the equation above, we obtain
%\be
%S_{T\bar T} = 
 2\pi \sqrt{\frac{c  E_L (R +2 \mu E_R) }{12\pi}} + 2\pi \sqrt{\frac{c  E_R(R+ 2\mu E_L)}{12\pi}}
\ee
where the relation between $E^{[0]}$ and $E, P$ was obtained by inverting \eqref{sqroot},  $E_{L,R} \equiv (E \pm P)/2$ and, for simplicity, we have dropped the $\mu$ label for the deformed energies. Note that the  high-energy behaviour is Hagedorn.

Finally, let us remind the reader  that reality of the deformed ground state energy \eqref{sqroot} implies that $T\bar T$ - deformed CFTs can only be  defined on cylinders whose circumference satisfies

\be
R \geq R_{min} = \sqrt{ \frac{2\pi \mu c}{3}} \label{rmin}
\ee
The high-energy behaviour of the entropy implies in turn that the thermal partition function only makes sense below the Hagedorn temperature $T_H = R_{min}^{-1}$, which amounts to the same constraint. More generally, we have

\be
\lim_{E_{L,R} \r \infty} Z_{T\bar T} \approx e^{- \b_L E_L - \b_R E_R} e^{2\sqrt{\frac{2 \pi \mu c}{3} E_L E_R}} \leq e^{2 \sqrt{E_L E_R} (R_{min} - \sqrt{\b_L \b_R})}
\ee
where $\b_{L,R}$ are the left/right-moving temperatures, which satisfy $\b_L \b_R = R^2 |\tau|^2$.   The partition function is thus well defined provided also $R |\tau| > R_{min}$. One may check - by appropriately choosing the integer part of $\tau_1$ - that 
% 
%$\b > R_{min}$. \emph{What is the general condition for a tilted torus?}  If all is to make sense, 
modular transformations do not take us out of this regime. %\textbf{\emph{Agree?}}(yes)

 %  Note this guarantees that the energy in the twisted sectors will be real, since their reality condition requires $ Rw > $ RHS. \emph{Later?}

% Thus, modular transformations relate theories defined on circles of different radii, if we keep the dimensionful $T\bar T$ coupling fixed, as is natural to do. Alternatively, if one writes the partition function in terms of the dimensionless coupling $\mu/R^2$, the above transformation can be interpreted as relating theories on a circle of the  same radius, but different dimensionless couplings. This discussion extends to any theory with dimensionful couplings, as long as they are scalar, and the modular covariance properties follow from the transformation of these couplings under Weyl rescalings of the metric.  
%
%The modular transformations now connect different theories (with different values of $\lambda$), as expected from a QFT with dimensionful couplings. The torus partition function has the following modular property
%\begin{align}
%Z\bigg(\frac{a\tau + b}{c\tau+d},\frac{a\bar{\tau}+b}{c\bar{\tau}+d},\frac{\lambda}{|c\tau+d|^2}\bigg)&=Z(\tau,\bar{\tau},\lambda)
%\end{align} {\color{ForestGreen}(expansion in $\lambda$, modular forms?)}
% This sort of relation is expected to generalize to $T\bar T$ deformations of  other UV-complete QFTs, provided the transformation of all dimensionless couplings is properly taken into account. 

%\be
%Z= \sum e^{-\b E(\mu, R) + i \theta P }
%\ee

%In particular, the existence of the Wick rotation implies the partition function obeys

%\be
%Z(\b,R,\mu) = Z(R,\b,\mu)
%\ee

\subsubsection*{$J\bar T$ - deformed QFTs}

The  $J\bar T$ deformation, as well as all Smirnov-Zamolodchikov deformations involving $U(1)$ currents and the stress tensor, can be treated in an  entirely analogous manner. The only new element is that now the coupling has non-trivial transformation properties under diffeomorphisms, which need to be taken into account when discussing the modular invariance properties of the partition function. 

\begin{figure}[t]
    \centering
    \begin{minipage}{.45\textwidth}
      \centering\includegraphics[width=7.5cm]{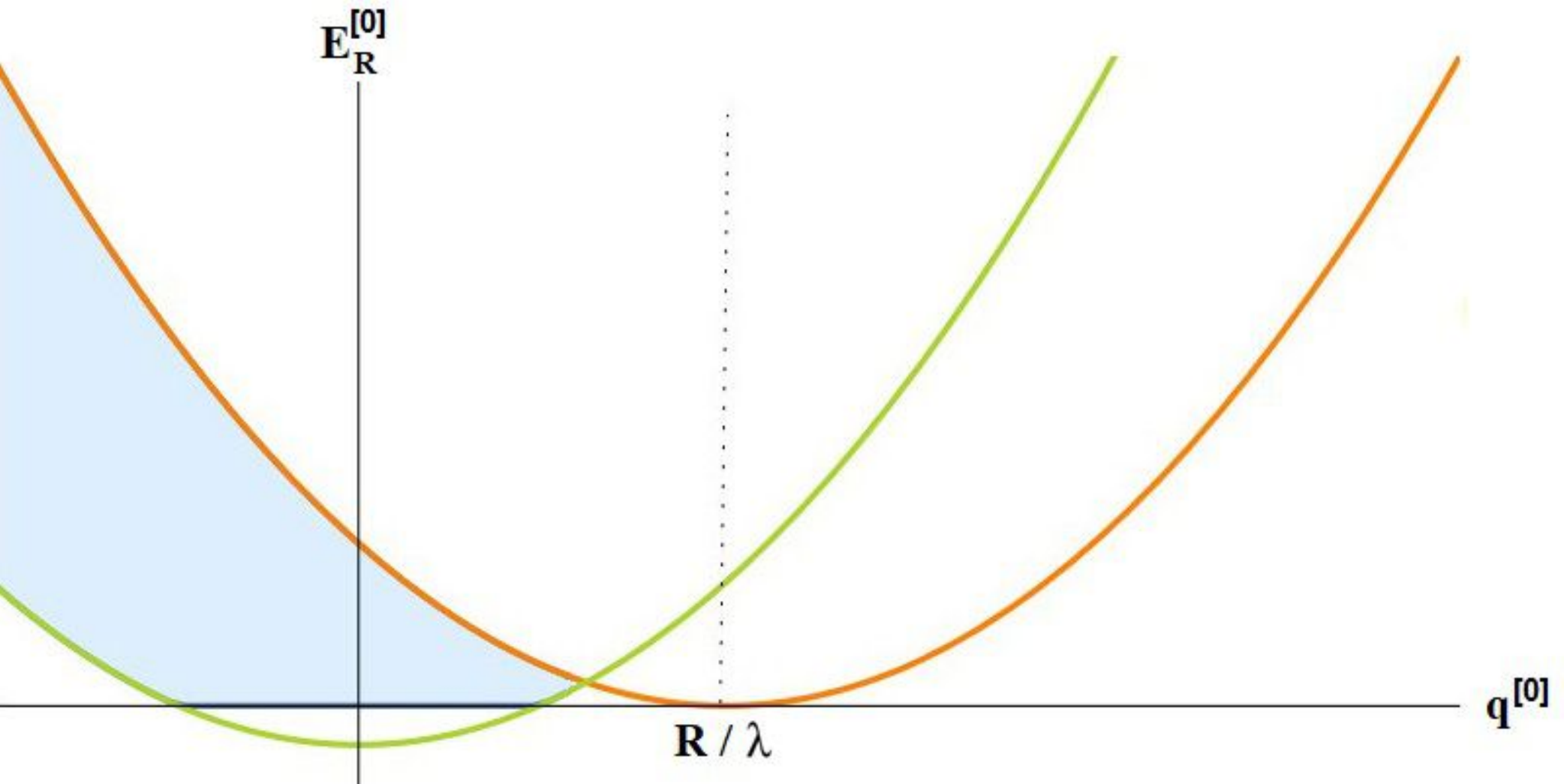}
   \subcaption{\footnotesize{The allowed region in the chiral case.}}
        \label{fig1a}
\end{minipage}
\hspace{1.2cm}
\begin{minipage}{.45\textwidth}
 \centering\includegraphics[width=7.5cm]{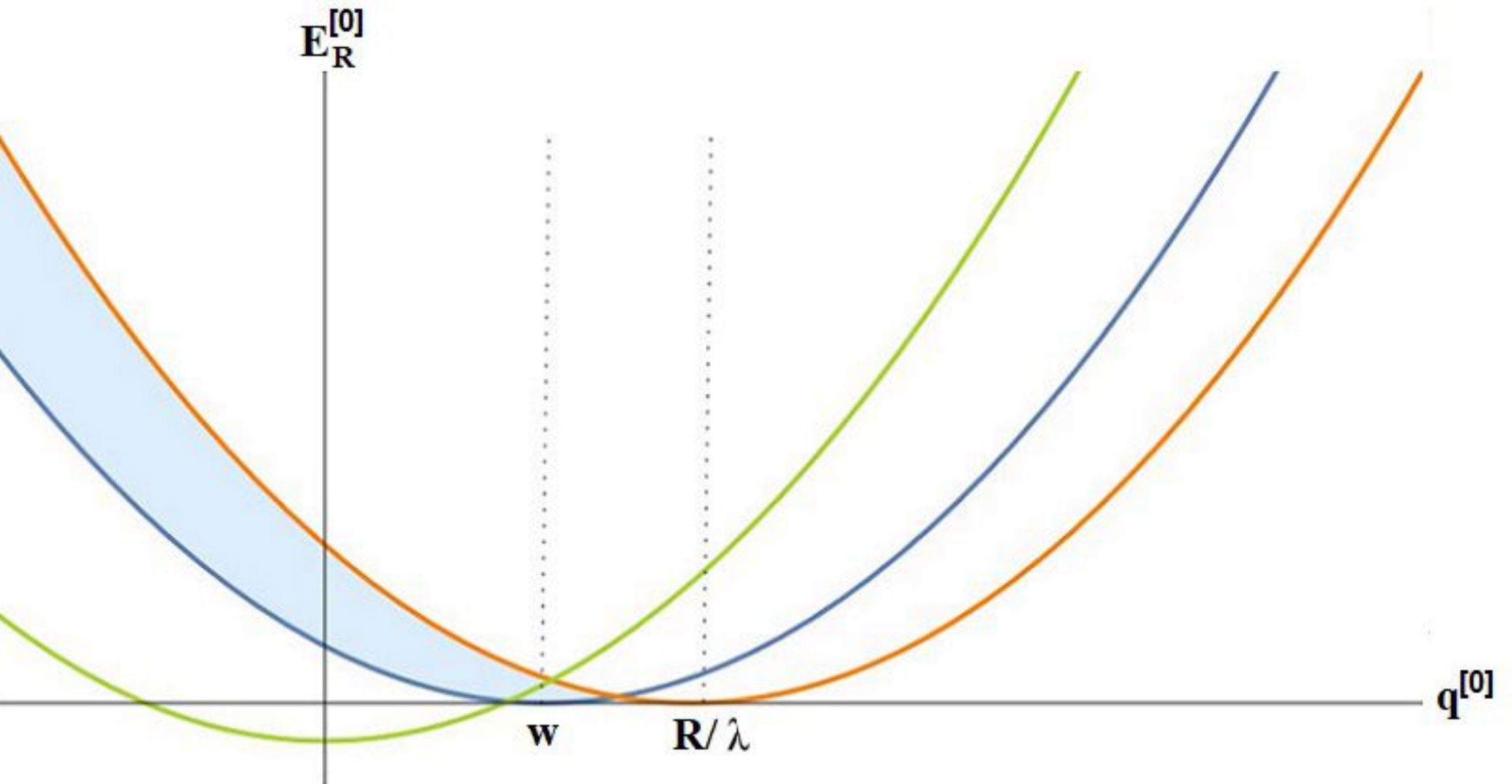} 
 \subcaption{\footnotesize{The allowed region for non-chiral $U(1)$ current. }}
      \label{fig1b}    
      \end{minipage}
    \vspace{3mm}
    \caption{\footnotesize{Range of  undeformed right-moving energies (shaded region) that lead to real energies in  $J\bar T$ - deformed CFTs and are allowed by  the  cosmic censorship bounds (green/blue parabolae). Note this range  extends to infinite energies.}}
     \label{figallowed}
\end{figure}

We start our discussion with a  general $JT^a$ deformation of  a two-dimensional QFT, defined via the flow equation 
\be
\p_{\l^a} S = \int d^2 x \,  T^{\a}{}_{ a}  \e_{\a\b} J^{\b}
\ee
The coupling parameters $\l^a$ are vectors with dimensions of length; these deformations thus break Lorentz invariance. The spectrum of the deformed QFT coupled to certain background fields is again simply related to the undeformed spectrum in a shifted  background   \cite{Frolov:2019xzi,Anous:2019osb}

\be
E_n^{[\l^a]}( R, v, a^\s) = E^{[0]}_n \left( R-\l^\s q_n^{[0]}, \frac{v R - \l^t q_n^{[0]}}{R-\l^\s q_n^{[0]}}, a^\s + \frac{\l^\s (P_n+v E_n^{[\l^a]})-\l^t E_n^{[\l^a]}}{R-\l^\s q_n^{[0]}}\right) \label{jtadefeng}
\ee
%which is obtained by coupling the deformed QFT to a 
where $v$ is a background vielbein, $q^{[0]}_n$ is the undeformed $U(1)$ charge of the state, and  $a^\s$ is a background gauge field, 
which may be set to zero at the end of the computation. 
%
 %where we may set the final background fields $a^a, v$ to zero at the end. 
%
The $J\bar T$ deformation corresponds to the case when $\l^a$ is a null vector with $\l^t = \l^\s =  \l$ (or, equivalently,  $\l^{\bar z} = 2 \l, \l^z=0$). If the seed theory is a CFT, then the $J\bar T$ deformation has the special property of preserving   locality and conformal invariance on the left-moving side, which leads to great simplifications in the study of the deformed theory.  The deformed spectrum is obtained by applying \eqref{jtadefeng} to a seed CFT, and is best expressed  in terms of the deformed right-moving energy 
%
%
%
%%For $J\bar T$, one either works with a chiral current and then uses spectral flow to fix the flow equation for the corresponding charge or, better, one couples to background fields and solves that system instead.
%
%The flow equations for the energies and chiral $U(1)$ charges are (cylinder $x\sim x+R$, which convention to keep?):
%\begin{align}
%\frac{\partial E(\alpha,R)}{\partial\alpha}&=-Q\bigg(\frac{\partial E}{\partial R}+\frac{P}{R}\bigg) \hspace{1cm} \frac{\partial Q(\alpha,R)}{\partial\alpha}=-\frac{k}{4\pi}R\frac{\partial E_R}{\partial R}
%\end{align}
%where $k$ is the level of the $U(1)$ Kac-Moody algebra. It follows that:
%\begin{align}
%E_R-\frac{2\pi Q^2}{k R}=const \hspace{1cm}Q=Q_0 + \frac{\mu k}{4\pi}E_R
%\end{align}
%
%
%For a $J\bar T$ - deformed CFT, the solution particularizes to 
%
\be
E_R^{[\l]} (R) = \frac{4\pi}{\l^2 k}\left( R-\l q^{[0]} -\sqrt{\left(R-\l q^{[0]}\right)^2-\frac{\l^2 k E_R^{[0]} R}{2\pi}} \right)\,% ,\hspace{0.9cm}E_{L,R}\equiv\frac{E\pm P}{2} 
\label{defengjtb}
\ee
where  $E_R^{[0]}, q^{[0]}$ are the right-moving energies and $U(1)$ charges of the corresponding state in the undeformed CFT, and we have dropped the label `$n$' on  the eigenstates. The left-moving $U(1)$ charge also changes non-trivially with $\l$, and is given by 
%
%{\color{ForestGreen}\emph{Change:}
%The right-moving energy satisfies:
%\begin{align}
%E_R&=E_R^{(0)}+\frac{\lambda Q_0}{R} E_R +\frac{\lambda^2 k}{8\pi R}E_R^2
%\end{align}
%The charge changes:
\be \label{chargeflow}
q^{[\l]}=q^{[0]}+\frac{\lambda k}{4\pi}E_R^{[\l]}=\frac{1}{\lambda}\bigg(R-\sqrt{\left(R-\lambda q^{[0]}\right)^2-\frac{\lambda^2 k R E_R^{[0]}}{2\pi}}\bigg)
\ee
Note that the deformed spectrum will become imaginary if the states in the undeformed CFT have large right-moving energy at fixed $q^{[0]}$, a behaviour that resembles that of $T\bar T$-deformed CFTs with $\mu <0$. At the same time, reality of the deformed energy results into an upper bound\footnote{One may increase $q^{[\l]}$ beyond the limiting value $R/\l$ by choosing a different branch of the square root. A similar behaviour was found for $T\bar T$ - deformed CFTs with $\mu <0$ \cite{Guica:2019nzm}. } on $q^{[\l]}$, suggesting it is the latter that should be held fixed as $E_R^{[0]}$ is taken to be large. The relationship between the deformed and undeformed right-moving energies at fixed $q^{[\l]}$ is given by %\textbf{\emph{Check!}} {\color{ForestGreen}(I checked)}

\be
E_R^{[\l]} =  \frac{4\pi}{\l^2 k}\left( \sqrt{\left(R-\l q^{[\l]}\right)^2+\frac{\l^2 k E_R^{[0]} R}{2\pi}} - \left(R-\l q^{[\l]} \right)\right) \label{ERql}
\ee 
From \eqref{chargeflow},
 $q^{[\l]}$ will be real provided
the undeformed dimensionless energies $ R E_R^{[0]}/2\pi$  lie below the parabola $(R/\l-q^{[0]})^2/k$ depicted in figure \ref{figallowed}, which still allows access to infinite energies. In addition, there are lower bounds on the allowed energy.  For example, if the seed CFT also posesses a right-moving $U(1)$ symmetry,  the cosmic censorship bound on the right-moving side $R E_R^{[0]}/2\pi \geq  (\bar q^{[0]})^2/k$ indicates it should lie above the parabola  $ (q^{[0]}-\mathrm{w})^2/k $, where $\mathrm{w} \equiv q^{[0]}-\bar q^{[0]}$ is the winding of the state, which is to be held fixed as we vary $q^{[0]}$.  As long as  $R> \l\mathrm{w}$ (for $\l >0$), there is always a sliver in the $E_R^{[0]}$, $q^{[0]}$ plane so that both conditions are satisfied.  % given by \ref{defengjtb} is large. 

%
% Reality of the deformed energies implies $E_R^{(0)}$ should be below the parabola $(R/\l-Q_0)^2$, whereas unitarity \emph{Correct?} tells us it should be above the parabola $\bar Q_0^2= (Q_0-w)^2$, where we assume $w$ %
% is held fixed

%$E^{[0]}{}_{\! R}$

The allowed values of $E_R^{[0]}$ are further restricted by the cosmic censorship bound on the left-moving energy and charge, which requires that $R E_R^{[0]}/2\pi \geq (q^{[0]})^2/k -  P R/2\pi$. It is easy to check that for $\mathrm{w} < R/\l$, there is always a region, depicted in figure \ref{fig1b},  that extends to infinite energies and obeys all three constraints. If the $U(1)$ current is chiral, then the second constraint is replaced by positivity of the energy%{\color{ForestGreen}shifted by $c/24$}
, and we are in the situation of figure \ref{fig1a}.  
%
% For $\mathrm{w}>0$, the slope of the green parabola is smaller than the slope of the blue one, so this constraint will only affect the region in the $E_R^{[0]}>0$, $q^{[0]}>0$ quadrant. Hence, for any value of the momentum, a sliver extending to $E_R \r \infty$ will always exist for negative values of $q^{[0]}$.
  %
  Within the  allowed region, we will be interested in the regime  where $E_R^{[0]}$ is large and  $q^{[0]}$ is large and negative, which corresponds via \eqref{ERql} to a large deformed  right-moving energy.

%In any case, a sliver extending to $E_R \r \infty$ always appears to exist. \textbf{\emph{True?}}

  % \emph{Draw proper plot and colour allowed region!} {\color{ForestGreen}(also momentum parabola comment)} see 

%\begin{figure}[h]
%    \centering
%  \includegraphics[width=7.5cm]{JTballowed1.pdf}
%  \caption{\footnotesize{
%  }}
%  \label{parabolaJT}
%\end{figure}

 The full deformed spectrum may be understood as a spectral flow by the right-moving Hamiltonian, as discussed in \cite{Guica:2020eab}. This observation also extends to the spectrum of  $SL(2,\mathbb{R})_L$ conformal dimensions on the plane, which in $J\bar T$ - deformed CFTs are well-defined thanks to the fact that the theory enjoys full left conformal invariance. This spectrum may be obtained by applying an infinite boost to \eqref{defengjtb}, and reads, as a function of the right-moving energy, now denoted $\bar p$ \cite{Guica:2019vnb}

\be
h^{[\l]}(\bar p)  =  h^{[0]} +  \frac{\l}{2\pi } q^{[0]} \bar p + \frac{\l^2 k}{16 \pi^2} \bar p^2 \;, \;\;\;\;\;\;\; q^{[\l]}(\bar p)  =  q^{[0]} + \frac{\l k}{4\pi} \bar p \label{jtbardefdim}
\ee
where $h^{[0]},  q^{[0]}$ are the left-moving conformal dimension and charge in the undeformed CFT. These dimensions can also be obtained  via conformal perturbation theory \cite{Guica:2019vnb}. Note this spectrum is manifestly real, indicating that the problems associated with the imaginary energy states disappear in infinite volume,
in agreement with their physical interpretation put forth in \cite{Cooper:2013ffa}. % for the imaginary energy states in finite size. 

%\emph{Discuss spectral flow?}. }
The torus partition function of a $J\bar T$ - deformed CFT is given by 
\begin{align}
Z_{J\bar{T}}\left(\tau,\bar{\tau},\nu, \frac{\l}{R}\right)&=\sum_{n} e^{- \tau_2 R E_n^{[\l]}(R)+i \tau_1 R P_n +2\pi i \nu q_n^{[\l]}(R)}
\end{align}
where $\nu$ is the chemical potential that couples to the chiral $U(1)$ current. We  wrote the coupling $\l$ as an argument - rather than a label - because it changes under diffeomorphisms, due to its vectorial nature.  Since  imaginary energy modes are present for any value of the radius, it is currently not well understood to what extent this partition function is well defined; however, the fact that the theory admits a non-perturbative definition \cite{Anous:2019osb} yields hope that its study is meaningful.

 The modular transformation properties of this partition function were discussed in \cite{Aharony:2018ics}. Since $\l^a$ transforms as a vector, $\l^{\bar z}$, under modular transformations, which has the same transformation properties as $R \bar z$, it follows that the dimensionless combination $\l/R$ transforms exactly as $\bar z$
\begin{align}
\frac{\lambda}{R}\mapsto\frac{\lambda}{R(c\bar{\tau}+d)}
\end{align}
Consequently, the dimensionful deformation parameter changes $\lambda\mapsto \frac{\lambda |c\tau+d|}{(c\bar{\tau}+d)}$. A similar argument can be used to derive the well-known transformation properties of the chemical potential\footnote{The chemical potential is related to a background gauge field, $a^z$,  that couples to the left current  as  $\nu = \beta a^{z} = \tau_2 R a^z$. Invariance of the action under diffeomorphisms implies that $a^z$ transforms in the opposite way from $R \bar z$ which, using \eqref{transfzzb}, leads to $ R a^z \mapsto (c \bar \tau +d) R a^z$. Taking into account the transformation of $\tau_2$, we find the above result.  } $\nu$ 

\be
\nu \mapsto \frac{\nu}{c\tau + d} 
\ee
With this in mind, the partition function has the standard anomalous transformation under diffeomorphisms of the torus %\emph{Check factors!}
\begin{align} \label{JTbarmodprop}
Z_{J\bar{T}}\bigg(\frac{a\tau+b}{c\tau +d},\frac{a\bar{\tau}+b}{c\bar{\tau} +d},\frac{\nu}{c\tau+d},\frac{\lambda}{R(c\bar{\tau}+d)}\bigg)=\exp\bigg(\frac{2 i
\pi k c\nu^2}{c\tau+d}\bigg)Z_{J\bar{T}}\left(\tau,\bar{\tau},\nu,\frac{\lambda}{R}\right)
\end{align}
One may also consider a slightly redefined partition function, which is invariant under these transformations \cite{Kraus:2006nb}
\be \label{modinvJTbar}
Z_{J\bar T}^{inv}\left(\tau,\bar{\tau},\nu, \frac{\l}{R}\right)\equiv e^{\frac{\pi k \nu^2}{\tau_2}}Z_{J\bar T }\left(\tau,\bar{\tau},\nu, \frac{\l}{R}\right)
\ee%
This transformation law  can be readily extended to QFTs that may have various couplings that transform non-trivially under Lorentz transformations. %In particular, one would expect these conclusions to also apply to the more general $T\bar T_s$ by the currents associated with integrability, for which the flow equation was not sufficient to fix the deformed energies in terms of the undeformed one. It is interesting to note that in those cases the results of \cite{Aharony:2018bad} that modular covariance {\color{ForestGreen}(they call it invariance)} plus the assumption that the spectrum only depends on the initial energies implies that the only solution is the $T\bar T$ - deformed spectrum show that in these cases the deformed energy of a given state cannot just  depend on the initial one. % However, supposing that the finite-size spectrum of one such deformed theory is known, then our results on the symmetric product orbifold should equally appply. 

Finally, let us discuss the density of states. The entropy is again estimated 
by  using the fact that the number of states does not change in fixed units. 
%
%in terms of the undeformed energies and charges. The CFT entropy is given by Cardy's formula with the energies replaced by the spectral-flow-invariant ones. 
One may distinguish two cases: if the current in the seed CFT is not chiral, then 
%
% in fixed units. In terms of the deformed energies and charges, one finds
%For $J\bar T$ there is a max number of states in finite size, which is related to the presence of CTCs.  \emph{Write down $J\bar T$ entropy formula!}
%
\bea
S_{J\bar T} (E,q) & = & S_{Cardy}(E^{[0]},q^{[0]})=2\pi \sqrt{\frac{c}{6}\bigg(\frac{R E_L^{[0]}}{2\pi}-\frac{q^{[0] \,2}}{k}\bigg)}+2\pi \sqrt{\frac{c}{6} \left(\frac{R E_R^{[0]}}{2\pi}-\frac{\bar q^{[0]\,2}}{k}\right) } \nonumber \\
&& \hspace{0.1cm}=\; 2\pi \sqrt{\frac{c}{6}\bigg(\frac{R E_L}{2\pi}-\frac{q^2}{k}\bigg)}+2\pi \sqrt{\frac{c}{6}\bigg(\frac{(R-\l \mathrm{w}) E_R}{2\pi}-\frac{\bar q^2}{k}\bigg)}
\eea
where we assumed, as explained above, that $q \equiv q^{[\l]}$ and $\bar q\equiv \bar q^{[\l]}$ are to be fixed in the deformed theory, as well as their difference, $\mathrm{w}$. Note that the difference from the standard Cardy formula in presence of $U(1)$ charge is rather minimal. If, on the other hand, the current $J$ is chiral, then   effectively $\bar q^{[0]}=0$, and we obtain instead
%
%Since the combination $\frac{R E_L}{2\pi}-\frac{Q^2}{k}$ is invariant along the flow, one can write \emph{Re-check!}
\begin{align} \label{entropyJTbar}
S_{J\bar{T}}&=2\pi \sqrt{\frac{c}{6}\bigg(\frac{R E_L}{2\pi}-\frac{q^2}{k}\bigg)}+2\pi \sqrt{\frac{c}{12\pi} \bigg(E_R(R-\lambda q)+\frac{\lambda^2 k E_R^2}{8\pi}\bigg)}
\end{align}
Taking $E_R $ large with $q$ fixed, one finds Hagedorn behaviour at large energies. The above formula can be alternatively rewritten in terms of $q^{[0]}$ using \eqref{chargeflow}, but then the limit of  large $E_R$  with   $q^{[0]}$ fixed is problematic because the  square roots become imaginary.
%\subsection{Brief review of symmetric product orbifolds on a cylinder}

\subsection{Torus partition function of general symmetric product orbifold QFTs}

In this subsection, we  review and slightly generalize the well-known group-theoretical derivation \cite{Bantay:1998fy} of 
 %facts about symmetric product orbifold in two dimensions, in particular the 
  the torus partition function of a  symmetric product orbifold of two-dimensional QFTs. %from modular invariance, following the  approach of .  
  While this discussion is usually particularised to symmetric orbifolds of \emph{C}FTs, we  point out that the derivation only  mildly depends on the conformal property of the seed. We can thus apply this method to general two-dimensional QFTs%whose torus partition function is well-defined
, and in particular to $T\bar T$ and $J\bar T$ - deformed CFTs.

  We thus consider a  two-dimensional QFT  %[with a discrete, gapped spectrum]
   on a cylinder of circumference $R$. This will be referred to as the \emph{seed} QFT and will  be denoted as  $\mathcal{M}$. The theory obtained by taking a $N$-fold tensor product $\mathcal{M}^N$ admits a natural action of the permutation group, $S_N$. Quotienting it by the permutation group, one obtains the symmetric product orbifold theory, denoted as  $\mathcal{M}^N/S_N$. 
  
  The Hilbert space of $\mathcal{M}^N/S_N$ is organized into twisted sectors \cite{Dijkgraaf:1996xw}, labeled by the conjugacy classes, denoted $[g]$, of $S_N$
\vspace{2mm}
\be
\H (\mbox{\footnotesize{$\mathcal{M}^N/S_N$}})= \oplus_{[g]} \H^{[g]} \label{hsdecomp}
\ee
Each $S_N$ conjugacy class  is entirely specified by the lengths $(n)$ and multiplicities $(N_n)$ of the cycles of the permutation, %which are the same for all the permutations within the conjugacy class, 
with $\sum_n n N_n = N$.  
Within each conjugacy class, one keeps the states invariant under the centralizer (a.k.a commutant) of $g$, which does not depend on the chosen representative.
%is the same for all elements of $[g]$.  The %centralizer is given in (2.4) of Verlinde, so 
The resulting structure of the factors $\H^{[g]}$ is

\be
\H^{[g]} = \otimes_{n>0}  \left( \H^{\mathbb{Z}_n}\right)^{N_n}\!\!/S_{N_n} %\;, \;\;\;\;\;\; \sum_n n N_n =N
\ee
where $\H^{\mathbb{Z}_n}$ is the Hilbert space assciated with a $\mathbb{Z}_n$  cyclic orbifold of the seed QFT, % {\color{ForestGreen}, which already takes into account the projection on integer momentum}\emph{Correct? Projected onto integer momentum?}, 
and the symmetrization is performed  with respect to all cycles of the same length $n$.  The untwisted sector of this Hilbert space, which corresponds to the conjugacy class of the identity, is simply $( \H_{seed})^N/S_N$. 
The twisted sectors are characterized by the basic fields having twisted boundary conditions around the spatial cycle of the cylinder.
States belonging to different twisted sectors are orthogonal, as is clear from 
%
% Within each twisted sector, the lowest energy state is called the twisted vacuum; in a CFT, it can be thought to be obtained by the action of a twist operator on the untwisted vacuum.  The rest of the CFT states can be constructed by acting with operators on the twisted vacuum. Since different twisted vacua are orthogonal, it follows that states from different sector are orthogonal to each other. This property is also true more generally from 
the direct sum structure \eqref{hsdecomp}.

The twisted sectors can be understood by mapping to  corresponding covering spaces. %, where the basic fields  are subject to standard boundary conditions. 
 This is particularly simple to implement for the torus partition function, as the relevant covering spaces are again tori, allowing one to express the partition function of the orbifold QFT solely in terms of the seed partition function.
The first results on the torus partition function (or, rather, elliptic genus)  of a symmetric product orbifold were obtained in \cite{Dijkgraaf:1996xw} using string-theoretical methods. In a series of articles \cite{Bantay:1998fy, Bantay:1999us, Bantay:2000eq} that built upon this  work, an explicit  group-theoretical construction of  the torus partition function of a symmetric product orbifold of a generic CFT  in terms of that of the seed  was given.
Importantly, this derivation - detailed below -  does not involve conformal invariance, but only relies on the modular invariance of the seed partition function.

\subsubsection*{Review and slight generalisation of Bantay's formula}

The basic idea is the following: the partition function of the symmetric product orbifold QFT receives contributions from the different twisted sectors of the theory. Rather than considering fields with twisted boundary conditions on the original torus - denoted $\mathcal{T}^2$ - one can equivalently work with fields with standard boundary conditions on a covering space of the torus. The latter %(with no punctures) 
are
%
% This is 
% and each of them is associated to 
  unramified coverings with $N$ sheets -  not necessarily connected -  
 for which the monodromy group - which encodes how the various sheets permute as one goes around a loop in base space - is a subgroup of the permutation group, $S_N$ \cite{hatcher}. Permuting the sheets of the covering space under the monodromy action of the fundamental group corresponds to permuting the copies in the symmetric product orbifold, thus implementing the action of $S_N$ in the QFT in a geometrical fashion.

 Connected components of such covering spaces are associated to orbits of elements of the set $\{1,2,...,N\}$ under the action of $S_N$. We generically denote these orbits by\footnote{For example, for $N=5$ and the monodromy group generated by the permutations $(12)$ and $(345)$, namely $\langle(12),(345)\rangle=\{e,(12),(345),(354),(12)(345),(12)(354)\}\subset S_5$, the orbit of e.g. the element $3$ is $\xi=\{3,4,5\}$. 
 %Certain references, for example \cite{Bantay:1998fy}, {\color{ForestGreen}take into account the orbits of all elements, case in which $\{3,4,5\}$ is counted three times. This overcounting needs to be compensated by an overall factor of $1/|\xi|$, where $|\xi|$ is the number of elements in the orbit $\xi$. We will be counting each distinct orbit $\xi$ once.}
 } $\xi$. By the Riemann-Hurwitz theorem%(which restricts the possible coverings of a Riemann surface by another Riemann surface)
 , each such connected component is a torus, denoted  $\mathcal{T}^2_{\xi}$, on which the seed theory lives and which covers the base $\mathcal{T}^2$ $|\xi|$ times. %Consequently, its area is $|\xi|$ times larger than the area  of the base $\mathcal{T}^2$. This information about the K\"{a}hler structure is not needed when discussing SPOs of CFTs, where the partition function only depends on the modular parameter, but will be needed for our more general case. 

%(this part only geometrical, later permutations)

 Each covering torus $\mathcal{T}^2_{\xi}$ can be written as the quotient of the complex plane by its fundamental group $\pi_1(\mathcal{T}^2_{\xi})\cong\mathbb{Z}\oplus\mathbb{Z}$, which is a subgroup of index\footnote{The index of a subgroup $H  \subset G$ is the number,  $|G/H|$, of cosets of $H$ in G.} $|\xi|$ of the fundamental group of the base torus $\pi_1(\mathcal{T}^2)\cong\mathbb{Z}\oplus\mathbb{Z}$ or, equivalently, a sublattice of index $|\xi|$ of the lattice associated to the base torus. These subgroups are labeled by three integers
$m_{\xi},r_{\xi},\ell_{\xi}$, with $\ell_{\xi}>r_{\xi}\geq 0$ and $m_{\xi}\ell_{\xi}=|\xi|$. Given a basis of generators for  $\pi_1(\mathcal{T}^2)$, usually denoted as the $a$ and $ b$ cycles, these integers determine the generators of $\pi_1(\mathcal{T}^2_{\xi})$ - namely, the $a_{\xi},b_{\xi}$ cycles - as
\begin{align}\label{loopcov}
a_{\xi}&=\ell_{\xi} a \hspace{2cm} b_{\xi}=r_{\xi} a + m_{\xi} b 
\end{align}
Hence, the modular parameters of the covering tori can be written as
\begin{align}\label{complexstr}
\tau_{\xi}&=\frac{m_{\xi}\tau+r_{\xi}}{\ell_{\xi}}
\end{align}
In addition, if the length of the $a$-cycle on the base torus is $R$, it follows that the  length of the $a_\xi$-cycle on the covering torus $\mathcal{T}^2_{\xi}$ is
\be \label{radius}
R_\xi = \ell_\xi R
\ee
An explicit example of the covering tori can be found in the pedagogical exposition of \cite{Haehl:2014yla}. The area of the covering torus is $R_\xi^2 (\tau_\xi)_2 = \ell_\xi m_\xi R^2  \tau_2 = |\xi| R^2 \tau_2$, in agreement with the fact that it covers the base torus $|\xi|$ times. 
%\emph{Drawing or refer to Mukund!}
%
%With this parametrization/in this basis \ref{complexstr}, $R_{\xi}$ is the length of the $a_{\xi}$-cycle of the covering torus $\mathcal{T}^2_{\xi}$, $\ell_{\xi}$ times larger than the length of the $a$ cycle of the base torus(cite section 2.1, maybe drawing there?). 
This size information will be important when discussing the partition function of
a non-conformal QFT, such as  $T\bar{T}$-deformed CFTs, whose partition function depends explicitly on the length, $R$, of the $a$-cycle. 

% {\color{ForestGreen} on the cylinder which is obtained by cutting the torus along the a-cycle, whose length thus becomes the circumference of the cylinder }. \emph{Keep?}
%
%\emph{Out?} {\color{ForestGreen}(I would keep it)}[
Note that in the above, we have made  a specific choice of parametrization, i.e. choice of basis of the generators of the fundamental group of the base and the covering tori. % We will discuss in the end what happens under a change of basis, namely under the automorphisms of the fundamental group, the modular transformations $SL(2,\mathbb{Z})$.} \emph{Keep?}
However, this choice should be immaterial as long as the quantities we compute are modular invariant. % (or  covariant at the worst). We return to this point later in this section. ] 

Let us now reformulate these geometric data in group-theoretical language. The covering spaces discussed above are in one-to-one relation with the homomorphisms $\phi:\pi_1(\mathcal{T}^2)\rightarrow S_N$.
Using %the one-to-one correspondence between the covering spaces that we consider and homomorphisms $\phi:\pi_1(\mathcal{T}^2)\rightarrow S_N$, 
this correspondence, one can rewrite the covering space data in terms of permutations. Any such homomorphism is fully specified by two commuting permutations, $\phi(a),\phi(b)\in S_N$ that generate $\phi(\pi_1(\mathcal{T}^2))\subset S_N$, corresponding to the choice of the two loops that generate $\pi_1(\mathcal{T}^2)$. From the  perspective of the QFT on the base $\mathcal{T}^2$, $\phi(a)$ and $\phi(b)$ correspond to the monodromies acquired by the fields as they circle around the $a$ and $b$-cycle. The covering tori $\mathcal{T}^2_{\xi}$ correspond to the orbits $\xi$ under the action of $\phi(\pi_1(\mathcal{T}^2))\subset S_N$. Each covering torus is determined by its fundamental group which, as explained in \cite{Bantay:1998fy}, is isomorphic to the stabilizer associated to the orbit

\be
S_{\xi}=\{\phi(x)\in \phi(\pi_1(\mathcal{T}^2))\hspace{0.1cm}|\hspace{0.1cm}\phi(x)\xi^*=\xi^*,\forall\xi^*\in\xi\}\cong \pi_1(\mathcal{T}_{\xi}^2)
\ee
 Intuitively,  the elements of the stabilizer act by definition as identity on $\xi$, thus mapping each sheet of the associated covering space into itself. Under this trivial monodromy action, all loops in the base space are lifted to loops in the covering space, i.e. elements of the fundamental group of the covering. In particular, for the choice \eqref{loopcov}, the generators of $\pi_{1}(\mathcal{T}^2_{\xi})$ are mapped by this isomorphism into %{\color{blue}[generators of $S_{\xi}$. \emph{Careful!} For the choice \eqref{loopcov}, these are given by]}
%
%\emph{Obvious they are given by this? E.g.: for the particluar case of $S_N$ and its orbits, they are: 
\vspace{1mm}
\be
\phi(a)^{\ell_{\xi}} \text{ and }\phi(a)^{r_{\xi}}\phi(b)^{m_{\xi}}
\ee
providing a group-theoretical interpretation of the integers $m_{\xi},r_{\xi},\ell_{\xi}$ that determine the complex structure of the covering tori: $m_{\xi}$ is the number of $\phi(a)$ orbits in $\xi$, $\ell_{\xi}$ is their common length and $r_{\xi}$ is the smallest nonnegative integer such that $\phi(a)^{r_{\xi}}\phi(b)^{m_{\xi}}$ belongs to the stabilizer\footnote{%(Thus, when we want later to restrict at single cycle length $ w$ sectors, we put $\ell_{\xi}= w$ so $m_{\xi}=1$)
%
%{\color{blue}
%[Let us relate these integers to the corresponding orbit $\xi$. A choice of generators of $\pi_1(T_{\xi}^2)$, $a_{\xi},b_{\xi}$ gives a choice of generators of $S_{\xi}$, $\phi(a)^{\lambda_{\xi}},\phi(a)^{r_{\xi}}\phi(b)^{\mu_{\xi}}$, where $\mu_{\xi}$ is the number of $\phi(a)$ orbits in $\xi$, $\lambda_{\xi}$ is their common length and $r_{\xi}$ is the smallest nonnegative integer such that $\phi(a)^{r_{\xi}}\phi(b)^{\mu_{\xi}}$ belongs to the stabilizer.
%
Let us give an example: in $S_5$, we consider again the covering space associated to the homomorphism $a\mapsto \phi(a)= (345),b\mapsto \phi(b)=(12)$. Clearly, it has two connected components. We first consider the one associated to the orbit $\xi=\{3,4,5\}$, that should give a covering space with $|\xi|=3$ sheets. The corresponding stabilizer is $S_{\xi}=\{e,(12)\}\cong \mathbb{Z}_2$. The number of $\phi(a)$ orbits in $\xi$ is 1 and its length is 3, which means $m_{\xi}=1,\ell_{\xi}=3$ and $r_{\xi}=0$, leading to a modular parameter $\tau_{\xi}=\tau/3$ for the covering torus. Note that $\phi(a)^{\ell_{\xi}}=(345)^3=e,\phi(a)^{r_{\xi}}\phi(b)^{m_{\xi}}=(12)$ are the elements of $S_{\xi}$. Similarly, for the orbit $\xi=\{1,2\}$, the stabilizer is $\{e,(345),(354)\}$. The number of $\phi(a)$ orbits is $m_{\xi}$=2 and their common length is $\ell_{\xi}=1$ because $\phi(a)$ acts as identity on 1 and 2; again $r_{\xi}=0$, implying $\tau_{\xi}=2\tau$. }. See e.g. \cite{Haehl:2014yla} for more examples.

%Thus, using the isomorphism, $\tau_{\xi}$ is completely determined by $S_{\xi}$. The important remark is that the isomorphism is unique up to modular transformations ($\tau_{\xi}$ and $\frac{p\tau_{\xi}+q}{c\tau_{\xi}+d}$ give the same complex structure), so the  physical quantities (e.g., the partition function) computed with a certain choice need to be modular invariant in order to be well-defined (in other words they should not depend on our choice of generators of the fundamental group).

Putting everything together, one can express the partition function of the symmetric product orbifold on a torus with modular parameter $\tau$ and length of the $a$-cycle $R$ in terms of the seed partition function on tori of different modular parameters \eqref{complexstr} and radii \eqref{radius} as \cite{Bantay:1998fy}
\begin{align} \label{bantay1}
Z^{S_N}(\tau,\bar{\tau},R)&=\frac{1}{N!}\sum_{\phi:\pi_1(\mathcal{T}^2)\rightarrow S_N}\prod_{\xi \text{ orbit}} Z^{seed}(\tau_{\xi},\bar{\tau}_{\xi},R_\xi)
\end{align}
where  
%$\tau_\xi$ and $R_\xi$ are given in \eqref{complexstr}, and 
we suppressed for now the possible dependence of the partition function on other parameters. % than the modular parameter of the corresponding tori {\color{ForestGreen} and the characteristic length}. T
This should be sufficient for constructing the partition function of a symmetric product orbifold of arbitrary Lorentz-invariant QFTs. The Lorentz-breaking case will be discussed when treating symmetric orbifolds of $J\bar T$ - deformed CFTs. 

This formula can be further massaged by considering the %grand canonical partition function/ 
generating function %\emph{Notation changed!}
\begin{align}\label{generating}
\sum_{N=0}^{\infty}p^N Z^{S_N}(\tau,\bar{\tau},R)=\exp{\sum_{n=1}^{\infty}p^n \mathcal{Z}^{(n)}},\hspace{0.5cm}\mathcal{Z}^{(n)}=\frac{1}{n}\!\!\!\!\!\!\!\sum_{\;\;\;\;\;\;\;\left.\T^2_\xi \right|_{ |\xi|=n} } \!\!\!\!\!\!\!Z^{seed}(\T^2_\xi)
\end{align}
where the sum in $\mathcal{Z}^{(n)}$ runs over connected covering space of $\T^2$ with $n$ sheets, which are the tori $\T^2_\xi$ discussed previously, with $|\xi| = n$. Collecting the coefficient of $p^N$ on the right-hand-side of the first sum  in\eqref{generating}, the formula \eqref{bantay1} for the partition function of the $\mathcal{M}^N/S_N$ orbifold theory can be written more compactly as %\emph{Have you checked this?} %. Extracting the coefficient of $p^N$ from the right-hand-side of \eqref{generating}, one obtains the following  formula for the partition function of the symmetric product orbifold \emph{Factors!}
\begin{align}\label{bantay2}
Z^{S_N}(\tau,\bar{\tau},R)=\frac{1}{N!}\sum_{x\in S_N}\prod_{\xi\text{ orbit }}|\xi| \mathcal{Z}^{(|\xi|)}
\end{align}
As noted in \cite{Bantay:2000eq}, in the  above formula the sum runs over all permutations in $S_N$, which is much simpler to handle than the previous sum  \eqref{bantay1} over homomorphisms, namely over pairs of commuting permutations in $S_N$. Since twisted sectors correspond to permutations up to conjugation, \eqref{bantay2} provides an easy way to read off  the contributions of the different sectors.

The individual contributions $\mathcal{Z}^{(n)}$ are given explicitly by
%and we considered also possible $R$ dependence, but suppressed other possible parameters. Thus, we can write 
\begin{align} \label{Znsum}
\mathcal{Z}^{(n)}=\frac{1}{n}\sum_{\ell|n}\sum_{0\leq r<\ell} Z^{seed}\bigg(\frac{n\tau}{\ell^2} +\frac{r}{\ell},\frac{n\bar{\tau}}{\ell^2} +\frac{r}{\ell}, R \ell\bigg) 
\end{align}
The above formula gives the contribution of all  sets of equal-length cycles  whose lengths sum to $n$, where $\ell$ is the length of the cycles and $n/\ell$ gives the number of cycles of that length. Choosing $\ell=1$, we obtain the contribution to this term of the states from the untwisted sector (in the form of $n$ identical copies of the same state in the seed QFT), 
%
% i.e. $n$ cycles of length 1,
  while choosing $\ell=n$ we obtain the contribution of the twisted sector of a single cycle of length $n$.

\subsubsection*{Comments on modular invariance}

As we already discussed, well-definiteness of the torus partition function of the seed QFT requires it to be modular invariant  in the generalised sense we reviewed. We would now like to show that modular invariance  of the symmetric orbifold partition function \eqref{bantay2} automatically follows from that of the seed. 
%
%Let us come back to the modular properties of this torus partition function. 

%The derivation of the Bantay formula uses a certain choice of parametrization, %namely basis of generators of the fundamental group of the tori. {\color{ForestGreen}The choice of basis for the base torus determines the basis for the covering tori. Since the complex structure is defined only up to modular transformations, which change this basis, the partition function of the symmetric product orbifold on the base torus and the partition function of the seed on the covering tori should be ``modular invariant", possibly in the generalised sense explained next.} In other words, we should be able to put the theories on the corresponding tori. Let us first focus on the modular properties of the seed theory. 

In the simplest case of a CFT seed, the partition function depends only on the modular parameter of the torus. %and it is modular invariant. When the seed partition function depends also on $R$, or more precisely on scalar couplings, the ``modular invariance" property of the seed is
It is then not hard to recognise the `connected' $\mathcal{Z}^{(n)}$ contribution as the action of the $n^{th}$ Hecke operator, $T_n$, on the seed partition function

\be \label{heckeac}
\mathcal{Z}^{(n)}_{\scriptscriptstyle{CFT}}=T_n Z^{seed}_{\scriptscriptstyle{CFT}}%(\tau,\bar{\tau})
\ee 
%
%Let us now turn to the modular properties of the symmetric product and explain how they follow from those of the seed theory, at least in the cases discussed in this article. For this purpose, we use
where, by definition, the action of a   Hecke operator  $T_n$   on a modular form of weight $(\kappa,\bar \kappa)$ %\footnote{{\color{ForestGreen}This is the notation widely used for the weight and should not be mistaken for the level of the $U(1)$ KM algebra.}} 
is
 \begin{align}
T_n f (\tau,\bar \tau)=\frac{1}{n} \sum_{\substack{r,\ell\in\mathbb{Z},\hspace{0.1cm}\ell|n\\0\leq r<\ell}} \frac{1}{\ell^{\kappa+\bar \kappa}} f\bigg(\frac{n\tau}{\ell^2} +\frac{r}{\ell},\frac{n\bar{\tau}}{\ell^2} +\frac{r}{\ell}\bigg)
\end{align}
and produces another modular form of the same weight. The seed partition function is modular invariant,  i.e. it  is simply a modular form of weight zero. Then,  \eqref{heckeac} implies that $\mathcal{Z}^{(n)}_{\scriptscriptstyle{CFT}}$, and thus the full symmetric orbifold partition function, is also modular invariant.
%
%If the seed partition function depends only on $\tau,\bar{\tau}$ and is modular invariant i.e. $k=\bar{k}=0$, as in the case of a CFT, we can rewrite the Bantay formula directly in terms of Hecke operators, recognizing 
%
%{\color{blue}[Since Hecke operators map modular functions to modular functions, we see explicitly that the modular invariance of the symmetric orbifold partition function follows from that of the seed CFT. ]}

More generally, if the theory only possesses scalar dimensionful couplings, the partition function would depend on the dimensionless combinations built from these couplings and $R$. If this partition function allows for a Taylor expansion in terms of this dimensionless coupling, as is the case for e.g. $T\bar T$-deformed CFTs 
\begin{align}\label{connectedcom}
Z^{seed}_{T\bar T}\left(\tau,\bar{\tau},\frac{\mu}{R^2}\right)=\sum_{\kappa=0}^\infty \left(\frac{\mu}{R^2}\right)^{\kappa} Z_{\kappa}(\tau,\bar{\tau})
\end{align}
%Therefore, for each $\ell$, the sum over $k$ can be written as $\ell^k$ times that dimensionless coefficient, resulting into an effective $R\ell$ factor.
then, as already discussed in \cite{Aharony:2018bad}, the coefficients of this expansion are all modular forms of weight $(\kappa,\kappa)$. Acting with the $n^{th}$ Hecke operator on each term and then resumming yields precisely \eqref{Znsum} 
\begin{align}
T_n Z^{seed}_{T\bar T}\left(\tau,\bar{\tau},\frac{\mu}{R^2}\right)&:=\frac{1}{n}\sum_{\kappa} \left(\frac{\mu}{R^2}\right)^{\kappa}\!\!\! \sum_{\substack{r,\ell\in\mathbb{Z},  \ell|n\\0\leq r<\ell}} \frac{1}{\ell^{2\kappa}} Z_{\kappa}\bigg(\frac{n\tau}{\ell^2} +\frac{r}{\ell},\frac{n\bar{\tau}}{\ell^2} +\frac{r}{\ell}\bigg)=\nonumber \\
&=\frac{1}{n}\!\! \sum_{\substack{r,\ell\in\mathbb{Z}, \ell|n\\0\leq r<\ell}}  \!\!\!\! Z_{T\bar T}^{seed}\bigg(\frac{n\tau}{\ell^2} +\frac{r}{\ell},\frac{n\bar{\tau}}{\ell^2} +\frac{r}{\ell}, \frac{\mu}{R^2 \ell^2}\bigg)
\end{align}
\vskip-3mm
\noindent Hence, we can again write $\mathcal{Z}^{(n)}=T_n Z^{seed}(\tau,\bar{\tau},R)$, where the action of the Hecke operator on the full partition function is defined as the action on each coefficient  in the series expansion in $\mu/R^2$. The modular invariance of the seed then implies the modular invariance of the symmetric product. The same reasoning applies when the couplings have non-trivial transformation properties; examples will be given in the following section, where we will be discussing in detail the case of $J\bar T$-deformed CFTs with a chemical potential. %, where Jacobi, rather than modular forms, appear.

  $T\bar T$ - deformed CFTs are, in a certain sense, the next simplest case  to consider beyond just CFT, since the fact that the coupling $\mu$ has a negative mass dimension allows for an expansion in terms of standard modular forms of positive weight. More generally, there is no reason to expect that the partition function would be analytic in the given coupling, and therefore the above  argument using the Taylor expansion would not hold. It is, nevertheless, possible to argue for modular invariance of $\mathcal{Z}^{(n)}$
 directly from the modular properties of the seed partition function: the $T$ transformation $\tau \r \tau+1$ of the base QFT simply reshuffles the terms in the sum \eqref{connectedcom}, whereas the $S$ transformation $\tau\r -1/\tau$ can be undone by a modular transformation of the covering tori, together with a reshuffling of the terms in the sum, as argued in \cite{Bantay:1997ek} for the CFT case. Including the radius dependence is straightforward\footnote{ The $S$ transformation on the base torus maps $\tau_{\xi}\mapsto\tau'_{\xi}=(r_{\xi}\tau-m_{\xi})/(\ell_{\xi}\tau)$ and $R_{\xi}\mapsto R'_{\xi}=|\tau|\ell_{\xi}R$. Using equation (12) of \cite{Bantay:1997ek}, one can show that $\tau'_{\xi},R'_{\xi}$ are related by a modular transformation to $\tilde{\tau}_{\xi}=(m^*_{\xi}\tau-r^*_{\xi})/\ell^*_{\xi},\tilde{R}_{\xi}=\ell^*_{\xi} R$, where $\{\ell^*_{\xi},m^*_{\xi},r^*_{\xi}\}$ are integers with $\ell^*_{\xi}m^*_{\xi}=|\xi|$ and $0\leq r^*_{\xi}<\ell^*_{\xi}$, that parametrize the orbit $\xi$, just like $\{\ell_{\xi},m_{\xi}, r_{\xi}\}$. The relation between the two parametrization is explained in \cite{Bantay:1997ek}. Using the modular invariance of the seed partition function, it follows that the sum in \eqref{Znsum}  is invariant under the $S$ transformation of the base torus.}.

\subsubsection*{General features of the spectrum of symmetric product orbifolds }

The spectrum of the symmetric product orbifold can be readily extracted from the partition function \eqref{bantay2}, by giving it a Hilbert space interpretation %\emph{Notation! Look back at Wei.}

\be
Z^{S_N} (\tau,\bar \tau, R) =  \sum_n d_n e^{- \beta E_n + i P_n \theta}  \;, \;\;\;\;\; \b = R \tau_2 \;, \;\; \theta = R \tau_1
\ee
where we have now explicitly included the ventual degeneracies, $d_n$, of the energy levels. We would like to express the finite-size energies and momenta  $E_n, P_n$ of the symmetric product orbifold, as well as $d_n$, in terms of those of the seed QFT, denoted $E_n^{(s)},P_n^{(s)}, d_n^{(s)}$

\be
Z^{seed}  (\tau,\bar \tau, R) = \sum_n d_{n}^{(s)} e^{- \beta E_{n}^{(s)}  + i P_{n}^{(s)} \theta} 
\ee
As a warm-up, it is useful to first work out the contribution to the partition function  of  the twisted sector associated to a single cycle of length $w$, which we will refer to as the $w$-twisted sector. %[one twisted sector of length $ w$ which] 
It corresponds to the $\ell=w$ contribution to $\mathcal{Z}^{(w)}$ and will be denoted % \emph{Notation! Precise definition of twisted sector?}
 $Z^{(w)}$

\begin{align}
Z^{ (w)} \equiv \left. \mathcal{Z}^{( w)} (\tau,\bar{\tau},R) \right|_{\ell= w} = \frac{1}{ w} \sum_{0\leq   r <  w} Z_{seed} \left(\frac{\tau+ r}{ w}, \frac{\bar{\tau}+ r}{ w}, R  w\right) \label{wtwz}
\end{align}
%Writing the seed partition function as a Boltzmann sum, from which we read the energies and momenta of this sector on a cylinder of radius $R$ (we can also add chemical potential and read off the charges)
%
One immediately notes that in this sector, the contributions of the seed partition are evaluated at the same inverse temperature,  $\b = R \tau_2$, as that of  the full orbifold, even though the length of the spatial circle  
%
% is the same as in the untwisted {\color{ForestGreen}(seed?)} one, and thus the  energies that appear are the same as the seed energies, but
 %
 is  $ w$ times larger. This implies that 
 
\be \label{energies}
E_n^{( w)} (R) = E_n^{(s)} ( Rw)
\ee
where $E_n^{( w)}(R)$ represent the finite-size energy levels in the $ w$ - twisted sector. Note this result follows without any use of conformal invariance, but only of the modular invariance properties of the partition function we have been assuming throughout this section.  For the case of a CFT, the energies on the cylinder can be related to the conformal dimensions of the corresponding operators on the plane via the usual conformal map, which yields

%Note that in a CFT, this readily gives the expected twisted sector conformal dimensions, since 

\be \label{cftenergy}
E^{( w)}_{n, \,\scriptscriptstyle{CFT}} (R) = \frac{2\pi(\Delta_n^{( w)} - \frac{c  w}{12})}{R} \;, \;\;\;\;\;\;\;\; E^{(s)}_{n, \,\scriptscriptstyle{CFT}} ( w R) = \frac{2\pi (\Delta^{(s)}_n - \frac{c}{12})}{ Rw} 
\ee 
%{\color{blue}(I put the CFT label because it appears in 2.55)}
where $c$ is the central charge of the seed CFT. Note that the gap above the ground state ($\Delta^{(s)}=0$) is, as is well-known,  $ w$ times smaller  in the twisted sector than in the untwisted one. 
 The $cw/12$ shift between the energy and the dimension in the twisted sector follows from the fact that the effective central charge of the latter is $c  w$. 
%\emph{Can this be obtained without the Schwarzian map?} 
%(\emph{Btw, I don't think the Casimir energy gets multiplied by $ w$}).
Combining \eqref{energies} and \eqref{cftenergy}, we obtain
\be  \label{conformaldim}
\Delta^{( w)}_n = \frac{\Delta_n^{(s)}}{ w} + \frac{c}{12} \left(  w - \frac{1}{ w} \right)
\ee
which reproduces the known result for the twisted sector operator dimensions \cite{Klemm:1990df}.   Note that in the above, conformal invariance was only used to translate the cylinder energies into operator conformal dimensions, but is otherwise not needed to derive \eqref{energies}, which holds equally well in a non-conformal theory.

Let us now also take into account the momentum dependence of the $ w$-twisted sector partition function \eqref{wtwz}. The quantity $\theta = \tau_1 R$ being the same as in the seed, we again have

\be \label{intergermom}
P^{( w)}_n(R) = P^{(s)}_n ( w R) = \frac{2 \pi p_n}{ w R} \;, \;\;\;\;\; p_n \in \mathbb{Z}
\ee
where $p_n$ is the integer-quantized momentum of the corresponding state in the seed QFT. The above
appears to imply that the twisted-sector  momentum may be a fractional multiple, $p_n/ w$, of the inverse radius, which would be inconsistent with modular invariance. This is resolved by the sum present in \eqref{wtwz}, since for every energy-momentum eigenstate in the seed, the full contribution to $Z^{(w)}$ is 
\be
\frac{1}{ w} \sum_{  r=0}^{ w-1} e^{-\b E_n + i (\theta +  r R) P_n} = \frac{1}{ w}  e^{-\b E_n + i \theta P_n} \sum_{ r=0}^{ w-1} e^{2\pi i  r \{\frac{p_n}{ w}\} }= % \frac{1}{ w}  e^{-\b E_n + i \theta P_n}  \frac{1-e^{2\pi i  w  \{\frac{s}{ w}\} }}{1-e^{2\pi i  \{\frac{s}{ w}\}}} =
 e^{-\b E_n + i \theta P_n}\, \delta (p_n=0 \; \mbox{mod} \;  w)
\ee
where in the second step we noted that only the fractional part of $p_n/ w$ contributes, and in the third we trivially summed the  geometric series. Thus, the momentum in the twisted sector is an integer, as expected, and only seed momenta that are multiples of $ w$ will end up contributing to the $ w$-twisted sector partition sum. %In other words, $s=\a  w$, and thus $P^{seed} ( w R) = 2\pi \a/R$. 

To summarize, each state in the seed QFT gives rise to a state in the $ w$ - twisted sector, whose  energy and momentum are 

\be
E^{( w)} (R) = E^{(s)} ( w R) \;, \;\;\;\;\; P^{( w)} (R) = P^{(s)} ( w R) \;\;\;\;\; \mbox{iff} \;\;\; \;\;P^{(s)} (R)  \in \frac{2 \pi}{R}  w\,  \mathbb{Z} \label{finalspec}
\ee
In particular, the degeneracies of these states are the same, provided the constraint on the momentum is satisfied.

%[{\color{blue}The above discussion concerns the contribution of a single cycle of length $ w$, corresponding to the $\ell= w$ contributions to $Z^{( w)}$. The terms with $\ell \neq  w$ correspond to the contribution of $ w/\ell$ cycles of length $\ell$.}] {\color{ForestGreen}
The contributions of the terms with $\ell \neq  w$ to  $\mathcal{Z}^{( w)}$ - namely, of  sectors with $ w/\ell$ cycles of length $\ell$ - can be analysed in an analogous manner. We find

%.  An entirely analogous analysis show  the spectra are given by  (the sum over $ r$ tells us $s$ must be a multiple of $\ell$

\be
E^{( w)}_n (R) =  \frac{ w}{\ell} E_n^{(s)} (\ell R) \;, \;\;\;\;\; P^{( w)}_n (R) = \frac{ w}{\ell}  P^{(s)}_n (\ell R) \;\;\;\;\; \mbox{iff} \;\;\;\;\; P^{(s)} (R) \in \frac{2 \pi}{R} \ell \, \mathbb{Z}
\ee
These states correspond to $ w/\ell$ identical copies of the same state from the $\ell$ - twisted sector, in agreement with the selection rule on the momentum.

The full spectrum  of the symmetric orbifold  is given by putting together these elements inside the partition function. It it useful to work out  explicity the full partition function \eqref{bantay2} for the the simplest example $N=2$, as higher $N$ work qualitatively similarly.  In this case, 
there are only two sectors, one untwisted and one 2-twisted. Applying Bantay's formula \eqref{bantay2}, we have 
\begin{align}
&Z^{S_2} (\tau,\bar \tau, R) =  \frac{1}{2} (\mathcal{Z}^{(1)})^2 +  \mathcal{Z}^{(2)}= \\
&= \;\frac{1}{2} Z_{seed}^2 (\tau,\bar{\tau},R) + \frac{1}{2}  Z_{seed}(2\tau,2\bar{\tau},R)+ \frac{1}{2}  Z_{seed} \bigg(\frac{\tau}{2},\frac{\bar{\tau}}{2} , 2 R\bigg)+ \frac{1}{2}  Z_{seed} \bigg(\frac{\tau+1}{2},\frac{\bar{\tau}+1}{2}, 2 R\bigg)\nonumber\\[2pt]
&= \;\frac{1}{2}\sum_{m,n} d_m^{(s)} d_n^{(s)} e^{-\b (E_m^{(s)}+E_n^{(s)}) + i \theta (P_m^{(s)}+P_n^{(s)})} + \frac{1}{2}  \sum_m d_m^{(s)} e^{-2\b E_m^{(s)} + 2i \theta P_m^{(s)}} + \sum_m d_m^{(s)} e^{-\b E_m^{(2)} + i \theta P_m^{(2)}} \nonumber
\end{align}
where in the last term we have used our previous result on the twisted sector spectrum and allowed momenta. The degeneracies of the various states simply follow from the seed degeneracies, with the given restriction.  The first two terms contribute to the untwisted sector partition function, as can be seen by further massaging them into
\be
\left. Z^{S_2}\right|_{untw}  = \sum_{m<n}  d_{m}^{(s)} d_{n}^{(s)} e^{-\b (E_m^{(s)} +E_n^{(s)} ) + i \theta (P_m^{(s)} +P_n^{(s)} )}   + \sum_m \frac{d_{m}^{(s)} (d_{m}^{(s)}+1)}{2}\, e^{-2\b E_m^{(s)}  + 2i \theta P_m^{(s)} }
\ee
The contributing  states belong to the  symmetrized tensor product $(\mathcal{H}_{seed})^2/S_2$ and they take the form $(|E_n\rangle |E_m\rangle+|E_m\rangle |E_n\rangle)/\sqrt{2}$ for the first term, and  $|E_m\rangle |E_m\rangle$ for the second. The degeneracies precisely correspond to those in the symmetrized tensor product of seed Hilbert spaces. %, i.e. $d_{m,s} d_{n,s}$ for $m \neq n$ and  $d_{m,s}(d_{m,s}+1)/2$ for the equal energy states. 
Note that integer degeneracies are obtained only after including all 
 contributions to the partition function. In the twisted sector, the degeneracies are the same as in the seed, subject to the projection \eqref{intergermom}. 

All higher $N$ cases work similarly. %, the explicit  $N=3$ analysis is given in eqn. $(26)$ of \cite{Hashimoto:2019hqo}, where we should remember to keep track of the radius dependence in order for the result to apply to general QFTs. One can easily work out 
 The full energy spectrum is given by sums of the form%. {\color{ForestGreen} For an arbitrary state in a  sector given by $[g]\in S_N$} \emph{Notation!}
\be \label{sumenergycycles}
\sum_{cycles} E^{( w_i)} (R)  \;, \;\;\;\;\;\; \sum_{cycles} P^{( w_i)} (R)
\ee
which run over all the cycles in the various conjugacy classes $[g]$.  %and depends only on their length which is the same within the conjugacy class. 
 The ground state energy in each sector varies from   $-\pi c N/(6 R)$ in the  untwisted sector to  $- \pi c/(6 N R)$ in the maximally twisted one. %, and  in general is given by  $-c/12 \sum N_n/n$. %One may wonder how this state is defined on the cylinder. Perhaps some wavefunctional of the (periodic?) fields at $t=0$ that picks up a funny factor as one goes around the circle? 

\subsection{Spectrum  of $T\bar T$ and $J\bar T$ symmetric product orbifolds } \label{ttjtspectrum}
We would  now like to apply these considerations to the specific examples of interest,  namely $T\bar T$ and $J\bar T$ - deformed CFTs.

\subsubsection*{$T\bar T$ - deformed CFTs}

As reviewed in section \ref{seedTTJT}, the partition function of a  $T\bar T$ - deformed CFT - a Lorentzian QFT with a single dimensionful coupling, $\mu$ -   
%
% case is a trivial application of our general considerations to the case of a partition function that depends on one dimensionless parameter $\mu/R^2$. 
%For a seed $T\bar{T}$ deformed CFT, the partition function 
depends not only on $\tau,\bar{\tau}$, but also on $R$ through the dimensionless combination $\mu/R^2$. This partition function is modular invariant in the generalised sense \eqref{TTbarmodinv}.

The partition function of the symmetric product orbifold of $T\bar T$ - deformed CFTs is obtained via a trivial application of Bantay's formula \eqref{bantay2} to this seed, where the `connected' contributions $\mathcal{Z}^{(n)}$ are given by particularizing \eqref{Znsum} to the specific dependence on $R$ of the $T\bar T$ seed partiton function. Explicitly, 
%
%As explained in the previous section, under a modular transformation $\tau\rightarrow \frac{a\tau+b}{c\tau+d}$, the radius changes $R\rightarrow |c\tau +d|R$. Hence, the modular invariance property required for the Bantay formula to hold is:
%\begin{align}
%Z^{(\mu)}_{seed}\big(\tau,\bar{\tau},R\big)=Z_{seed}^{(\mu)}\bigg(\dfrac{a\tau+b}{c\tau + d},\dfrac{a\bar{\tau}+b}{c\bar{\tau} + d},|c\tau+d| R\bigg) 
%\end{align}
%which is precisely the modular property of $T\bar{T}$ deformed CFTs \ref{modularttbar} of \cite{}. 
%%Hence, the results above apply and the torus partition function of single-trace $T\bar{T}$ can be written as
%
%\be \label{TTbantay}
%Z^{S_N}(\tau,\bar{\tau},\mu/R^2)=\frac{1}{N!}\sum_{x\in S_N}\prod_{\xi  \text{ orbit}}|\xi|\mathcal{Z}^{(|\xi|)}
%\ee
%{\color{ForestGreen}
%with the quantity appearing in the lhs 
\be
\mathcal{Z}^{(n)}_{T\bar T}=\frac{1}{n}\sum_{\ell|n}\sum_{0\leq r<\ell} Z_{T\bar T}^{seed}\bigg(\frac{n\tau}{\ell^2} +\frac{ r}{\ell},\frac{n\bar{\tau}}{\ell^2} +\frac{ r}{\ell} ,\frac{\mu}{\ell^2R^2}\bigg) 
\ee
 As explained in the previous section, the modular invariance of $\mathcal{Z}^{(n)}$ follows from that of the seed partition function. It can be made particularly evident by   rewriting  $\mathcal{Z}^{(n)}$ in terms of Hecke operators. This result is in full agreement with the previous worldsheet computations \cite{Hashimoto:2019wct,Hashimoto:2019hqo} and the recent derivation \cite{Apolo:2023aho}.

As explained in our general analysis, this allows us to obtain the spectrum in the various twisted sectors. In particular, the energies  in the $ w$ - twisted sector are given by %\footnote{Note that a particular class of states, with $E^{(w)}(R)=P^{(w)}(R)$, do not flow, just like in the case of double-trace $T\bar{T}$.} 

\be \label{spoTTbar}
E^{( w)}_{T\bar T} (R) = E^{(s)}_{T\bar T} ( Rw) = \frac{R  w}{2\mu} \bigg(\sqrt{1+\frac{4\mu E^{(s)}_{\scriptscriptstyle{CFT}} ( Rw)}{  w R}+\frac{4\mu^2 \left(P^{(s)} ( w R)\right)^2}{R^2  w^2}}-1\bigg)%, \hspace{0.1cm} P^{( w)}(R)=\frac{2\pi\alpha}{R},\alpha\in\mathbb{Z}
\ee
where  we have opted to sometimes use the subscript  `CFT' to denote the undeformed fields, either in the seed or in the symmetric orbifold. The momenta are given by \eqref{finalspec}, which includes the projection. % as in our general discussion (%\be
 One may further plug in the expression \eqref{cftenergy} for $E^{(s)}_{\scriptscriptstyle{CFT}} (w R)$ in terms of the conformal dimensions in the seed CFT, obtaining perfect agreement with the spectrum previously worked out in the literature \cite{demisethesis,Hashimoto:2019wct,Hashimoto:2019hqo}; note this brings additional powers of $ w$ to the denominators. Alternatively, we may use \eqref{energies} to replace $E^{(s)}_{\scriptscriptstyle{CFT}} ( w R)$ by $E^{( w)}_{\scriptscriptstyle{CFT}} (R)$, and interpret \eqref{spoTTbar} instead as the solution to a universal flow equation in the twisted sector with an effective parameter $\mu/ w$, as was previously observed in \cite{Apolo:2023aho}.   
The full spectrum of the symmetric orbifold of $T\bar T$ - deformed CFTs is given by sums over this kind of terms, as in \eqref{sumenergycycles}, and is thus entirely determined by the spectrum of the seed undeformed CFT. Note that since the twisted sectors are equivalent to the seed theory on a cylinder of radius $R  w$, the torus partition function of the symmetric orbifold is well-defined provided the seed is, namely if the circumference of the torus   satisfies \eqref{rmin}. 

%[{\color{blue} This is in complete agreement with the string worldsheet approach \cite{} and a more recent approach \cite{} that  requires generalised modular invariance of the orbifolded theory. (also before) }]
 
Note the deformed spectrum may also be obtained directly from the flow equation. As usual, first order perturbation theory implies that
\be
\p_\mu E = \langle n | \sum_{I=1}^N T_I \bar T_I | n \rangle% = \langle n | \sum_i H_i K_i - P_i^2 | n \rangle \stackrel{?}{=} \sum_i \langle H_i \rangle  \langle K_i \rangle - \langle P_i \rangle^2 
\ee
In the untwisted sector, $E=\sum_I E_I$, where each $E_I$ obeys the $T\bar T$ flow equation in the given copy. 
%where in the second step we have dropped the total commutators with the hamiltonian, and in the third we have assumed that $|n\rangle$ is an eigenstate of $H_i$ (\emph{not sure this is true/makes sense}). In any case, 
In the twisted sectors, one may uplift the flow equation to the covering space, which is a cylinder of circumference $R  w$. Since the right-hand-side of the flow equation is inversely proportional to the radius, it 
%
% the flow equation for the total energy (in that sector)  
 will pick up an overall factor of $1/ w$

\be
\p_\mu E^{( w)} = \frac{1}{ w} (E^{( w)} \p_R E^{( w)} - P^{( w)2}/R) \label{effflowwsect}
\ee
The solution will be given by the usual $T\bar T$ solution, but with $R \r R  w$ or, equivalently, $\mu \r \mu/ w$. This agrees  with  \eqref{spoTTbar}, provided one  takes into account  the fact that the undeformed energies and momenta are already in the twisted sector, and thus are related via \eqref{finalspec} to the ones of the seed. 

% appears to show some factors of $ w$ are missing, one should note that the seed energy and moemntum used should already be those of the symmetric product, where there is an additional division by $ w$. 

 \subsubsection*{$J\bar{T}$-deformed CFTs}

The case of $J\bar{T}$-deformed CFTs is more interesting, since the dependence on the couplings must be explicitly included in the partition function, as they transform non-trivially under modular transformations. This concerns both the $J\bar T$ coupling, $\l$, and the external chemical potential, $\nu$, for the left-moving charge.   In addition, the partition function of the seed  is not modular invariant, but instead transforms \eqref{JTbarmodprop}  as a Jacobi form of weight $(0,0)$ and index $(k,0)$, where $k$ is the level of the $U(1)$ Kac-Moody algebra.  %{\color{ForestGreen}(this is true only for CFT with chemical potential, otherwise we expand in powers of the coupling and the coefficients are Jacobi forms of different weights, not 0)}
%\emph{Factor of 2!?}

%
%
%
%, the partition function with chemical potential depends not only on $\tau,\bar{\tau}$, but also on the dimensionless combination $\lambda/R$, which changes under modular transformations as argued in section 2.1:
%\begin{align}
%\frac{\lambda}{R}\mapsto \frac{\lambda}{R(c\bar{\tau}+d)}
%\end{align}
%Hence, the modular invariance property is: 
%\begin{align}
%Z\bigg(\frac{a\tau+b}{c\tau +d},\frac{a\bar{\tau}+b}{c\bar{\tau} +d}, \frac{\nu}{c\tau+d},\frac{\lambda}{R(c\bar{\tau}+d)}\bigg)=\exp\bigg(\frac{2 i
%\pi k c\nu^2}{c\tau+d}\bigg)Z\left(\tau,\bar{\tau},\nu, \frac{\lambda}{R}\right)
%\end{align}
%which is precisely \ref{modularjtbar}.
Our goal is to understand the dependence on the parameters $\l/R$ and $\nu$ of the seed theories on the covering tori. Let us first treat the case of  the left-moving chemical potential $\nu$. As explained, $\nu = \beta a^{z} = \tau_2 R a^z$, where $a^z$  is the gauge field that couples to the chiral left current. This coupling is held fixed when placing the seed theory on a covering torus; as a result, the 
%
%
%transforms oppositely from $R \bar z$, namely $R a^z \r R a^z (c\bar \tau+d)$.
%Putting the theory on a circle of radius $R\ell$ multiplies this coefficient by $\ell$. (Also note that, in an expansion of the partition function in $R a^z$, the coefficients of the various powers are modular forms of weight $(0,k)$; this can then be written in terms of Hecke operators, as explained in the previous subsection). The
 chemical potential $\nu_\xi$ on the covering tori is  given by 

\be
\nu_\xi = (\tau_\xi)_2 R_\xi a^z_\xi = \frac{n}{\ell^2} \tau_2 R  \ell a^z = \frac{n}{\ell} \nu
\ee
On the other hand, the dimensionless coupling $\l/R$ simply picks up the factor of $\ell$ that follows from dimensional analysis, $\l$ itself being the same. 

% Allowing for the mild violation of modular invariance for the seed because of the chemical potential,
The partition function of the symmetric product orbifold of $J\bar T$ - deformed CFTs is again given by   Bantay's formula \eqref{bantay2}, where the individual contributions read

%Ignoring for now the lack of modular invariance, a simple application of Bantay's formula yields a partition function of the form \eqref{bantay2}, where the individual contributions read

%
% For $\nu=0$, the Bantay formula holds and we can express the torus partition function for the symmetric product of $J\bar{T}$ deformed CFT as
%\be
%Z^{S_N}(\tau,\bar{\tau},\nu, \lambda/R)=\frac{1}{N!}\sum_{x\in S_N}\prod_{\xi  \text{ cycle}}Z^{(|\xi|)}_{seed}
%\ee
%with
\be
\mathcal{Z}^{(n)}_{J\bar T} \left(\tau,\bar \tau, \nu, \frac{\l}{R}\right)=\frac{1}{n}\sum_{\ell|n}\sum_{0\leq r<\ell} Z_{J\bar T}^{seed}\bigg(\frac{n\tau}{\ell^2} +\frac{ r}{\ell},\frac{n\bar{\tau}}{\ell^2} +\frac{ r}{\ell}, \frac{n \nu}{\ell} ,\frac{\lambda}{\ell R}\bigg)  \label{ZnJTbar}
\ee
Given the modular transformation properties \eqref{JTbarmodprop} of the seed partition function, 
we would now like to show that the partition function of the symmetric orbifold transforms in the same manner, but with $k \r N k$, as follows from the fact that the level of the $U(1)$ current in the symmetric product is $N$ times larger than that of the seed. Remember from \eqref{modinvJTbar} that the seed partition function differs from a modular-invariant one by a factor of $\exp (\frac{\pi k \nu^2}{\tau_2})$.  On the covering tori, we have
%
%{\color{ForestGreen}Let us now comment on the modular properties of the SPO. Each $\mathcal{Z}^{(n)}$ differs from a modular-invariant partition function by an exponential factor of 
%
\be
\frac{k \nu_\xi^2}{(\tau_\xi)_2}  = \frac{ n k \nu^2 }{\tau_2}
\ee
which is $\ell$ - independent. Thus, each $\mathcal{Z}^{(n)}$ will differ from a modular-invariant contribution by the exponential of such a factor. 
 Since the symmetric orbifold partition function is a sum of products $\prod_n (\mathcal{Z}^{(n)})^{N_n}$ and $\sum_n n N_n =N$, we immediately note that the lack of modular invariance of each term in the sum in \eqref{bantay2}  is $k N \nu^2/\tau_2$, which is the same for every possible partition of the integer $N$. Thus, the transformation properties of  the seed partition function under modular transformations determine those of the symmetric  orbifold one, which transforms as in \eqref{JTbarmodprop}, but with $k \r N k$. 
 %
% is picking an overall $e^{k N \nu^2/\tau_2}$ phase. Note that this corresponds to the chemical potential being the same as in the seed, but $k = N k_{seed}$, as expected. 
%
%By setting $\lambda=0$, the discussion above particularizes to a SPO CFT with chemical potential, case in which we can check the results: the partition function of the seed is a Jacobi form of weight $(0,0)$ and index $k$, while the partition function of the SPO is a Jacobi form of weight $(0,0)$ and index $Nk$.
%
This connection %between the modular property of the seed and that of the SPO 
can be made explicit by rewriting the result using Hecke operators, whose action can also be defined on the
%
%. For this purpose, one needs the action of a Hecke operator on a 
Jacobi forms of weight $(0,\kappa)$ %$(0,p)$ 
and index $(k,0)$ relevant to $J\bar T$ as \cite{jacobibook}
\begin{align}
T_n \phi(\tau,\bar{\tau},\nu)=\frac{1}{n} \sum_{\substack{ r,\ell\in\mathbb{Z},\ell|n\\0\leq  r<\ell}} \frac{1}{\ell^{\kappa}}\phi\bigg(\frac{n\tau}{\ell^2} +\frac{ r}{\ell},\frac{n\bar{\tau}}{\ell^2} +\frac{ r}{\ell},\frac{n\nu}{\ell}\bigg)
\end{align}
and yields a Jacobi form of the same weight and index $(n k,0)$. 
Expanding the $J\bar{T}$-deformed CFT partition function in a Taylor series in $\lambda$, the coefficient of $\lambda^{\kappa}$ is $R^{-\kappa}$ times a $(0,\kappa)$ Jacobi form of index $(k,0)$. Thus, \eqref{ZnJTbar}  can be  written as %, similarly to CFT and $T\bar{T}$-deformed CFT:
\begin{align} \label{jthecke}
\mathcal{Z}^{(n)}_{J\bar T}=T_n Z_{J\bar T}^{seed}\left(\tau,\bar{\tau},\nu,\frac{\lambda}{R}\right)
\end{align}
while the whole partition function is given by the right-hand side of \eqref{bantay2}.

Let us now understand the consequences of this formula for the spectrum of single-trace $J\bar T$ - deformed CFTs. We focus first on the $ w$-twisted sector, for which  $\ell= w$ and thus $\nu_\xi =\nu$, implying that the spectrum of left-moving charges is the same as in the seed.  According to our general formula \eqref{energies}, the right-moving energies in the $ w$-twisted sector read
%
%, for which we restrict to  
%\begin{align}
%Z_{( w)}&:=\frac{1}{ w}\sum_{ r=0}^{ w-1}Z_{seed}\bigg(\frac{\tau+ r}{ w},\frac{\bar{\tau}+ r}{ w},\frac{\lambda}{R  w}\bigg)
%\end{align}
%The same reasoning applies regarding the integer momentum and we obtain
%\begin{align}
%Z_{( w)}&=\sum_{E_m, RP^{seed}\in  w\mathbb{Z}} e^{-2\pi\tau_2 R E_m} e^{2\pi i \tau_1 R P_{m}^{seed}/ w}
%\end{align}
\begin{align}
E_{R, J\bar T}^{( w)} (R)&= E_{R, J\bar T}^{(s)} ( Rw) =\frac{4\pi }{k\lambda^2}\left(R  w-\lambda q^{[0]}-\sqrt{\left(R  w-\lambda q^{[0]}\right)^2-\frac{\lambda^2 k}{2\pi} R  w\, E_{R,\scriptscriptstyle{CFT}}^{(s)}(R  w) }\right)\nonumber
\end{align}
\be
 q^{( w)}=q^{[0]}+\frac{\lambda k}{4\pi}E_{R,J\bar T}^{( w)}(R) \label{chtransf}
\ee
where $q^{[0]}$ is the charge in the undeformed seed and  $P^{( w)}(R)$ is given as before by \eqref{finalspec}, which entails a selection rule on the seed momenta. One can rewrite \eqref{chtransf} in terms of the seed conformal dimensions by plugging in the explicit expressions \eqref{cftenergy} for the CFT finite-size energies. Alternatively, one can reinterpret $E_{R,\scriptscriptstyle{CFT}}^{(s)} ( Rw)$  as $E_{R,\scriptscriptstyle{CFT}}^{( w)} (R)$
%
%while $E_R^{CFT}( Rw)=E_R^{CFT}(R)/ w$ because on the cylinder of radius $ Rw$, the CFT energies are divided by $ w$. The left moving energy is
%$E^{( w)}_L=E^{( w)}_R+P^{( w)}$, where $P^{( w)}=P^{seed}/ w$, but only for the values of $P^{seed}$ for which %this is a multiple of $1/R$,
%
%where $E_m$ are the energies of the seed theory on a cylinder of radius $ Rw$ and the momenta are $P_{m}^{seed}/ w$ with
 %$R P_{m}^{seed}\in 2\pi  w \mathbb{Z}$. %This implies that also the right moving energy is the one of the seed on a cylinder of radius $R\mapsto R  w$:
%Pulling out a factor of $ w$ from the paranthesis, and reinterpreting $E_R^{CFT}(R  w) = E_R^{( w), CFT}$, 
and view this expression  as the solution to the $J\bar T$ flow equation in the $ w$ - twisted sector, where the flow parameter is effectively $\l/ w$ and the effective $U(1)$ level is $k  w$. %Also, in the $ w$-twisted sector, the fractional Kac-Moody algebra has level $k^{( w)}=kw$. \emph{Understand why fractional expected.} 
 The above formula  matches the worldsheet analysis of  \cite{Apolo:2018qpq,Apolo:2019yfj,Chakraborty:2019mdf}\footnote{In \cite{Apolo:2018qpq,Apolo:2019yfj} a different convention for the winding is used, such that $ w_{here}=- w_{\text{there}}$. The charges in \cite{Apolo:2018qpq} are also related to ours by $\bar{q}=-q^{[0]},\bar{Q}=-q^{( w)}$  and the definitions of left and right are exchanged. %(and $E_L$ and $E_R$ are exchanged because they are doing $T\bar{J}$ instead of $J\bar{T}$)
  }.

%{\color{blue}[This appears to reproduce the worldsheet analysis. \emph{\textbf{Check}! Are there additional operators that one should see in the string worldsheet? } (need to specify that in states we need to sum over charges from these cycles, for ex in  w twisted sector we still have N- w seed charges contributing)??]}

For the contributions to $\mathcal{Z}^{( w)}$ that have $\ell\neq  w$, note that the chemical potential on the covering torus is $ w/\ell$ times that of the full symmetric orbifold. This implies that the chages in this sector are $ w/\ell$ times the seed ones. This is in agreement with the fact that the states contributing to these terms take the form $(\otimes | E^{(\ell)}, q^{(\ell)}\rangle)^{ w/\ell}$.

% For example, the untwisted sector term comes from $\mathcal{Z}^{(N)}$ with $\ell=1$, so $q=N q^{(s)}$, in agreement with the fact that states are of the form $\underbrace{|E,q\rangle|E,q\rangle\ldots |E,q\rangle}_{N \text{ times}}$.

%{\color{blue}
%The situation seems more interesting for certain contributions to e.g. the untwisted sector, e.g. those with $\ell=1$ , where $\nu=  w \nu_{seed}$. This implies that the charges are also $ w$ times the seed ones. This makes perfect sense, as these sectors are nothing but $|E\rangle|E\rangle\ldots |E\rangle$ ($ w$ timess).  \emph{Keep?}

Finally, since the single-trace $J\bar T$ deformation also preserves the left conformal symmetry on the plane, we may again compute the corresponding left conformal dimensions, following %the . As explained in \cite{Guica:2019vnb}, we can obtain the spectrum of left-moving conformal dimensions on the plane from the spectrum on an infinitely boosted cylinder, on which the left conformal invariance is restored. Following 
 the same steps as in the double-trace analysis of \cite{Guica:2019vnb}. We obtain %\emph{Notation!}
\begin{align} \label{conformaltwisted}
h^{( w)}_{J\bar T}(\bar p)&=h^{( w)}_{\scriptscriptstyle{CFT}}+\frac{\lambda \bar p \, q^{[0]}}{2\pi  w} + \frac{k \lambda^2 \bar p^{2}}{16\pi^2  w} = \frac{h^{(s)}_{J\bar T}(\bar p)}{ w}+\frac{c}{24}\bigg( w-\frac{1}{ w}\bigg) 
\end{align}
where $h^{( w)}_{\scriptscriptstyle{CFT}}$ is the undeformed conformal dimension in the $ w$-twisted sector,  related to that in the seed CFT via \eqref{conformaldim},  $h^{(s)}_{J\bar T}(\bar p)$ is the momentum-dependent conformal dimension \eqref{jtbardefdim} in the $J\bar T$ seed QFT,   and $\bar p \leftrightarrow E_R^{( w)}$ is the right-moving energy on the boosted cylinder.  Overall, we  obtain the standard CFT formula \eqref{conformaldim} for the orbifolded conformal dimensions, now taking into account the fact that the seed left-moving conformal dimension has been modified to \eqref{jtbardefdim}.  
%formula can be readily obtained by noting that at fixed right-moving energy, the twisted sector energy formula can be obtained by the replacement $\l \r \l/ w$ and $k \r k  w$ in the seed one. 
Another possible interpretation of this formula 
%
%As we show in (see after Discussion), in the undeformed theory we have that the seed charge is the same as the charge in the $ w$-twisted sector $Q_0:=q^{( w)}$. 
% Hence, we can write everything in terms of $ w$-twisted sector quantities:
%\begin{align}
%h^{( w)}(\lambda)&=h^{( w)}+\frac{\lambda p^{( w)}q^{( w)}}{2\pi  w} + \frac{k^{( w)} \lambda^2 p^{( w)2}}{16\pi^2  w^2}
%\end{align}
is  as fractional spectral flow \cite{Martinec:2001cf,Chakrabarty:2015foa} with parameter $\lambda\bar{p}/ w$ in the $ w$ - twisted sector, where the level is $k^{( w)}= k  w$ . The left-moving charge is simply given by  \eqref{chtransf} with $E_R^{( w)} \r \bar p$. 
%
%We would like to understand what happens to the charge in the $ w$-twisted sector of the $J\bar{T}$ SPO. If the result above is correct, then we would expect that:
%\begin{align}
%Q^{( w)}(\lambda)=q^{( w)}+\frac{\lambda k^{( w)} \bar{p}}{4\pi  w } = Q_0+\frac{\lambda k \bar{p}}{4\pi } 
%\end{align}
%where $Q_0$ is the charge of the undeformed seed. %So the question is: if $Q(\lambda)=Q_0+\frac{\lambda k}{4\pi}E_R$ in the seed, can we prove that in the $ w$-twisted sector of the SPO we have $Q^{( w)}=Q_0+\frac{\lambda k}{4\pi}E_R/ w$?
This observation will be important for constructing the single-trace $J\bar T$ correlation functions   in section \ref{strcorrf}.

\subsection{Comments on the entropy %of single-trace $T\bar T$ and $J\bar T$- deformed CFTs
}

Given the partition function of the symmetric orbifold, the density of states can be readily extracted from it. In this section, we comment upon the entropy of both single-trace $T\bar T$ and $J\bar T$
 - deformed CFTs, as well as its relation to that of the respective double-trace deformations.

\subsubsection*{$T\bar T$ - deformed CFTs}

 In the  $T\bar T$ case, the entropy of both the single-trace and double-trace deformation has been discussed in detail 
in the recent work \cite{Apolo:2023aho}, so we will be brief. %The features found in that work are directly analogous to those in SPOs of 2d CFTs.
 For simplicity, we will set $P=0$. 

The analysis of \cite{Apolo:2023aho} closely follows that of \cite{Hartman:2014oaa} for the case of two-dimensional CFTs. One of their results is  that the entropy of a large $N$ symmetric product orbifold of $T\bar{T}$ - deformed CFTs presents two regimes\footnote{Note that our conventions differ from those in \cite{Apolo:2023aho} by $\mu_{here}=2\pi \mu_{there}$ and also $R_{here}=2\pi R_{there}$.}  % \emph{Check!}{\color{ForestGreen}(we checked)}

\be
S_{SPO\, of \, T\bar T }(E) = \left\{\begin{array}{ccc} R(E-E_{vac}) & \mbox{for} & E_{vac} \lesssim E<E_c \\[4pt] 2\pi \sqrt{\frac{c^{(s)} E}{ 6\pi} (R N+\mu E)}& \mbox{for} & E>E_c \end{array}  \right. \label{entropyaaa}
\ee
with a sharp transition between them. Here

\be \label{expreEc}
E_c = - \frac{E_{vac}}{1+ 2 \mu E_{vac}/NR} \;, \;\;\;\mbox{with}\;\;\;\;\; E_{vac} = \frac{N R}{2\mu} \left(\sqrt{1- \frac{2\pi \mu \, c^{(s)}}{3 R^2}} -1\right)
\ee
%and $c$ the central charge of the seed theory. 
and, in this subsection only,  $c^{(s)}$ is the central charge of the seed CFT. 

\begin{figure}[t]
    \centering
    \begin{minipage}{.41\textwidth}
  \centering
    \includegraphics[width=6.2cm]{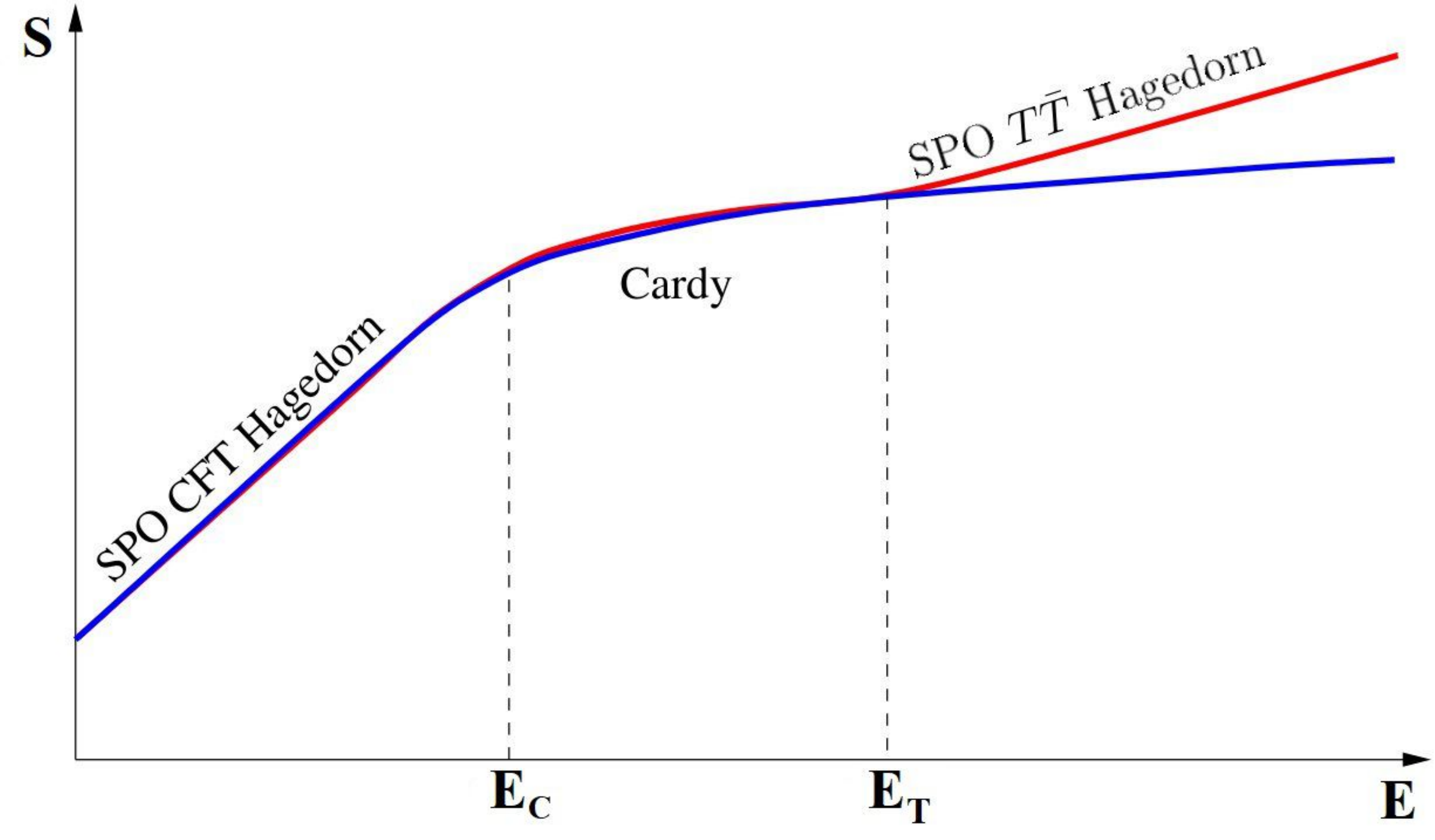} 
     \subcaption{\footnotesize{ The deformation parameter $\mu$ is small enough so that there is a well-separated Cardy regime.}}
    \label{fig2a}
\end{minipage}%
\hspace{1.7cm}
\begin{minipage}{.41\textwidth}
  \centering
    \includegraphics[width=6.1cm]{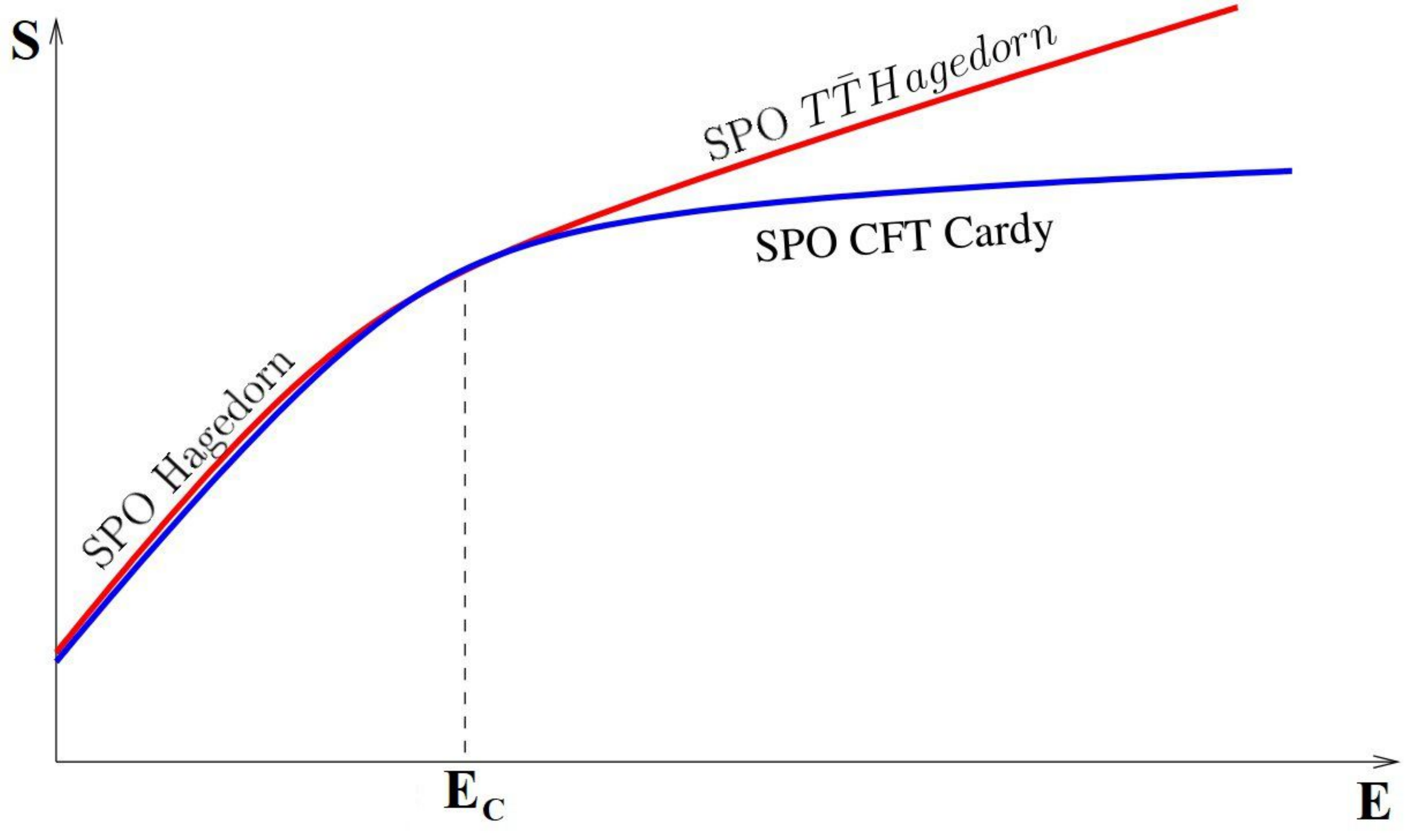} 
    \subcaption{\footnotesize{
  When $\frac{4R^2}{3\pi c}<\mu\leq \frac{3R^2}{2\pi c}$ , there is a direct transition between the two Hagedorn regimes.}}
  \label{fig2b}
    \end{minipage}
    \vspace{3mm}
    \caption{\footnotesize{Plot of the entropy as a function of energy for a large $N$ symmetric orbifold of $T\bar T$ - deformed CFTs (red), as compared to the entropy of a symmetric orbifold of CFTs with the same central charge (blue). 
   % 
  %   plot demonstrates the entropy versus energy plot for an undeformed, large $N$ SPO CFT whereas the red curve demonstrates the same for single trace $T\bar{T}$ deformed large $N$ SPO. %\footnote{Since their slopes are very close in this range ($\sqrt{x}$ for $x \lesssim 1$), we have enhanced their difference for visualisation purposes. } 
% \emph{Keep?}  Note the intercept is slightly higher than in CFT ($-E_{vac}>1/2$); also, $E_c>1/2$.
}}
     \label{fig2}
\end{figure}

Thus, the behaviour of the entropy is Hagedorn in an intermediate range of energies and then transitions to the universal $T\bar T$ behaviour (Cardy $\r$ Hagedorn) at high energies.  Note that, since the partition function of single-trace $T\bar T$ - deformed CFTs  only  makes sense on a circle of circumference $R>R_{min}$, with $R_{min}$  given in \eqref{rmin}, the slope of the high-energy Hagedorn regime is always less than the slope of the intermediate Hagedorn one. It is interesting to ask whether the two Hagedorn regimes need to be separated by a Cardy one. The crossover between Cardy and  Hagedorn behaviour in the universal regime $E>E_c$ occurs at an energy scale $E_T \sim NR/\mu$. The ratio of this scale\footnote{In units of $N \pi c/3R$, the energies are $E_T = 2/x$, $E_c = (1/\sqrt{1-x}-1)/x$ and $E_{vac} =  (\sqrt{1-x}-1)/x$.} to $E_c$ is
\be\label{xdef}
\frac{E_T}{ E_c} = \frac{R N}{\mu E_c } = %\frac{6 R^2  x }{2\mu \pi  c (1/\sqrt{1-x}-1)} = 
\frac{2 \sqrt{1-x} }{1-\sqrt{1-x}} \;, \;\;\;\; \;\;\;\;x \equiv \frac{2\pi \mu\, c^{(s)}}{3R^2} \leq 1 
\ee
%which implies that $E_c R= \frac{\pi N c}{3} \cdot  \frac{3  R^2}{2 \pi \mu c} (1/\sqrt{1-x}-1)$. 
which is a monotonously decreasing function of $x \propto \mu$, from infinity at $\mu=0$ to zero at the maximum allowed $\mu$ for that compactification radius ($x=1$). For $x<8/9$, the transition from the Cardy to the second Hagedorn regime occurs after the high-energy universal regime sets in (see figure \ref{fig2a}).  However, when  $8/9<x\leq 1$, then the Hagedorn term dominates from the beginning, and we have a Hagedorn to Hagedorn transition (see figure \ref{fig2b}). This regime is possible precisely because 
the value of $E_c$ at which the universal regime kicks in depends on $\mu$; otherwise, the above ratio would be $4/x$, which never becomes less than one in the given range. 
% The two Hagedorn slopes differ by a factor $\sqrt{x}$, which is very close to one in the situation envisaged in the second plot, so we have enhanced for visual effect. 

%The minimum value of this is now zero, so if we consider this bound, then we can 
%have a direct transition from Hagedorn to Hagedorn without Cardy.

%Figure 1 demonstrates the entropy versus the energy plot. The blue plot exhibits the entropy in the undeformed large $N$ SPO whereas the red curve demonstrates the the same for single trace $T\bar{T}$ - deformed large $N$ SPO. Note the behaviour is extremely different from that of the double-trace $T\bar T$ deformation of the SPO CFT studied in \cite{}. 

 As shown in \cite{Hartman:2014oaa} (henceforth HKS) using modular invariance, in a two-dimensional CFT with a large central charge and a sparse light spectrum, the entropy is universally given by Cardy's formula for energies $E> \pi c/6 R$, and satisfies a Hagedorn upper bound for smaller energies, which is saturated by symmetric product orbifold CFTs. Closely following this analysis, \cite{Apolo:2023aho} (henceforth ASY) showed that a  similar statement holds in double-trace $T\bar{T}$-deformed CFTs with a large central charge and  an appropriately sparse light spectrum: the high-energy density of states is given by the universal formula \eqref{entropytt} for $E> E_c$, where $E_c$ is given by \eqref{expreEc} with $N=1$ and $c^{(s)}$ replaced by $c$. Below $E_c$, the entropy satisfies an upper  bound, given by
\begin{align}
S_{ ASY\; T\bar T \; bnd.}(E)&=R(E-E_{vac}), \hspace{0.5cm} E < E_c %\;, \;\;\;\; E_c =  %E_{vac} = \frac{R}{2\mu} \left(\sqrt{1- \frac{2\pi \mu c}{3 R^2}} -1\right)
\end{align}
where $E_{vac}$ is given by \eqref{expreEc} with $N=1, c^{(s)} \r c$. 
From \eqref{entropyaaa}, one can see that the symmetric product orbifold of $T\bar{T}$ - deformed CFTs can be thought of precisely saturating the bound for  $c =N c^{(s)}$, provided we replace  $\mu_{s.tr} \r N \mu_{d.tr}$, as follows from the matching of the entropies in the high-energy universal regime $E>E_c$. Thus, the red curves in the plots above can also be interpreted as upper bounds on the entropy of double-trace $T\bar T$ - deformed CFTs with central charge $N c$, coupling $\mu/N$ and a certain sparseness condition on their light states.

Another way to obtain an upper bound on the density of states of a $T\bar T$ - deformed CFT is to use the fact that in such a theory the degeneracies are identical, at leading order, to those in the seed CFT, but they are measured in  a different variables. The degeneracies of a CFT with a sparse light spectrum satisfy, as discussed, the HKS bound \cite{Hartman:2014oaa}, which is saturated by a symmetric orbifold CFT. Simply plugging in 
 the relationship between the undeformed and deformed energies into this universal  bound, one obtains  %  entropy as a function of the undeformed energy and then replacing in it the undeformed energy by the deformed one. 
%The result takes the form %\emph{Factor of $N$ below looks wrong - Check!!}
\begin{equation}\label{dtentspo}
S_{T\bar T \, of \, HKS\, bnd.} (E)= \left\{\begin{array}{ccc} E(R+ \mu E) + \frac{\pi c }{6}  & \mbox{for} & E < E'_c
\\[4pt]
2\pi \sqrt{\frac{c E}{ 6\pi} (R+\mu E)} & \mbox{for} & E > E'_c 
\end{array} \right.
\end{equation}
where $E'_c$ is given by
\begin{equation} \label{defEcprim}
E'_c=\frac{R}{2\mu}\left(\sqrt{1+ \frac{2 \pi \mu c}{3 R^2}}-1\right)
\end{equation}
%with $x$ introduced in \eqref{xdef}. 
For $c=N c^{(s)}$, this represents the entropy of a double-trace $T\bar T$ - deformed   symmetric product orbifold CFT, which was studied in \cite{Chakraborty:2022xmz}. From \eqref{dtentspo} it follows that  this  exhibits an intermediate super-Hagedorn regime in the microcanonical ensemble, smoothly crossing over to the usual $T\bar{T}$ Hagedorn behaviour at high energies. The specific heat in the intermediate super-Hagedorn regime is negative, and the system exhibits a first-order phase transition when  coupled to a heat bath. 

\begin{figure}[h]
    \centering
  \includegraphics[width=7cm]{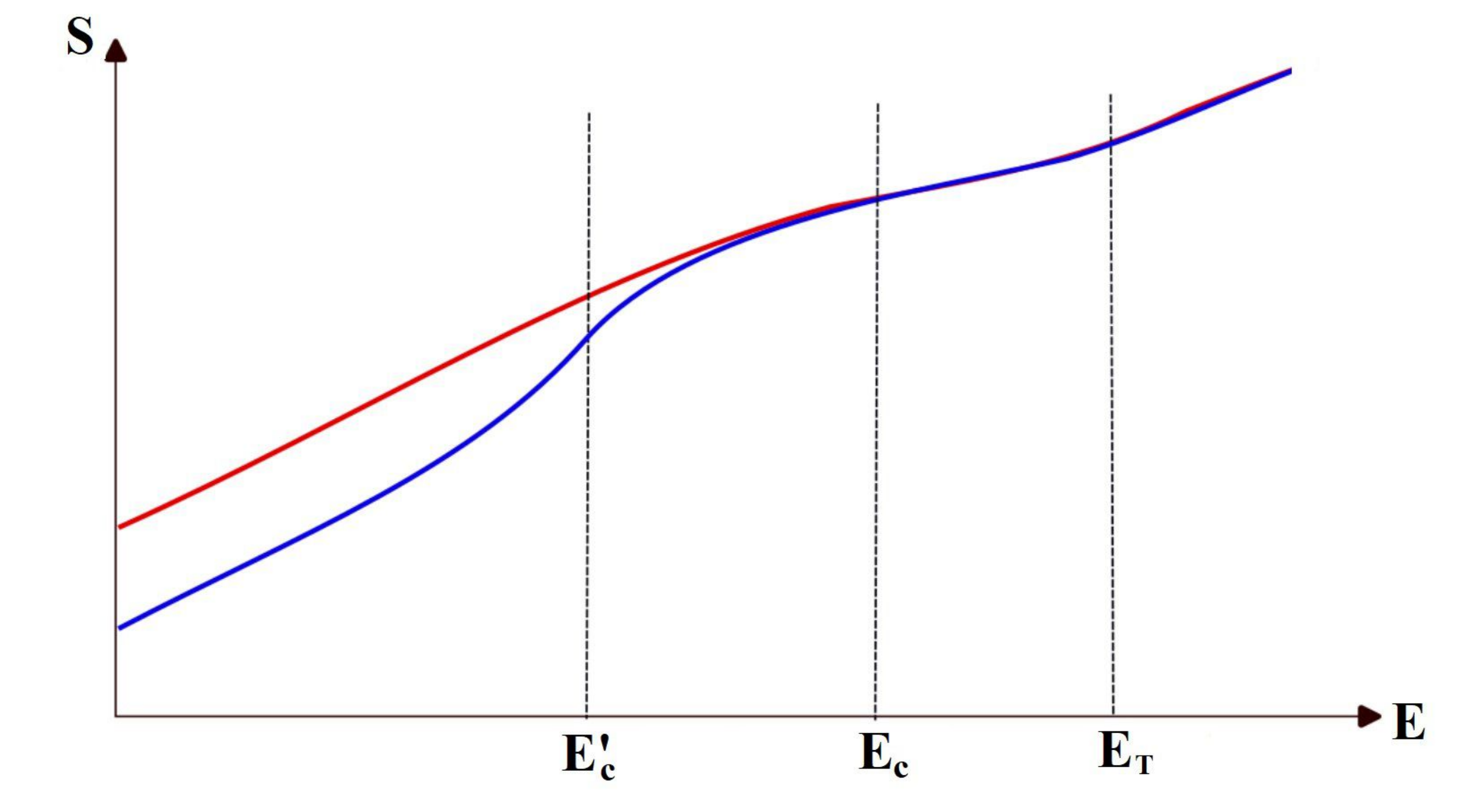}
  \caption{\footnotesize{Comparison of the ASY (red) and the $T\bar T$ deformation of the HKS (blue) bounds on the entropy of $T\bar T$ - deformed CFTs with a large central charge. The latter bound is stronger due to the associated stricter sparseness condition on the light states. These plots can also be interpreted as the entropy of a single-trace (red) and double-trace (blue) $T\bar T$ - deformed entropy of a symmetric product orbifold CFT. 
  %
  %The upper (red) plot represents the bound  on the spectrum of a double-trace  $T\bar T$-deformed  CFT with a sparse light spectrum. The symmetric product orbifold of $T\bar{T}$-deformed CFTs saturates this bound provided its coupling is $N\mu$. The  lower (blue) plot shows the spectrum of the a double-trace $T\bar T$ deformation of the symmetric orbifold CFT at large $N$, which  clearly lies below the bound. %The fact that the intercept $-E_{vac}$ of the red curve is larger than $\pi c N/6 R$ \emph{Correct?} is important for the bound to be obeyed. % For $E>E_c'$ (with $E_c'<E_c$) the spectrum of a double-trace $T\bar T$ deformation of the SPO CFT is given by the universal formula \eqref{entropyaaa}
  }}
  \label{compare}
\end{figure}

Let us now check whether this super-Hagedorn intermediate regime is consistent with the bound derived in \cite{Apolo:2023aho}. From \eqref{defEcprim} and \eqref{expreEc} with $N=1$, $c^{(s)} \r c$,  it is easy to check that $E'_c<E_c$%\footnote{Remember that while comparing single-trace with the double trace results one needs to consider $\mu_{s.tr} = N \mu_{d.tr}$ and also the central charge of the SPO is of the form $N c^{seed}$.}
, so the universal high-energy regime  is reached by the  $T\bar{T}$ - deformed HKS bound  before the prediction \eqref{entropyaaa} of the ASY bound. By evaluating \eqref{entropyaaa} and \eqref{dtentspo} at $E_c'$, we obtain:
\begin{equation}\label{compareent}
S_{ ASY\;  T\bar T \;bnd}(E'_c)>S_{T\bar T \, of \,HKS\, bnd} (E'_c).
\end{equation}
One can check that this is also the case  at $E=0$.
Since the double-trace $T\bar T$ entropy is monotonously increasing in this interval, and at the end of the super-Hagedorn regime the double-trace $T\bar{T}$ entropy is smaller than the AS bound, we can conclude that
\begin{equation}
S_{ASY\, T\bar{T}\, bnd}(E)>S_{T\bar T \, of \, HKS} (E) , \ \ \ \forall E\leq E_c,
\end{equation}
%
%From \eqref{defEcprim} and \eqref{expreEc}, it is easy to see that $E'_c<E_c$, so the universal high-energy regime in the double-trace $T\bar{T}$ deformed symmetric orbifold CFT is reached before the one that follows from the bound \eqref{entropyaaa}. Also, comparing \eqref{entropyaaa} and \eqref{dtentspo}, it follows  that  \textbf{\emph{??? Explain!}}
%
%it is interesting to check whether this super-Hagedorn intermediate regime is consistent with the bound derived in \cite{Apolo:2023aho}. 
%
% \emph{Explain!} On the other hand, since the bound \eqref{entropyaaa} is valid for an arbitrary CFT (with sparse spectrum), deformed by double trace $T\bar{T}$, one would expect the entropy of a double-trace $T\bar{T}$ deformed symmetric orbifold CFT to lie inside the bound. This is enough to conclude
%\begin{equation}
%S(E)>S_{T\bar T \, on \, SPO} (E) , \ \ \ \forall E\leq E'_c,
%\end{equation}
 %namely,  the entropy of the double-trace $T\bar{T}$ - deformed  symmetric orbifold CFT at large $N$ indeed satisfies the bound \eqref{entropyaaa}. }
 %
 as depicted in figure  \ref{compare}. To obtain this result, it was important that the transition between the two regimes in \eqref{entropyaaa} is given by \eqref{expreEc}, as opposed to simply replacing the CFT energies by the $T\bar T$ - deformed ones. %{\color{ForestGreen}universal regime($T\bar{T}$(CFT))$\neq$ $T\bar{T}$(universal regime(CFT))}

%
%{\color{blue}
%{The plot for the double-trace deformation is obtained by taking the CFT plot, which has $S= R E$ for $0?<E_0< 2\pi c N/12 R$ and Cardy above, and rewriting  $E_0$ in terms of the $T\bar T$ - deformed energy. One obtains \emph{Factors!}
%
%\be
%S_{T\bar T \, on \, SPO} = \left\{\begin{array}{ccc} E(R+ \mu E) + \frac{\pi c N}{6}  & \mbox{for} & E < E_c
%\\
%2\pi \sqrt{R+\mu E^2} & \mbox{for} & E > E_c
%\end{array} \right.
%\ee
%where $E_c = 1/x (\sqrt{1+ N x} -1)$, with $x<1$. \emph{Show always samller then previous $E_c$ and discuss. }
%
%
%Double trace $T\bar{T}$ deformation of large $N$ SPO CFTs    exhibits an intermediate super-Hagedorn regime in microcanonical ensemble smoothly crossing over to the usual $T\bar{T}$ Hagedorn at high energies. The specific heat in the intermediate super-Hagedorn regime is negative which exhibits first-order phase transition when the coupled to a heat bath. One may wonder whether this super-Hagedorn intermediate regime is consistent with the bound proven in \cite{Apolo:2023aho}. 
%A careful analysis shows that the spectrum of the  double-trace $T\bar T$ deformation of the SPO CFT satisfies it (see figure \ref{compare}). \textbf{\emph{Can you please include a sketch of the calculation and a proper discussion? The figure cannot go in like this, so please fix it! }}
%
%}}

\vspace{5mm}

%%It is interesting that Hagedorn to Hagedorn exists only if we use Wei's bound. If instead we used the naive CFT bound $E_c= ER > \pi c  w/6$ , given that Hagedorn dominance is for $ER > R^2  w/\mu ( > 2\pi  w c/3) $, we find  that at the minimum (naive) limit of validity of this formula, the ratio of the
%Cardy term $R  w$ to the Hagedorn term is
%
%\be
%\frac{R  w}{\mu E_{min}} = \frac{6 R^2 }{\mu \pi c } \geq 4
%\ee
%Thus, we should transition (after the orbifold) from a regime with Hagedorn growth with slope $2\pi R$ to a Cardy regime, which lasts at least $4 E_{min}$, and then to the $T\bar T$ Hagedorn growth with slope $\sqrt{2\pi c \mu/3}$. \emph{Since the physics is somewhat different, should ascertain Wei is correct.}

%If we use the less naive limit of validity of the universal growth, namely 

\noindent Thus, the entropy bound obtained by $T\bar T$ - deforming the HKS bound for a CFT with a large central charge and a sparse light spectrum is tighter than that obtained in \cite{Apolo:2023aho} directly from modular invariance in the $T\bar{T}$ - deformed CFT. This can be explained by the fact that the sparseness condition imposed in \cite{Apolo:2023aho} on the light spectrum of $T\bar{T}$ - deformed CFTs is less restrictive than the $T\bar{T}$ deformation of the HKS sparseness condition on the seed CFT, which results in a less strong bound at intermediate energies. Given that these bounds can also be interpreted as the entropy of a single-trace and, respectively, double-trace $T\bar T$ - deformed  SPO CFT at appropriately scaled parameters, the above relation can also be written as

\begin{equation}
S_{SPO \, of \, T\bar{T}}(E)>S_{T\bar T \, of \, SPO} (E) , \ \ \ \forall E\leq E_c,
\end{equation}
We explicitly note that, while the entropies of the deformed theories agree at large energy, they differ  in the  intermediate regime. 

%Thus, it follows that the entropy of the double-trace $T\bar{T}$ deformed symmetric orbifold CFT at large $N$ lies inside the bound \eqref{entropyaaa} saturated by the symmetric product of $T\bar{T}$ deformed CFT.

\subsubsection*{$J\bar T$ - deformed CFTs}

Let us now  discuss the single-trace $J\bar{T}$ deformation, concentrating on the  case of a purely chiral $U(1)$ current, for concreteness. As discussed in section \ref{seedTTJT} for the double-trace case, there exists a sliver of energies in the undeformed CFT, given in figure \ref{figallowed}, for which the deformed  energies are real and can become arbitrarily high; in the chiral case, the high-energy behaviour of the entropy in the right-moving sector is Hagedorn. The starting point of the Hagedorn regime may be determined, as in our previous discussion, by translating the onset of the universal regime in CFT to $J\bar T$ variables. 
%
% is harder to determine in $J\bar T$ - deformed CFts, as for this one usually uses modular invariance, but in $J\bar T$ the partition function is harder to make sense of due to the inevitable presence of imaginary energies. \emph{In $T\bar T$, is the starting point of the universal behaviour before or after the naive prediction from deformed CFT? }
%
We will henceforth assume that the  energies in the deformed theory  are sufficiently high, so that the Hagedorn formula \eqref{entropyJTbar} applies, and deduce the high-energy behaviour of the entropy in single-trace $J\bar T$ - deformed CFTs from the fact that the twisted-sector degeneracies are directly determined from the seed $J\bar T$ degeneracies  via \eqref{ZnJTbar}.

Let us concentrate on the contribution  of a single twisted sector of length $n$, which is given by the   $\ell=n$ term in \eqref{ZnJTbar}. %  $\mathcal{Z}^{(n)}_{J\bar T}$ . 
 We will moreover only write the  right-moving piece of the entropy, as the  left-moving one is identical to that of a CFT with a fixed charge. Since in this sector, the degeneracy is the same as that of the seed at the same right-moving energy and charge, but on a cylinder  $n$ times larger, we find

\be
S_R^{(n)}(E_R,q) = 2\pi\sqrt{\frac{cE_R}{12\pi} \left( (n R-\l q) +\frac{ \l^2 k E_R}{8\pi} \right)}
\ee
It is interesting to compare the degeneracy of the maximally-twisted sector, $S^{(N)}_R (E_R, q)$, to that of the untwisted one with the same total energy and charge

\be
S_R^{untw}(E_R,q) =2\pi  N \sqrt{\frac{c}{12\pi}\left( \frac{ E_R}{N} (R- \frac{\l q}{N}) + \frac{\l^2 k E_R^2}{8\pi N^2}\right) } = 2\pi\sqrt{\frac{cE_R}{12\pi} \left( (N R-\l q) +\frac{ \l^2 k E_R}{8\pi} \right)}% \sim \sqrt{\mu c} \, E
\ee
where we have assumed that, on average, each copy posesses an equal share of the energy and charge, as this distribution maximizes the entropy. Thus, we find the same leading behaviour across sectors, similarly to the case of CFTs  and $T\bar T$ - deformed CFTs. 

In order to establish which sector dominates, one would need to consider subleading corrections  to the entropy, which can be analysed by translating the CFT results  \cite{Hartman:2014oaa} to deformed $J\bar T$ variables. In addition, one would need to ascertain that all sectors may in principle compete in the given energy range: for example, if the energies in the maximally twisted sector are real, then one should check that so are the contributing energies from the untwisted one, at least on average.  In the maximally twisted sector, the reality constraint on the undeformed (twisted) energy and charge is

\be
\frac{(q^{[0]})^2}{k  } - \frac{P R N}{2\pi} \leq \frac{R N E_R^{[0]}}{2\pi} \leq \frac{1}{k } \left( \frac{R N}{\l} - q^{[0]}\right)^2
\ee
where we have simply translated the double-trace constraints discussed in section \ref{seedTTJT} to a cylinder of circumference $R N$.  This should be compared with the reality condition for a state in the seed CFT that has, on average, energy $E_R^{[0]}/N$, momentum $P/N$ and charge $q^{[0]}/N$; it is straightforward to check that the two conditions are the same. We conclude that if we concentrate on a  range of energies and charges in the undeformed symmetric product orbifold CFT  such that the deformed energies in the untwisted sector are real, the ones in the maximally twisted sector will also be real, and viceversa.

It would be interesting to understand the validity of these formulae at lower energies, as well as to establish alternate universal bounds based on modular methods, analogous to  that of \cite{Apolo:2023aho} for $T\bar T$. Given the relation \eqref{jthecke} between  the full $\mathcal{Z}^{(n)}_{J\bar T}$  and generalised Hecke operators, one would expect its behaviour to be universal and fixed by modular invariance. However, such arguments  are harder to invoke in $J\bar T$ - deformed CFTs, despite the formal modular covariance \eqref{JTbarmodprop} of the partition function, due to the generic presence of imaginary energy states. Should one be able to apply such methods, it would be interesting to understand  the energy at which different twisted sectors enter the universal regime. Note also that, as  can be seen from \eqref{ZnJTbar}, $\mathcal{Z}^{(n)}_{J\bar T}$   not only captures the contributions from the $n$-twisted sector, but also special contributions from sectors of lower twist; however, these other contributions %are significantly subleading at the energies {\color{ForestGreen}(more about this?)}.
%
%One can recognize this entropy to be that of $T_N Z_{seed}$, just like in the CFT case. For a CFT, one can deduce the universal behaviour from modular invariance, which is problematic in the case of $J\bar{T}$ deformed seed restricted to real energies. Nevertheless, one can see that in $T_N Z_{seed}$, 
%the contributions
 with $\ell<  N$ are exponentially subleading, since their degeneracy is given by $d( E \ell/ N, R \ell)$ i.e. the degeneracy  of tensor products of \emph{identical} states of lower energy. %Hence, just like for the symmetric product of CFTs, the entropy comes from the maximally twisted sector. This discussion concerned so far the high-energy behaviour. One expects though that at large $N$, the entropy of the symmetric product orbifold of $J\bar T$ - deformed CFTs is universal over a much larger range.

Finally, if the $U(1)$ current $J$ is non-chiral, then the results are very similar to those in a CFT with both left- and right-moving $U(1)$ charges, one simply needs to ensure that the undeformed energies lie in the correct regions, given in figure \ref{fig1b}, to yield real deformed energies and charges.

\section{Flow of the states and symmetries}\label{symflow}

One of the most remarkable properties of the $T\bar T$/$J\bar T$ deformation of a two-dimensional CFT is that it preserves the Virasoro - and, if present, Kac-Moody - symmetries of the undeformed theory, despite rendering it non-local. These symmetries were studied from both a holographic and  field-theoretic perspective in \cite{Guica:2020uhm,Guica:2021pzy,Guica:2022gts,Georgescu:2022iyx,Guica:2020eab,Guica:2019nzm,Bzowski:2018pcy}. The most rigorous way to date to ascertain their presence
%
%Currently the best method for understanding the symmetries in $T\bar T$ and $J\bar T$ - deformed CFTs 
is by transporting the  symmetry generators of the undeformed CFT along the irrelevant flow\footnote{For $T\bar T$ - deformed CFTs, this perspective clears up a number of ambiguities present in the earlier analysis \cite{Guica:2020uhm}.}. %\emph{Review lierature a bit}. 

In this section we show,  as intuited in \cite{Guica:2021pzy}, that an analogous analysis  predicts the existence of Virasoro ($\ltimes$ Kac-Moody) symmetries in single-trace $T\bar T$ and $J\bar T$ - deformed CFTs.  For completeness, we start this section with a brief review of these  symmetries in double-trace $T\bar T$ and $J\bar T$ - deformed CFTs, then discuss the flow operator and the symmetries of the single-trace variant. In addition, we show that the fractional Virasoro and Kac-Moody generators present in the undeformed symmetric  orbifold CFT can also be defined in the deformed theories and are conserved.

\subsection{Brief review  of the  symmetries of $T\bar T$ and $J\bar T$ - deformed CFTs} \label{sgenerators}

The simplest way to show that $T\bar T$ and $J\bar T$ - deformed CFTs possess Virasoro (and, if initially present, also Kac-Moody) symmetries is by transporting the original Virasoro and Kac-Moody generators along the $T\bar T$/$J\bar T$ flow. The operator used to define the transport is precisely the one that controls the flow of the energy eigenstates under the $T\bar T$/$J\bar T$ deformation.

Concretely, since the $T\bar T/ J\bar T$ deformations are adiabatic, one can define for each a flow operator
\be \label{eigenflow}
\partial_{\lambda}|{n_{\lambda}}\rangle=\X 
|n_{\lambda}\rangle
\ee
where $| n_{\lambda}\rangle$ are the energy eigenstates in the deformed theory with generic coupling $\l$, which in this section can  represent either the coupling $\mu$ of $T\bar T$, or $\l$ of $J\bar T$. The flow operator is determined via %(instantaneous)
first order perturbation theory of the states $|n_\l\rangle$, and can be shown to satisfy
\begin{align} \label{Hflowcom}
[H,\X]&= \int d\s \, \mathcal{O}_{T\bar{T}/J\bar T}-diag
\end{align}
%and the analogous relation for $J\bar{T}$.
where ``diag'' stands for  the diagonal elements of the integrated $T\bar T$/$J\bar T$ operator in the energy eigenbasis. If the energy levels are degenerate, as is always the case if we start from a CFT, then it represents the matrix elements     of the $T\bar T$/ $J\bar T$ operator on the union of the  degenerate subspaces. These matrix elements are only non-zero on the diagonal, as follows from the fact that the $T\bar T$ and the $J\bar T$ deformation do not break any existing degeneracies. 
 The operators $\X_{T\bar T}$, $\X_{J\bar T}$  are known  explicitly, at least in the classical limit \cite{Guica:2020eab,Guica:2021pzy,Guica:2022gts,Kruthoff:2020hsi}, though their specific expression is not needed for the argument that follows.  
 
  The flow operator can be used to define the generators $\widetilde{L}_m^\l,\widetilde{\bar{L}}{}^\l_m$ (and their Kac-Moody counterparts $\widetilde J_m^\l, \widetilde{\bar{J}}{}^\l_m$) as solutions of the flow equations \cite{LeFloch:2019rut,Guica:2021pzy} 
\begin{align} \label{floweqvir}
\partial_{\lambda}\widetilde{L}^{\lambda}_m&=[\X,\widetilde{L}^{\lambda}_m]\hspace{1cm}\partial_{\lambda}\widetilde{\bar{L}}{}^{\lambda}_m=[\X,\widetilde{\bar{L}}{}^{\lambda}_m]
\end{align}
with the initial condition  $\widetilde{L}^{0}_m=L_m^{\scriptscriptstyle{CFT}},\widetilde{\bar{L}}{}^{0}_m=\bar{L}_m^{\scriptscriptstyle{CFT}}$, etc. % where $L_m,\bar{L}_m$ are the Virasoro generators of the undeformed CFT.
 %In $J\bar{T}$ - deformed CFTs, one can similarly define the flowed affine generators $\tilde{J}_m^{\lambda},\tilde{\bar{J}}_m^{\lambda}$, which at $\lambda=0$ coincide with  the CFT $U(1)$ Kac-Moody generators. 
 From this definition it follows that at any point along the flow, the algebra of these generators will consist of two commuting copies of the  Virasoro ($\ltimes$ Kac-Moody) algebra,  with the same central charge as that of the undeformed CFT. 

So far, this is just a definition. In order for these operators to generate  symmetries of  $T\bar T$ and $J\bar T$- deformed CFTs, one needs to show that they are conserved.  
%
% However, using the special properties of the $T\bar T$ / $J\bar T$ deformation, and more precisely the universality of the deformed spectrum, one can show that these operators are also conserved, and thus correspond to symmetries of these theories.
  The conservation equation  for  the Schr\"{o}dinger  picture operators  reads
%
%These operators are introduced in the Schrodinger picture. To them, one can add some explicit time dependence such that their Hamiltonian picture version is conserved, in the sense:
\begin{align}\label{conservation}
\frac{\partial\widetilde  L_m^\l}{\partial t}+\frac{i}{\hbar}\,[H,\widetilde L_m^\l]=0
\end{align}
%where $L_{H,S}$ denote the Heisenberg and, respectively, (the operators discussed above are introduced in the Schr\"{o}dinger picture).
 The commutator of the flowed generators with the Hamiltonian can be computed from first principles using  the universality of the spectrum  of $T\bar T$/$J\bar T$- deformed CFTs,  and takes the simple form \cite{Guica:2021pzy}
\be \label{defalfa}
[\widetilde{L}^{\lambda}_m,H]=\alpha_m \widetilde{L}^{\lambda}_m \;, \;\;\;\;\;\;\;[\widetilde{\bar L}^{\lambda}_m,H]=\bar \alpha_m \widetilde{\bar L}^{\lambda}_m
\ee
where $\a_m, \bar \a_m$ are operator-valued  functions that depend on the Hamiltonian and other conserved charges. Their expressions for the two types of theories we consider are %\emph{\textbf{Re-check conventions!}}

\bea
 && \hspace{-4mm} T\bar{T}\;\;:\hspace{0.2cm}\alpha_m (H, P) = \bar{\alpha}_m(H,-P)\,  = \,\frac{1}{2\mu}\left(\sqrt{(R+2\mu H)^2+\frac{4\mu m \hbar(R+2\mu P)}{R}+\frac{4 \mu^2 m^2\hbar^2}{R^2}} -(R+2\mu H)\right)\nonumber  \\[3pt]
%\bar{\alpha}_m(H,P) &=& \alpha_m(H,-P)\nonumber \\[2pt]
 && \hspace{-4mm} J\bar{T}\;\;:\hspace{0.2 cm}\alpha_m = \frac{m \hbar}{R} \, , \;\;\;\hspace{0.3cm}\bar{\alpha}_m (Q)=2\frac{R-\lambda Q -\sqrt{(R-\lambda Q)^2-\hbar k m \lambda^2}}{k\lambda^2}\;, \;\;\;\;\;\; Q \equiv \tilde J_0 + \frac{\l k}{2} H_R \label{expralm}
\eea
where  $\hbar$ is Planck's constant. Note that in $J\bar T$-deformed CFTs,  $\a_m$ is a $c$-number, which is related to the fact that these theories are local on the left-moving side.   The conservation equation \eqref{conservation} is immediately satisfed if we assign the following explicit time dependence to the Schr\"{o}dinger picture generators %(note ordering is important)
%
%The explicit time dependence can be written generically:
\begin{align}\label{timedep}
\widetilde{L}^{\lambda}_{m}(t)&=e^{i\alpha_m t}\widetilde{L}^{\lambda}_{m}(0) \hspace{1cm}\widetilde{\bar{L}}^{\lambda}_{m}(t)=e^{i\bar{\alpha}_m t}\widetilde{\bar{L}}^{\lambda}_{m}(0)
\end{align} 
where $\widetilde{L}^{\lambda}_{m}(0)$ are the solutions to the flow equation \eqref{defalfa}. The   Kac-Moody generators are treated in an entirely analogous manner.
%
%in the case of $J\bar{T}$,  are the $\tilde{L}_m^{\lambda}$ discussed above and the same for right-movers and \emph{Write more compactly}
%
%and similarly for $\bar{\alpha}_m$. %Using this, it is easy to show that for \ref{timedep}, the conservation equation \ref{conservation} is satisfied.
%
%[A crucial property \emph{Why crucial?} is that for both deformations, $\alpha_m,\bar{\alpha}_m$ commute with the Hamiltonian, since they depend only on $H,P$ and $Q$ for $J\bar{T}$.] 
Thus, the operators we  constructed correspond to conserved charges of the theory, and the Virasoro($\ltimes$ Kac-Moody) algebra  they obey represents its symmetry algebra.  Note this conservation argument is the same as the one used in standard two-dimensional CFTs; the only difference is that now $\a_m, \bar \a_m$ are operators, whereas in the CFT case they were $c$-numbers. 
 
The above argument, while valid at the full quantum-mechanical level, is somewhat abstract, and does not lead to an intuitive picture of the action of these symmetries. It is also not clear whether  this basis of generators is the  one that acts most naturally  on the fields in the theory. For example, in the case of $J^1 \wedge J^2$  deformation of two-dimensional CFTs - an exactly marginal deformation of the Smirnov-Zamolodchikov type - the Virasoro generators obtained via an analogous flow argument can be explicitly shown to \emph{differ} from the Virasoro generators of standard conformal symmetries \cite{Guica:2021fkv}.

In  both $T\bar T$ and $J\bar T$ - deformed CFTs,  this question can be addressed very concretely at the classical level, by explicitly constructing the flow operators. In $J\bar T$ - deformed CFTs, which are conformal in the standard sense on the left-moving side, one again finds  that the flowed left-moving Virasoro and Kac-Moody generators $\widetilde L_n, \widetilde J_n$  \emph{differ} from the Virasoro - Kac-Moody generators $L_n, J_n$ of left conformal  and affine transformations. This difference may be characterised as  a ``spectral flow by $\l H_R$", where $H_R$ is the right-moving Hamiltonian %\emph{\textbf{Re-check factors $2\pi$!!!}} {\color{red}(I agree with these factors)}

\be \label{physicalgen}
L_n=\widetilde{L}_n+\frac{\lambda H_R \widetilde{J}_n}{R}+\frac{\lambda^2 k H_R^2}{8\pi R}\, \delta_{n,0}\hspace{1cm} J_n=\widetilde{J}_n+\frac{\lambda k}{4\pi}H_R\, \delta_{n,0}
\ee
where we have dropped the $\lambda$ index  on the generators 
 and assumed that the explicit classical relation can be extended to the full quantum level. Note that in our conventions, the Virasoro generators are dimensionful; in particular $L_0 = H_L, \bar L_0 = H_R R_v /R$ .
  A similar relation holds for the right-movers %\emph{\textbf{Factors $2\pi$!!!}}
\be
 \bar{L}_n=\widetilde{\bar{L}}_n+\frac{\lambda :H_R\;\, \widetilde{\!\!\bar{J}}_n:}{R}+\frac{\lambda^2 k H_R^2}{8\pi R}\, \delta_{n,0} \hspace{1.5cm} \bar{J}_n=\;\widetilde{\!\!\bar{J}}_n+\frac{\lambda k}{4\pi}H_R\, \delta_{n,0} \label{rmphysgen}
\ee
 At least classically, the %left-moving generators implement standard conformal and affine $U(1)$ transformations, whereas the 
 right-moving generators $\bar L_m$ implement  infinitesimal field-dependent coordinate transformations, as one may note from their expression %\textbf{\emph{Check!}} 
\be
\bar L_m = \frac{R_v}{R} \int d\s\, e^{- 2\pi i m \hat v } \H_R  \;, \;\;\;\;\;v \sim \s-t - \l \phi  
\ee 
where  $\phi$  is roughly  the bosonisation of the $U(1)$ current with its zero mode  removed - see \cite{Guica:2021pzy} for the full expression - $R_v$ is the  (field-dependent) circumference of the above  field-dependent coordinate and $\hat v = v/R_v$. The $\bar J_m$ implement similar field-dependent affine $U(1)$ transformations. As discussed at length in \cite{Guica:2021fkv}, it is the $L_n$, $\bar L_n$ - rather than their tilded counterparts -   that act naturally on the  operators in the theory, %which are required to satisfy standard-looking Ward  identities with respect to them. Thus, in $J\bar T$ - deformed CFTs, it appears that these generators 
and which  should therefore be considered as the `physical' symmetry generators in the theory; in particular, it is the $L_m, \bar L_m$ that  have a simple integral expression in terms of the conserved currents in the theory, at least at the classical level. 
 The algebra of these generators follows from their definition \eqref{physicalgen} - \eqref{rmphysgen} and the fact that the tilded generators satisfy two copies of the Virasoro $\ltimes$ Kac-Moody commutation relations. One finds that the algebra of the left-moving generators $L_n,J_n$ is again  Virasoro $\ltimes$ Kac-Moody, while that of $\bar L_n, \bar J_n$  is a non-linear modification of 
the right  Virasoro $\ltimes$ Kac-Moody algebra that does not commute with the left generators. It has been worked out explicitly in \cite{Guica:2021pzy}   using identities such as 

\be \label{comLalpha}
[\bar L_m, \bar \a_n] = (\bar \a_{m+n} -\bar \a_m-\bar \a_n) \bar L_m
\ee
which follow from the special properties of the functions $\bar \a_m$ defined in \eqref{expralm}. 
%
%and similarly for the other generators. \emph{Check if this also holds in $T\bar T$!} 

The $T\bar{T}$ case appears to work similarly, though to date is less understood. In the classical limit,  the flowed generators take the form 
\begin{align}
\widetilde L_m^{cls}&= \frac{R_u}{R} \int d\s e^{2\pi i m\hat{u}}\mathcal{H}_L \hspace{1cm} \widetilde{\bar{L}}{}_m^{cls}=\frac{R_v}{R} \int d\s e^{-2\pi im\hat{v}}\mathcal{H}_R \label{clsttbgen}
\end{align}
where $\H_{L,R}$ are the left/right-moving Hamiltonian densities,  $R_{u,v} \equiv R+2\mu H_{R/L}$, % and $R_v=R+2\mu H_L$ represent the radii of the field-dependent coordinates $u,v$,  where 
$\hat{u}\equiv \frac{u}{R_u},\hat{v}\equiv \frac{v}{R_v}$ and $u,v$ are  the  $T\bar T$ field-dependent coordinates
 that emerge\footnote{More precisely, one can view $\widetilde L_m^{cls}$ as the Fourier modes of a current $\mathscr{H}_L(\s)$ that satisfies the flow equation $\p_\mu \mathscr{H}_L = [\X_{T\bar T}, \mathscr{H}_L] $ and is chiral. The full solution is given in \cite{Guica:2022gts}. Upon integrations by parts, the Fourier modes can be put in the form \eqref{clsttbgen}, where the expressions for $\hat u, \hat v$ simply follow from the flow equation. In this sense, the field-dependent coordinates are ``emergent''.}  from the solution to the  flow equation
\be \label{fielddepcoord}
u \sim  \s+t + 2 \mu \int^\s \!\H_R \;, \;\;\;\;\;\; v \sim \s -t +2\mu \int^\s \!\H_L
\ee
whose full definition can be found in \cite{Guica:2022gts}. Classically, \eqref{clsttbgen} generate field-dependent coordinate transformations. Note, however, that the 
%
%which identifies them as generators of certain field-dependent coordinate transformations. The natural 
 generators of the symmetries in the natural Fourier basis
   are, rather, the so-called ``unrescaled'' generators \cite{Guica:2022gts}  %\textbf{\emph{Check!}} {\color{red}(I think it's ok)}
\begin{align}
Q_m&=\frac{R\widetilde{L}_m}{R_u}\hspace{1cm} \bar{Q}_m=\frac{R\widetilde{\bar{L}}_m}{R_v} \label{relqlt}
\end{align}
%where .  However, due to the field-dependent factors, it follows that 
Their algebra  is given by  a non-linear modification of the Virasoro algebra  %{\color{ForestGreen}(mention cylinder for central charge term?)}
%\textbf{\emph{How did you get these $2\pi$ factors???}} {\color{red}(for the first term for ex $R^2/R_u^2[L_m,L_n]= \frac{2\pi(m-n)}{R} L_{m+n} \frac{R^2}{R_u^2}=\frac{2\pi (m-n)}{R_u}Q_{m+n}$ from the algebra of L on cylinder)}
%
\be \label{doubletralgebra}
[Q_m,Q_n]=\frac{2\pi \hbar}{R_u}(m-n)Q_{m+n}+\frac{8\pi\hbar \mu^2 H_R}{R R_H R_u}(m-n)Q_m Q_n + \frac{\pi^2 c\hbar\, m^3 }{3 R_u^2} \d_{m+n} + \O(\hbar^2)
\ee
where  $R_H=R+2\mu H$ and the $\O(\hbar^2)$ terms and higher can be computed once the full quantum relation between the $Q_m$ and $\widetilde L_m$ is given\footnote{An example of a fully quantum relation between the two that takes into account operator  ordering is given by the symmetrization of \eqref{relqlt}, by letting $\widetilde L_m= Q_m + \mu H_R Q_m + \mu Q_m H_R$. Note that the algebra of these $Q_m$ generators  is  entirely determined by their definition and the fact that the  $\widetilde L_m$ satisfy a   Virasoro algebra. It can be worked out using identities of the form  \eqref{comLalpha}, which also apply to the $T\bar T$ generators  for the appropriate choice of $\a_m$.}. This result was also confirmed holographically by the  analysis of \cite{Georgescu:2022iyx}.  %This algebra can also be generalised to the full quantum case (all orders in $\hbar$) by justifying a fully quantum definition of the $Q_m$ with respect to the $\widetilde L_m$. 

We end our review of the extended symmetries of $T\bar T$ and $J\bar T$ - deformed CFTs by discussing the charges associated to integrability, which have been known  to be preserved by  these deformations since the original work of \cite{Smirnov:2016lqw}. %Thus, one expects that in addition to the Virasoro symmetry generators, one also preserves the KdV charges, which 
In a CFT$_2$, these charges correspond to the so-called KdV charges \cite{Bazhanov:1994ft}, which are associated   to currents constructed from  higher powers of the (anti)holomorphic stress tensor. For example, the first two non-trivial KdV charges are given by  (for $R=2\pi$)
%{\color{ForestGreen} 

\be \label{kdvdef}
I_3 = 2 \sum_{n=1}^\infty L_{-n} L_n + f(L_0)% L_0^2 - \frac{c+2}{12}  L_0 + \frac{c}{24}\bigg(\frac{c}{24}+\frac{11}{60}\bigg)
\ee
\vskip-1mm
\be \label{kdvdef2}
I_5 = \!\! \sum_{n_1+n_2+n_3=0} :L_{n_1} L_{n_2} L_{n_3}: + \sum_{n=1}^\infty  \bigg(\frac{c+11}{6} n^2 -1 - \frac{c}{4}\bigg) L_{-n} L_n + \frac{3}{2} \sum_{n=1}^\infty L_{1-2n} L_{2n-1} +g(L_0)%-\nonumber\\
%&-\bigg(3k+\frac{1}{2}\bigg)L_0^2 +\frac{(8k+5)(12k+1)}{72} L_0 -\frac{k(42k+17)(36k%+7)}{1512}
\ee
%with $k=c/24$ (change notation).
where $f(L_0), g(L_0)$ are quadratic functions of $L_0$, whose explicit expressions are given e.g. in \cite{Maloney:2018hdg}, but are not relevant for our purposes.
As discussed in \cite{Guica:2022gts}, one can easily define the flowed KdV charges 
$\widetilde I_s$ by requiring that they satisfy a homogenous flow equation analogous to \eqref{floweqvir}; its solution  is
simply given by replacing $L_n \r \widetilde L_n$ in the expressions above.  %conjugating the original charges by the unitary transformation induced by the flow operator. A rescaled version of these charges satisfy 
 A rescaled version of these charges satisfies the flow equations discussed in \cite{LeFloch:2019wlf}.

 For this prescription to make sense, we should also show that the flowed KdV charges are conserved, i.e. they commute with the Hamiltonian. This commutator can be computed using \eqref{defalfa} and the commutation relation \eqref{comLalpha}, which turns out to also hold in $T\bar T$ - deformed CFTs for the corresponding  $\a_m$. One can check explicitly that the commutator of $H$ with any product of $\widetilde L_m$ whose indices sum to zero is proportional to $\a_0$, which vanishes. Since this is true term by term, % and the KdV charges are built from such terms,
  we conclude that all  the  KdV charges constructed via the flow remain conserved.

\subsection{Brief review of the symmetries of symmetric  orbifold CFTs}

Let us now review a few facts about the extended symmetries of symmetric product orbifolds of two-dimensional CFTs. These include the standard Virasoro symmetries, the fractional Virasoro generators that act in the twisted sectors, as well as higher spin symmetries \cite{Apolo:2022fya}, whose structure has yet to be fully understood.

 In this  subsection, we concentrate mostly on the Virasoro symmetries and their fractional counterparts, which we will  subsequently generalise to single-trace $T\bar T$ and  $J\bar T$ - deformed CFTs. Given that the action of the symmetric product orbifold CFT is simply the sum over individual CFT copies, the  (e.g., holomorphic) stress tensor of the theory (classically denoted  $\H_L^{[0]}$) is a sum over copies
\be
T(\s) = \sum_{I=1}^N  T^I(\s) \label{ststresst}
\ee
where we are working on a fixed time slice, say $t=0$.
%{\color{ForestGreen}(mention that we work with operators at $t=0$ so we only have $\s$ dependence) }
 The Virasoro generators correspond to the Fourier modes of the stress tensor, whose quantisation depends on the boundary conditions the latter obeys. In the untwisted sector, the stress tensor in each copy obeys periodic boundary conditions, so we can write 
\be
L_m = \sum_{I=1}^N L_m^I
\ee
Note that in our conventions, the Virasoro generators are dimensionful (with dimensions of mass) for reasons that will become clear in the sequel. %They differ from those of the previous section by a factor of $R$. \emph{Fix notation!}

In the twisted sectors, the boundary conditions relate operators from the different copies. %, and we will generically obtain fractionally-moded generators, which we now review. % {\color{ForestGreen} (In the case of the stress tensor, they will span fractional Virasoro algebras, which contain the usual Virasoro algebra of the full SPO. In the following, we will discuss arbitrary single-trace operators.)}
%
%Let us start by discussing fractionally-moded generators in  the undeformed symmetric orbifold CFT on a cylinder of circumference $R$. 
 We will focus on $\mathbb{Z}_ w$ cyclic orbifolds, which are the building blocks of the twisted-sector operators. Since we will only be interested in single-trace operators, we will consider  only one cycle of length $ w$; 
%since the twisted sectors of the SPO are constructed from them, by splitting into cycles of lengths $ w$ the corresponding permutations. More precisely, in each twisted sector of the SPO, the operators are obtained as $S_N$ symmetrized products of operators of $\mathbb{Z}_w$ orbifolds, which are fields with $\mathbb{Z}_ w$ - twisted boundary conditions. 
for simplicity, we will assume that the copies of the symmetric orbifold involved %in the cycle of length $ w$ 
are $1,..., w$,  in this order. The final expressions need of course to be symmetrized over all possible choices of the $ w$ copies, to ensure  $S_N$ - invariance.  Since we will be working on a fixed time slice, all dependence on the time coordinate will be dropped.

%The final $S_N$ - invariant expressions need of course to be symmetrized over all possible choices of the $ w$ copies.  

%{\color{ForestGreen}Since we are working on the cylinder, all the operators have time dependence, but we will set $t=0$ (work in Schrodinger picture instead of Heisenberg). It is easy to relate the results to those at different values of $t$.} % out of $N$. {\color{blue}Remember that in each twisted sector, the field in the theory acquire $\mathbb{Z}_n$ - twisted boundary conditions, where $n$ correspond to the length of the cycles of the permutation. The full operators need of course to be symmetrized with respect to the full permutation group. }

% On a cylinder, twisted sector operators  can be simply defined in terms of twisted boundary conditions.  Let us take for simplicity $N_n=1$. Then, in each $Z_n$ sector, each field simply has twisted boundary conditions (\emph{Does it need to be free?}), which implies its modes are fractionated by $1/n$.

%Operators in the $ w$-twisted sector are constructed using

We thus consider
 $ w$ copies, $\phi_I(\s)$, of a (bosonic)  field $\phi$ - now taken to be generic -  with the boundary condition
$\phi_I(\sigma+R)=\phi_{I+1}(\sigma), \; \forall I$ defined mod $N$. This twisted boundary condition can be thought of as being due to the insertion  of a $w$-twist operator  at Euclidean time  $\tau=-\infty$ on the cylinder. It is  natural to  consider the combinations 
\be
\phi^{(k)}(\s) = \sum_{I=1}^{w} \phi_I(\sigma)\, e^{- 2\pi i k (I -1)/  w} \;, \;\;\;\;\; \phi^{(k)}(\s + R) = e^{2\pi i k/ w} \phi^{(k)}(\s)\;, \;\;\;\; k \in [0,  w-1]
\ee
which transform diagonally under $\mathbb{Z}_{w}$.
Thus, each  field in the theory will give rise to $ w$ fields with twisted boundary conditions. The moding of each of them will be $n - k/ w$ for  $k \in [0, w-1]$ and $n \in \mathbb{Z}$; %{\color{blue},[ and one can define the twisted vacuum to be annihilated by the appropriate fractional mode number]}
 ultimately, this will result in a field with all possible fractional modes. %In this notation, the untwisted field corresponds to $k=0$ (integer moded). % {\color{blue}[simply has a fractional mode number that is an integer]}.
  The fractional modes can be expressed via a Fourier transform as 
\begin{align}
\phi_{m/ w}&= \int^R_0 d\sigma \left. e^{2\pi i\sigma m/ R  w} \phi^{(k)} (\sigma) \right|_{k = - m \, mod \,  w}= \int_0^R d\s \sum_{I=1}^{w} \phi_I(\s) \,  e^{2\pi i m (\sigma +(I-1) R)/ R  w} \label{relfracmcopies} 
\end{align}
%
%This relation can also be inverted to express the field $\phi_I(\s)$  as
%{\color{ForestGreen}(above k was defined positive, from 0 to  w-1)}
Conversely,  the modes of the field in a single copy can be obtained by inverting the  relation above %\emph{Check signs and factors!}
\begin{align}
\phi_I(\sigma)&=\frac{1}{ w R}\sum_{m=-\infty}^{+\infty} \phi_{m/ w} \, e^{-2\pi i m \sigma/ Rw}e^{-2\pi i m(I-1)/ w} \label{fcopytws}
\end{align}
%In particular, for the energy momentum tensor, this formula defines fractional Virasoro modes:
%\begin{align}
%T^{cyl}_I(\sigma)&=\frac{1}{w}\sum_{m=-\infty}^{+\infty} L^{cyl}_{m/w} e^{2\pi i m \sigma/ Rw}e^{-2\pi i m(I-1)/w}
%\end{align}
This is not quite an operatorial relation, for two reasons: first, it can only be used   when acting 
% but it may be used  as such provided one acts
 on states from the $ w$ -
  twisted sector; second, the left-hand-side 
 % 
%  . Also, there are issues with gauge invariance. {\color{ForestGreen} This
   is not an operator in the symmetric orbifold, because it is  not gauge-invariant. This relation  can nevertheless be used to construct operators that act on the twisted sector. For example, the action of the single-trace untwisted operator $\sum_I \phi_I(\s)$ on this sector is given by the Fourier sum where only integer modes appear, as all $m$ in \eqref{fcopytws} that are not multiples of $ w$ will be projected out by the sum over $I$. Even so, note that the integer modes in this sector are not associated to any particular copy, but rather correspond to Fourier modes of the entire sum over operators, which satisfies periodic boundary conditions by construction.

% {\color{blue}   For example, one can sum over copies $I$ the expression above to obtain an operator in the untwisted sector, the sum will force the modes to be integer, as we know it's the case for untwisted operators. However, we can use the expression above to introduce the fractional modes, which no longer belong to a copy, but to the whole set of operators entering the cycle. \emph{Explain!}  Note, in particular, that the zero mode of the field in copy $I$ is related to the global zero mode in this sector as $\phi_0/ w$. \emph{This last thing looks wrong.}}

The modes \eqref{relfracmcopies} can be related to the integer  modes on a cylinder of circumference $ Rw$. This is simply achieved by letting the coordinate $\tilde \s$ on this larger cylinder equal $ \tilde \s = \s + (I-1) R$ in the $I^{th}$ patch given by $\tilde \s \in [R(I-1), R I)$, and defining a field on this covering space via $\phi^{cov}(\tilde \s) = \phi_I(\s)$ on the given patch. One can easily see that on the $I^{th}$ patch, $\phi^{cov} (\tilde \s) = \phi_I(\s) = \phi_1(\s+(I-1) R), \, \forall I$, and thus $\phi^{cov}(\tilde \s)$ is simply the field of the seed QFT, defined on this larger cylinder with periodic boundary conditions. The fractional Fourier modes discussed above become simply the Fourier modes of the  field of the seed QFT on this larger cylinder 

\be
\phi_{m/ w} = \int_0^{R w} d\tilde \s \, \phi^{cov} (\tilde \s)\, e^{2\pi i m \tilde \s/w R}  = \phi^{cov}_m \label{relfracmseed}
\ee
Note that  so far no use  of conformal symmetry was made, but rather we simply sewed the various copies of the fields into a single copy on the covering space\footnote{The standard discussions of the covering map consider the CFT on the plane, and the map is of the form $z_{base}=t_{cov}^{w}$. The Fourier modes in radial quantization are integrated over a circle of  circumference $2\pi$. To relate these to the Fourier modes on the RHS of \eqref{relfracmseed}, which are defined on a circle $w$ times larger, one
%
%Thus, to relate $\phi_{n/w}$ to the standard Fourier  modes of the field on a cylinder of circumference $2\pi$, we
 needs to perform a conformal rescaling $\tilde \s \r  \s=\tilde \s/w $. Under it, if $\phi$ is a primary field of weight $h$, then $ \phi^{\{w R\}} (\tilde \s)= w^{-h} \phi^{\{R\}}(\s) $, where the superscript indicates the size of the cylinder on which the theory is defined. It follows that
$
\phi_{m/w} = \phi_m^{\{w R\}}= w^{1-h} \phi_m^{\{R\}}
$
where for $R=2\pi$,  $\phi_m^{\{R\}}$ are the standard Fourier modes of the field; this reproduces the expressions in the literature.  We emphasize the fact that the conformal transformation is not needed to relate the fractional modes of the twisted field to those of the seed on the covering cylinder, but only in resizing that covering cylinder back to the initial length $R$. For the non-conformal models we are interested in, we will prefer to work directly with the cylinder of size $ w R$. }.
%
%and $T^{cov}(\s^{cov}) = T_I (\s)$ in each corresponding range. \emph{Not sure it is right to call it $\s^{cov}$ yet.} One may easily see that $T^{cov}$ is nothing but the seed stress tensor by noting that $T^{cov} (\s + (I-1) R) = T_I(\s) = T_1 (\s-(I-1) R)$ \emph{Fix signs!}, so they all correspond to the same field in the interval $(0,  w R)$ 
%

This procedure may be applied to any field in the theory; in particular, it may be applied to the stress tensor, which results in an infinite set of fractional Virasoro modes $L_{m/ w}$. The algebra  of these modes  defined on the  $t=0$ circle of the cylinder can be obtained using \eqref{relfracmseed}

\begin{align}
[L_{\frac{m}{ w}},L_{\frac{n}{ w}}]&=[L_m^{cov},L_n^{cov}]= \frac{2\pi(m-n)}{ w R} L_m^{cov} + \frac{4\pi^2 c \,m^3}{12 R^2  w^2} \d_{m+n}=\frac{2\pi}{R} \cdot \frac{m-n}{ w } L_{\frac{m+n}{ w}}+ \frac{4\pi^2}{R^2}\cdot \frac{c \, m^3}{12  w^2} \delta_{m+n}
\end{align}
where we have used the fact that in our conventions, the generators are dimensionful, and thus explicit factors of $R$ appear in their algebra ($ Rw$ if we are on the covering space). The above is known as the fractional Virasoro algebra and, by construction, it is isomorphic to the Virasoro algebra of the seed CFT.
 This algebra may be brought to a more standard form by  rescaling the fractional generators by $R/2\pi$, or simply setting $R=2\pi$.  The reason that the central term  simply involves $m^3$ is that we work on the cylinder, where the eigenvalue of $L_0$ is shifted with respect to that on the plane by an amount proportional to the central charge in that sector.  
%%%It is also isomorphic, by the shift of the zero modes, to the usual fractional Virasoro algebra on the plane.
%
%e.g. the stress tensor when it acts on the twisted sector.

 The global Virasoro symmetry generators correspond to choosing $m \in  w \mathbb{Z}$ which, as we already explained, are the only modes that survive in the action of the gauge-invariant operator \eqref{ststresst} on this sector. 
The fractional Virasoro modes can be used to build fractional descendants, which may be Virasoro primaries under certain conditions \cite{Burrington:2018upk}.  %They can also be used to define the twisted vacua as the the seed vacuum on the covering space \cite{}.
Relatedly,  if in the seed CFT  on the covering space, two fields are related via the action of $L_{-n}$, then their images in the twisted sector will be related via $L_{-n/ w}$.

A similar discussion holds for the case of other symmetries of the seed CFT, such as an $U(1)$ affine symmetry. %Considering the fractional modes of the $U(1)$ current
%\begin{align}
%J_{m/ w}&=\int_0^R d\sigma \sum_{I=1}^ w J_I(\sigma)e^{2\pi i m (I-1)/ w} e^{2\pi i m\sigma/ w R}
%\end{align}
%we can lift the expression, just like before, to the cylinder of circumference $ Rw$ and obtain
%\begin{align}
%J_{m/ w}&=J_{m}^{cov}
%\end{align}
%The only difference wrt the previous computation is that $J(\sigma)=J(\sigma^{cov})/ w$ and this is why we have a $ w$ factor difference wrt to the Virasoro case.
%
The commutation relations of the fractional current modes are found to be
%
% Using the usual Kac-Moody algebra on the larger cylinder, the algebra of the fractional current modes is simply given by
\begin{align}
[J_{m/ w},J_{n/ w}]&=\frac{k}{2} m \, \delta_{n+m,0} \;, \;\;\;\;\;\; 
[L_{m/ w},J_{n/ w}]=-\frac{2\pi n}{ w R} \, J_{\frac{m+n}{ w}}
\end{align}
where, as before,  $k$ represents the $U(1)$ level of the seed CFT. 
Note that since the mode number is $m/ w$, the level of this algebra, as it appears in the position space OPE,  is $k^{( w)}= k  w$, in agreement with the fact that $ w$ copies of the seed CFT are involved in the computation. %Similarly, the commutation relations between the fractional Virasoro and Kac-Moody generators are
%\begin{align}\label{viru1com}
%[L_{m/w},J_{n/w}]&=\frac{n}{w}J_{\frac{m+n}{w}}
%\end{align}
%(factors of $i$?)
More generally, for a primary operator $\mathcal{O}$ from the seed CFT with left conformal dimension $h^{(s)}$ and $U(1)$ charge $q^{(s)}$, we find that %in the $w$-twisted sector its modes are given by  \eqref{}.
%\begin{align}
%\mathcal{O}_{m/w}&=\frac{\mathcal{O}^{cov}_m}{w^{h^{(s)}-1}}
%\end{align}
%{\color{ForestGreen}
the seed commutation relations on the covering cylinder descend to the following commutation relations with the fractional Virasoro and Kac-Moody generators on the base
\begin{align}
[L_{m/ w},\mathcal{O}_{n/ w}]&= \frac{2\pi}{R w}\bigg((h^{(s)}-1)m -n\bigg)\mathcal{O}_{(m+n)/ w}\\
[J_{m/ w},\mathcal{O}_{n/ w}]&=q^{(s)}\mathcal{O}_{(m+n)/ w}
\end{align}
where we have used the relation \eqref{relfracmseed} between the fractional modes on the base and Fourier modes on the covering cylinder.  
%
%\emph{We actually seem to need this in position space!}
%
These are nothing but the momentum-space commutation relations on the cylinder, particularized to fractional momenta. %; they  can be obtained by sending $R\rightarrow  R w$.

Note that the conformal dimension of the operator  is  $h^{(s)}$, corresponding to a standard untwisted-sector operator acting on a sector with twisted boundary conditions; in particular, its short-distance behaviour is governed by $h^{(s)}$. Just as for the energy-momentum modes, the twisted boundary conditions can be thought of as generated by the insertion of a $w$-twist operator at $\tau=-\infty$, which allows the modes of any operator acting on the cylinder to be fractional. 

We would like to draw a distinction - which will become important in section \ref{strcorrf} -
 between these operators, which act in the presence of a twist inserted at a different location, and genuine twisted operators, denoted  $\O^{(w)}$, which contain a  twist  at their own location. To obtain the Ward identities of the latter  with the Virasoro generators one may simply start on the plane, and then lift the result to the covering space using the standard map $z = z_0 + (t-t_0)^w$, where $z_0$ is the insertion point of the operator. Following the steps in \cite{Burrington:2018upk} and suppressing,  for simplicity, the right-moving labels, we find
%
 %In a certain sense, these can be thought of as standard operators acting on the cylinder, where a twist operator has been inserted at $\tau = - \infty$. They are different from twisted operators that contain the twist at their location.  The Ward identity of this operator is then computed as \cite{} \emph{Understand and explain!}

\be
[L_n,\mathcal{O}^{(w)} (z_0)] = \oint_{t_0} \frac{dt}{2\pi i}\, \frac{dz}{dt}  z^{n +1} \frac{1}{z'(t)^2} \left(T(t) - \frac{c}{12} \{ z, t\}\right) \O(t_0) 
\ee
Using the OPE on the covering space and the fact that 
the Schwarzian derivative is $\{z,t\}=\frac{1-w^2}{2(t-t_0)^2}$, the above reduces to
\begin{align}
[L_n,\mathcal{O}^{(w)} (z_0)] = \oint_{t_0} \frac{dt}{2\pi i}\, (z_0+(t-t_0)^w)^{n+1} \bigg[\frac{h^{(w)}\O(t_0)}{(t-t_0)^{w+1}}+\frac{\partial_z\O(t_0)}{t-t_0}\bigg] 
\end{align}
where $h^{(w)}$ is given in \eqref{conformaldim}. Integrating and redescending the result to the base space, we obtain
\begin{align}
[L_{n},\mathcal{O}^{(w)} (z)] = h^{(w)} (n+1)z^{n}\mathcal{O}^{(w)} (z) + z^{n+1}\partial_{z}\mathcal{O}^{(w)} (z)
\end{align}
where we have dropped the index `$0$' from the base space coordinate. In contrast with our previous computation, the conformal dimension which appears in the Ward identity is now $h^{(w)}$, due to the presence of a twist  at the location of $\O^{(w)}$. The covering space considered is also different from the larger cylinder previously used and, in particular, in this case the Schwarzian derivative does yield a non-trivial contribution.  
%The difference with our previous computation is that now there is a twist operator at the location of $\O^{(w)}$, which produces the standard Ward identity with the correct dimension. 
One may then use the plane to cylinder map on the base, $z=e^{2\pi\zeta/R}$ with $\zeta = \tau + i \s$, to obtain the standard cylinder Ward identities

\begin{align} \label{Wardidcft1}
[L_{n},\mathcal{O}^{(w)}(\zeta,\bar{\zeta})]&=e^{\frac{2\pi n \zeta}{R }}\bigg( \frac{2\pi n h^{(w)}}{R }\mathcal{O}^{(w)}(\zeta,\bar \zeta) + \partial_\zeta \mathcal{O}^{(w)}(\zeta,\bar \zeta) \bigg)
\end{align}
\begin{align} \label{Wardidcft2}
[J_{n},\mathcal{O}^{(w)}(\zeta,\bar{\zeta})] = q^{(s)} \, e^{\frac{2\pi n \zeta}{R }}  \mathcal{O}^{(w)}(\zeta,\bar \zeta)
\end{align}
These commutation relations correspond to the cylinder Ward identities for a (properly periodic) operator of conformal dimension $h^{(w)}$ and charge $q^{(w)}=q^{(s)}$. Note that we could have also considered the commutation relations of this twisted  operator with the fractional Virasoro modes, which would have simply amounted to replacing   $n\rightarrow n/w$ - or, equivalently, $R\rightarrow R w$ - in he expressions above. However, since the operator creating the twist is located at $(\zeta, \bar \zeta)$, and not at $-\infty$, these Ward identities would have only held provided no other operator was inserted between these two points, which is a very restrictve requirement. On the other hand, the commutation relations with the globally-defined Virasoro generators are entirely general. 
 The right-moving commutation relations take an analogous form.% For $n\rightarrow n w$ we obtain the usual Ward identities with respect to the full integer Virasoro symmetry of the SPO.

%{\color{blue}Note these are different from \eqref{}, the difference consisting in the insertion of the twist operator at the location $\zeta$ of the operator. Note we assume that $\O^{(w)}$ is a properly periodic operator. \emph{Not sure these should be  fractional Virasoro modes; in any case, we only need the integer ones}. }

\subsection{Flow of the states in single-trace $T\bar T$ and $J\bar T$-deformed CFTs}

To uncover the Virasoro and Kac-Moody symmetries of single-trace $T\bar T$ and $J\bar T$ - deformed CFTs, we follow exactly the same strategy as in the double-trace case. For this, we first need to understand the structure of the operator that drives the flow of the energy eigenstates via the single-trace analogue of \eqref{eigenflow}, where now $|n_\l\rangle$ represent the energy eigenstates of  the $T\bar T$/$J\bar T$ symmetric product orbifold. For simplicity, we restrict our discussion to the single-trace $T\bar T$ deformation; a virtually identical analysis holds in the $J\bar T$ case.

To determine the flow operator, it is sufficient to consider  first order  degenerate quantum-mechanical perturbation theory about the instantaneous energy eigenstates $|n_\l\rangle$. Ultimately, we will  show that the flow operator in single-trace $T\bar T$ - deformed CFTs simply corresponds to the standard $T\bar T$ flow operator on the covering space, namely on the cylinder of circumference $R w$. However, to arrive at this result, we first need to understand the degeneracies that are broken by the deformation, as these play a role in determining the flow operator.  To make the discussion self-contained, we start by 
reminding the reader a few  facts about degenerate perturbation theory, following e.g.  \cite{mitcourse}. % \emph{Explain a bit better.}

\subsubsection*{Brief review of degenerate perturbation theory}

Assume we have a quantum-mechanical system where a subspace - denoted $\H_n$ - of the Hilbert space is degenerate. In this subspace, the undeformed energy levels will be denoted as $|n^{(0)}, k\rangle$, where $k \in \{1,\ldots, dim_{\H_n}\}$, and their energy as $E_n^{(0)}$. The energy levels in the orthogonal part of the Hilbert space will be denoted as $|p^{(0)}\rangle$. The $\O(\l)$ change in the energy eigenstates and eigenvalues under a perturbation $\d H=\l  H^{(1)} + \l^2 H^{(2)}+\ldots$  is given by  %\emph{How does an underbrace under the parantesis look?bad}
\be \label{pertstate}
|n, k\rangle = |n^{(0)}, k\rangle + \l   \left( \sum_{p \neq n} |p^{(0)}\rangle\, \frac{\langle p^{(0)}| H^{(1)} | n^{(0)},k\rangle}{E_n^{(0)} - E_p^{(0)}}  + \sum_{l\neq k} a_{k,l} |n^{(0)}, l \rangle\right) + \O(\l^2)
\ee

\be  \label{pertenergy}
 E_{n,k}(\l) = E^{(0)}_n + \l \langle n^{(0)},k|  H^{(1)} | n^{(0)}, k\rangle+ \O(\l^2)
\ee
We note that the first order correction, denoted $ |n^{(1)},k\rangle $,  to the state corresponds simply to the action of the flow operator $\X$ defined as in \eqref{eigenflow} on $|n^{(0)},k\rangle$ at leading order in perturbation theory. We can then easily show that, at this order 

\be
[H,\X] |n^{(0)},k\rangle = H |n^{(1)}, k \rangle - E_n^{(0)}  |n^{(1)},k\rangle = - \sum_{p\neq n}  |p^{(0)}\rangle\,\langle p^{(0)}| H^{(1)} | n^{(0)},k\rangle
\ee
Since the state $|n^{(0)},k\rangle$ is arbitrary, we conclude that, to this order %\emph{Careful, you don't need the outside}

\be \label{comflowop}
[H,\X] = - (H^{(1)} - diag) %\;\;\;\;\;\;\; \mbox{outside} \;\; \H_D \label{gcommHX}
\ee
where `$diag$' represents the diagonal matrix elements of the deforming operator on the union, denoted $\H_D$, of all the degenerate subspaces. 
% is nothing but $\X|n^{(0)} \rangle$ discussed previously. We can immediately check that the relationship \eqref{HcXst} -  which implies that
Thus,  the commutator of $\X$ with $H$ is determined by the fully off-diagonal pieces of the perturbation, whether the degeneracy is lifted or not.

The basis $|n^{(0)},k\rangle$ and the coefficients $a_{k,l}$ depend on whether the perturbation breaks or not the degeneracy of the undeformed theory. If it does, then the basis $|n^{(0)},k\rangle$ must be \emph{chosen} so that $\d H$ is diagonal, and the $a_{k,l}$ - which correspond to the matrix elements of $\X$ that lie  inside the initially degenerate subspace - are determined by the $\O(\l^2)$ analysis to be  %\emph{Check! Better notation?} 

\be
a_{k,l} = \frac{1}{E_{n,k}^{(1)}-E_{n,l}^{(1)}} \left( \sum_p \frac{ H^{(1)}_{nl,p}  H^{(1)}_{p,nk}}{E_n^{(0)}-E_p^{(0)}} + \langle n^{(0)},l| H^{(2)} | n^{(0)}, k \rangle \right) \label{corrdgs}
\ee 
where  $H^{(2)}$ is an eventual $\O(\l^2)$ correction to the Hamiltonian %{\color{ForestGreen}(in the $H^{(2)}$ term the states should be undeformed, right?)}
 (not considered in the analysis of \cite{mitcourse}) and  $\ H^{(1)}_{p,nk} \equiv \langle p^{(0)}| H^{(1)} | n^{(0)},k\rangle $, $ H^{(1)}_{nk,p} = ( H^{(1)}_{p,nk})^\star$ and we have assumed, for simplicity, that the breaking of the degeneracy is complete. 
If, on the other hand, the perturbation does not break the degeneracy at $\O(\l)$, but rather at some higher order $b$, then the coefficients $a_{k,l}$ are determined by the analysis at $\O(\l^{b+1})$. 
If the degeneracy  is never broken, then one is free to choose any basis on the degenerate subspace. Note also that the matrix elements of the perturbing operator $\d H$ are diagonal on $\H_D$, either because the basis had to be chosen so that this holds, or because $\d H$ is proportional to the identity on this subspace, as is the case when the degeneracies are not broken. 
%Writing $|n^{1}\rangle = \X |n^{(0)} \rangle$, where $\X$ is the operator that drives the flow of the energy eigenstates to this order, its commutator with the Hamiltonian, which can be inferred from the above explicit expressions, satisfies
%
\subsubsection*{Degeneracies of single-trace $T\bar T$ - deformed CFTs}

The main lesson of the discussion above is that, whenever a deformation breaks an existing  degeneracy, the elements of the flow operator in the broken subspace of the Hilbert space are fixed. Thus, in order
to understand the 
% This is basically required by the commutator structure. The question is what is the 
structure of the single-trace $T\bar T$ flow operator, % itself within the degenerate subspace.
%As we  reviewed, the answer strongly depends on
we need to check  
 whether the deformation breaks - or not - the existing degeneracies. This can be determined by the exact knowledge of the spectrum of the deformed theory. For example, in the case of the double-trace $T\bar T$ deformation, the fact that the deformed energies are only functions of the undeformed ones implies that any degeneracy initially present in the CFT will not be lifted by the deformation. Consequently, the elements of
%
%In the case that the degeneracy isn't lifted (such as for ),
 $\X_{T\bar T}$  on the degenerate subspace of the corresponding double-trace $T\bar T$ - deformed CFT are not fixed. %, since their contribution to% \emph{Careful perturbative!}
%
%\be
%\langle n, l | [H, \X] | n , k\rangle = (E_{n,l}-E_{n,k}) \langle n,l|\X| n,k\rangle 
%\ee
%the commutator vanishes because the levels remain degenerate by assumption.
 We can in particular choose, as in \cite{Guica:2022gts}, the same expression for $\X_{T\bar T}$ that is given by the assumption of non-degenerate eigenstates; this amounts to a particular way to continue the arbitrarily-chosen basis of degenerate eigenstates from the undeformed CFT to the deformed one. %It is important to note though that the expression proposed in \cite{} used the fact that the $T\bar T$ operator only had non-zero elements on the diagonal of $\H_D$, and therefore, in order to use an expression such as \eqref{}, we need to check that the same is true of the single-trace $T\bar T$ operator, acting on the much larger Hilbert space of the SPO. \emph{Not correct-fix!} %(If not true, the sum will be over operators that differ from the double-trace $\X$ in each copy - I'm really not sure we get a constraint from here, though.)
 %
%An argument that may possibly ensure that the single-trace $T\bar T$ expectation value vanishes between different degenerate eigenstates is as follows. One can use analogous manipulations to \cite{} to arrive at the following operatorial relation 
% 
% \be
%\sum_i \p_\mu H_i + \mu H_i \p_\mu H_i + \mu \p_\mu H_i H_i = \sum_i  H_i^2-P_i^2 + [H_i, \; \cdot_i \;]
%\ee
%and then try to sandwich it between the two states to show that the matrix elements of $\p_\mu H$ must be zero. In the commutator, we should be able  to replace $H_i$ by $H$ (because it commutes with fields from other copies), which implies this term will not contribute to the matrix element between two degenerate energy eigenstates, be they from the twisted sector. If we could show that $\langle H_i^2 \rangle$ vanishes between two different (twisted-sector) energy eigenstates of the same energy, then we are almost done. The second and third term on the LHS combine into $\mu \p_\mu H_i^2$. \emph{Keep? Later? Footnote?} Try: if no accidental degeneracies, then maybe same $H \Rightarrow$ same $\sum_i H_i^2$, which would basically imply that the cross terms vanish. 
%
%Thus, %if ${}_{tw}\langle n, k| \sum_i T_i \bar T_i | n,l\rangle_{tw} =0\;, \forall l\neq k$ in the twisted sector   is satisfied, 
%then we are fine as long as $\mu \neq 0$, since perturbation theory does not fix $\X$ inside the degenerate subspaces as long as the perturbation does not break the degeneracy.
 
The case of the single-trace $T\bar T$ deformation is different from the double-trace one in that  - as remarked in \cite{Apolo:2023aho} - the degeneracy is partially lifted when we turn on the deformation. 
% 
%  This happens as we turn on $\mu$; for $\mu$ incrementally changed from non-zero value to another, we do not expect any further degeneracies to be broken.
%
%In \cite{}, the expression for $\X$ was worked out for double-trace $T\bar T$ - deformed CFTs. That expression  assumed that the $T\bar T$ operator only has non-zero elements along the diagonal of the degenerate subspace. This fact can be proven using equation (2.15) in that work, which reads ($\p_\mu H$ is basically the $T\bar T$ operator)
%
%\be
%\p_\mu H + \mu H \p_\mu H + \mu \p_\mu H H = H^2-P^2 + [H, \; \cdot \;]
%\ee
%Sandwiching this between two different but degenerate energy eigenstates, we find that the matrix element of the $T\bar T$ operator between them is zero, yielding a consistency check for those results.
%
%In single-trace $T\bar T$ and $J\bar T$ - deformed CFTs, the states and their energies are again fully determined by the flow operator and the SPO structure, we just need to determine the operator $\X$ that drives the flow of the energy eigenstates. The main difference with the flow driven by the double-trace deformation is, as remarked in \cite{}, that the double-trace deformation does not break any of the degeneracies present in the original CFT, whereas the single-trace one does.
%
This breaking of the degeneracies can be easily seen from the energy formula \eqref{sumenergycycles}
\be
E= \sum_{cycles} E^{( w_i)} \left(\mu, R, E_{\scriptscriptstyle{CFT}}^{( w_i)},P_{\scriptscriptstyle{CFT}}^{( w_i)}  \right)
\ee
where $E_{\scriptscriptstyle{CFT}}^{( w_i)}$ represents the undeformed CFT energy in that sector.

 Let us first discuss  the untwisted sector, where the the undeformed energies are of the schematic form  $\sum_{I=1}^N \Delta_I +n_I$, with $\Delta_I$ the primary operator dimensions (we omit writing the shift by $c/12$, and set the radius to one) and $n_I$ the total level of the descendants. One type of degeneracies are those within a single copy (fixed $n_I$), which are the same as in the double-trace case, and thus are not lifted. Another type of degeneracies are those among different copies (fixed $\sum_I n_I$) of the seed CFT. These degeneracies are generically broken when the deformation is first turned on, as the first order correction to the energy is $\mu \sum_I (\Delta_I+n_I)^2$. For generic operator dimensions of the seed CFT,  the $\sum \Delta_I n_I$ term completely breaks any initial degeneracy; if some of the $\Delta_I$ happen to coincide, then the $\sum n_I^2$ term will break the degeneracy. A similar discussion holds for the twisted sectors, where the energy at the CFT point is given by
$E = \sum_{i \in cycles} (\Delta_i + n_i)/ w_i $. Degeneracies within a single cycle will not be broken at any order in perturbation theory, because the flow equation \eqref{effflowwsect} within a single cycle is nothing but the standard $T\bar T$ flow equation with an effective parameter $\mu/w_i$. However, the degeneracies corresponding to different ways of distributing the energy among the different cycles, with $\sum_i n_i/ w_i$ fixed, will be broken once we turn on the single-trace $T\bar T$ perturbation. We will ignore the possibility of level crossing at finite $\mu$.

 Thus, the degeneracies are only broken when the deformation is first turned on.
 Denoting the degnerate subspace at $\mu=0$ by $\H_{D^{(0)}}$, and the smaller degenerate subspace at $\mu \neq 0$ by $\H_D$, then the matrix elements  
 of the single-trace flow operator on $\H_{D^{(0)}} \setminus \H_D$  are fixed and are given by \eqref{corrdgs}
%We should therefore check that this prediction is consistent with our proposal \eqref{strflop} on the broken subspaces
; the matrix elements on $\H_D$ can be chosen at will, since further tuning $\mu$ away from zero is not expected to break any additional degeneracies. For this statement to be true, we assume that the spectrum of the seed CFT is generic, i.e. the primary dimensions are arbitrary real numbers subject to the unitarity and all other consistency constraints. 

Note that the above result should be consistent with that obtained from \emph{instantaneous } perturbation theory $\l \r \l+\d \l$, in the limit $\l \r 0$. The expression for the instantaneous correction to the state vector is given by \eqref{pertstate}, with $|n^{(0)} \rangle$ and $E_n^{(0)}$ now representing the state and energy at a given finite $\l$, and the sum in the first term running over the states outside $\H_D$. As $\l \r 0$, the energy difference between some of these states (namely, those who have inter-cycle degeneracies at $\l =0$) becomes $\O(\l)$, which implies that the numerator of that expression will receive contributions from different orders in the $\l$ expansion. The structure of these terms is similar to the expected contribution \eqref{corrdgs}, and we expect that
%
%
%
%Let us write the deformed $T\bar T$ Hamiltonian  as $H= H^{(0)} + \l H^{(1)} + \l^2 H^{(2)} + \ldots$, and the deformed eigenvalues as $|n \rangle = |n^{(0)} \rangle + \l |n^{(1)} \rangle + \ldots $ which do not have any inter-cyce degeneracies, as long as $\l \neq 0$, but only intra-cycle ones (that we have supressed). In this case, the correction to the state - i.e., the matrix element of $\X^{(\l \neq 0)}$  - is determined by standard non-degenerate perturbation theory to take the general form  \emph{factor of $2$ problem?}
%
%\be
%%\d |n_\l\rangle \sim  \d \l  \frac{ (\langle m^0 |+\l \langle m^1 |)( H^{(1)} + 2 \l H^{(2)} )| (n^0+\l n^1) \rangle }{E_m-E_n} \sim 2 \frac{(\langle m^0|H^{(2)} | n^0\rangle +  \langle m^0 |H^{(1)} | n^{1} \rangle }{E_m-E_n}
%%\ee
%We note that in the limit $\l \r 0$, the energy difference between two states that are degenerate at $\l=0$, but not at $\l \neq 0$ is $\O(\l)$, which cancels the factor of $\l$ from the numerator. One may also explicitly check that among such states, $\langle n, l |H^{(1)} |n,k\rangle =0$. This agrees \emph{\textbf{up to a factor of 2??}} with the general predictions of degenerate perturbation theory.
 the matrix elements that get fixed on $\H_{D^{(0)}} \setminus \H_D$ precisely coincide with those predicted by general non-degenerate perturbation theory outside $\H_D$, and we need not worry about them. The elements inside $\H_D$ are not fixed, and can be chosen conveniently.

%{\color{ForestGreen}(2 possibilities, $\sum_I n_I/ w_I$=fixed with fixed set $\{ w_1,...\}$ (degeneracies within the same sector) or with different sets (degeneracies between different sectors). For the case between different sectors it is trivial that the coefficients below vanish because the perturbation is from the untwisted sector, so the matrix elements between different sectors are 0. We only need to check within a given sector.)}\emph{Check! Conjugate?} {\color{ForestGreen}(wrong unless untwisted sector)}

%If we could somehow show that the above is zero, then we would have shown that our proposed flow operator is the correct one. \emph{Show!} 

%
%
% A very simple way to see that only the single-trace $T\bar T$ deformation preserves the orbifold structure is to look at the action. 
%
%\bi
%\item check matrix elements of $T\bar T$ in the sum
%\item prove that $a_{lk}$ is zero, by showin descendants of two ortogonal states are orthogonal
%\ei
%%

The case of the single-trace $J\bar T$ deformation can be treated in an exactly analogous manner. Since the first order correction to the energy is $\l \sum_I q_I (\bar h_I + n_I) $, with $\bar h_I$ the undeformed right-moving dimensions, it follows that the degeneracy among the different cycles will generically be broken as soon as the deformation is turned on. The rest of our conclusions follow straightforwardly.

%, this applies at first order in $\l$ if the charge of the state is non-zero, if not one needs to wait until $\O(\l^2)$ for the degeneracy to be broken. \emph{Shall we do a more complete analysis?}

%
% The flow of the energy eigenstates is driven by the operator defined via
%
%\be
%[H, \X] = \O_{\sum T_i\bar T_i} - diag
%\ee
%which shoud hold as an operator equation quite generally \emph{Check!} barring degeneracies that we probably need to take into account. 

%The above discussion should be at the level of maxtrix elements. More simply, since $H = \sum_i H_i$ is single-trace and the RHS is single-trace, it follows that $\X = \sum_i \X_i$. \emph{Check its action on twisted sector states, too!}. Below is aslo comparison of the action of single-trace versus double-trace $T\bar T$ on untwisted sector states in the SPO. 

 %One immediately notes that states from different twisted sectors will not mix, because the deforming operator belongs to the untwisted sector. The same is true of the double-trace $T\bar T$ deformation. 

%It seems the answer is rather trivial, because a single-trace deformation will preserve the SPO structure. \emph{Is this the only type of deformation that will preserve it?}

\subsubsection*{The single-trace $T\bar T$ flow operator }

Let us now apply this to the single-trace $T\bar T$ deformation. Using \eqref{comflowop},  the matrix elements of the single-trace flow operator, schematically denoted as $\X_{T_I\bar T_I}$,  satisfy%\footnote{In this section, the flow operator for the single-trace case will be denoted simply as $\X$, while the notation $\X_{T\bar T}$ is reserved for the double-trace $T\bar T$ flow operator, which will serve as a building block for the former.}  %\emph{Notation? }[
 %One should deform by the integral of the $T\bar T$ operator, rather than integrating the correlation function, which introduces extra punctures. ]
%
\be
[H, \X_{T_I\bar T_I}] = \sum_I  \int d \s \, \O_{T_I \bar T_I} - diag \label{HcXst}
\ee
where `$diag$' now refers to the diagonal matrix elements of the integrated single-trace $T\bar T$ operator on $\H_D$. As discussed in the previous subsection, even though the single-trace $T\bar T$ operator itself is a sum over copies, its integrated version is not, except in the untwisted sector.    Since then also $H=\sum_I H_I$, it is natural that 

\be
\X_{T_I\bar T_I} = \sum_I \X^I_{T\bar T} \;\;\;\;\;\; \mbox{in the untwisted sector, outside} \;\; \H_{D}
\label{strflop}
\ee
 where $\X_{T\bar T}^I$ is the flow operator associated with a single copy of the $T\bar T$ - deformed CFT, which is worked explicitly in \cite{Guica:2022gts}, at least at the classical level. This relation only need to hold outside $\H_D$, which consists of only in-cycle degeneracies; however, as discussed, the elements of $\X_{T_I\bar T_I}$ can be chosen at will inside it.

  The fact that the flow operator takes this form in the untwisted sector can also be seen very explicitly by studying the flow equation for states in this sector, which are built as sums over all possible permutations $|n_\l^{\s(1)}\rangle \otimes \ldots |n^{\s(N)}_\l\rangle$  of $N$ fixed energy eigenstates,  $|n^I_\l\rangle$,  in the seed theory. Plugging this into the right-hand-side of \eqref{pertstate} and using the fact that $\d H$ consists of a sum over copies in this sector, we find that the instantaneous energy eigenstates $|m_\l\rangle$ (corresponding to   $|p^{(0)} \rangle$ in \eqref{pertstate}), which in this sector take the form  $|m_\l^1\rangle \otimes \ldots |m^N_\l\rangle$, fully symmetrized,  should ``click'' with $|n_\l\rangle$ in all its entries but one, where the particular copy of the $T\bar T$ operator acts. More precisely
 \be
 \p_\l |n_\l\rangle = \sum_{ m\neq n} |m_\l\rangle \frac{\langle m_\l|  \int \O_{T_I \bar T_I}  | n_\l \rangle}{E_n^\l - E_m^\l} = \sum_{perm} \sum_I |n^{1}_{\l}\rangle \otimes \ldots  \sum_{ m^{I}\neq n^{I}} |m^{I}_\l\rangle \frac{\langle m^{I}_\l | \int \O_{  T_I \bar T_I} | n^{I}_\l\rangle}{E_n^{I,\l}-E_m^{I,\l}} \otimes \ldots |n^{N}_{\l}\rangle
 \ee
%
% {\color{ForestGreen}
% \be
% \p_\l |n_\l\rangle = \sum_{ m\neq n} |m_\l\rangle \frac{\langle m_\l|  \int \O_{T_I \bar T_I}  | n_\l \rangle}{E_n^\l - E_m^\l} \propto \sum_{\substack{\s,\tau\in S_N}} \sum_I |n^{\s(1)}_{\l}\rangle \otimes \ldots  \sum_{ m^{\tau(I)}\neq n^{\s(I)}} |m^{\tau(I)}_\l\rangle \frac{\langle m^{\tau(I)}_\l | \int \O_{  T_I \bar T_I} | n^{\s(I)}_\l\rangle}{E_n^{I,\l}-E_m^{I,\l}} \otimes \ldots |n^{\s(N)}_{\l}\rangle
% \ee
 %{\color{ForestGreen}(state normalization $1/N!$)}
 where in the second step we % plugged in the explicit expressions for the twisted sector states and 
 used the fact that  the energies in the untwisted sector are just sums of the energies of the individual copies. Both sums run only over non-degenerate eigenstates; the sum over permutations should be properly normalised. % Since we are only interested in the structure of the states and flow operators, we will not write explicitly the normalization factors throughout this section.
  We have also set all ambiguities in the state due to (intra-cycle) degeneracies to zero; including them would  shift the individual copy contributions, without affecting the general structure. 
   Near $\l=0$, one could also check that the fixed matrix elements of $\X_{T_I\bar T_I}$ do not break the symmetric product orbifold structure, as would be implied by having $a_{k,l} \neq 0$ between different cycles; however, this is ensured by the fact that the $\l \r 0$ case can be embedded in the generic $\l \neq 0$ one, which does respect the symmetric orbifold structure. 
 %
  %, at the end of the day the denominator of \eqref{} will simply reduce to ... we obtain 
%
Since the terms in the intermediate sums may be written as $\p_\l |n^I_\l\rangle = \X_{T \bar T}^I | n^I_\l\rangle $, we find that  
%{\color{red}(fix notation, indices for copies up or down, $\l$ index; sum over permutations, we should put some $\s$ like $n_{\sigma(1)}$ etc; fix normalisation, if we want $\partial_{\l}| n_\l\rangle$ to have norm 1 I got $\frac{1}{\sqrt{N N!}}$)}
%
\be
\p_\l | n_\l\rangle = \sum_{perm} \sum_I |n_\l^{1}\rangle \otimes \ldots  \p_\l |n^{I}_\l\rangle \otimes \ldots |n^{N}_{\l}\rangle
\ee
%
% {\color{ForestGreen}(state normalization $1/N!$)}
%
%\be
%\p_\l |n_\l\rangle = \sum_{perm} \sum_I |n_1\rangle \otimes \ldots  \sum_{m_I\neq n_I} |m^I_\l\rangle \frac{\langle m^I_\l | \int d\s \O_{ T_I \bar T_I} | n^I_\l\rangle}{E_m^I-E_n^I} \otimes \ldots |n_N\rangle
%\ee
and so  the deformed state is just the tensor product of $T\bar T$ - deformed states - thus confirming the fact that single-trace $T\bar T$ preserves the symmetric orbifold structure\footnote{  By contrast, the action of the double-trace $T\bar T$ deformation on the symmetric  orbifold is given by  
%%
%\be
%\p_\l |n_\l\rangle \propto \sum_{\substack{\s,\tau\in S_N}} \sum_{I , J} |n^{\s(1)}_{\l}\rangle \otimes \ldots  \sum_{m^{\tau(I)}\neq n^{\s(I)}} |m^{\tau(I)}_\l\rangle \frac{\langle m^{\tau(I)}_\l |  T_I  | n^{\s(I)}_\l\rangle}{E_m^I-E_n^I} \otimes \ldots \otimes \ldots  \sum_{m^{\tau(J)}\neq n^{\s(J)}} |m^{\tau(J)}_\l\rangle \frac{\langle m^{\tau(J)}_\l |  \bar T_J | n^{\s(J)}_\l\rangle}{E_m^J-E_n^J} \otimes \ldots |n^{\s(N)}_{\l}\rangle
%\ee
%
\be
\p_\l |n_\l\rangle = \sum_{perm} \sum_{I , J} |n^{1}_{\l}\rangle \otimes \ldots  \sum_{m^{I}\neq n^{I}} |m^{I}_\l\rangle \frac{\langle m^{I}_\l |  T_I  | n^{I}_\l\rangle}{E_m^I-E_n^I} \otimes \ldots \otimes\!\!  \sum_{m^{J}\neq n^{J}} |m^{J}_\l\rangle \frac{\langle m^{J}_\l |  \bar T_J | n^{J}_\l\rangle}{E_m^J-E_n^J} \otimes \ldots |n^{N}_{\l}\rangle
\ee

which implies that $\mathcal{X}_{T\bar T \, on \, SPO}$ has a bilocal structure, which does not respect the symmetric product form of the state.} - and that the flow operator in this sector takes the form \eqref{strflop}.

 In the twisted sector, it is no longer true that the deforming operator is a sum over copies; nevertheless, an explicit expression can still be easily  obtained for $\X_{T_I\bar T_I}$, at least at the (semi)classical level, by following the steps of \cite{Guica:2022gts}. More precisely, we have
 %
% in this case the expression for $\X$ acting on the twisted sector corresponds to the descent to the base  of $\X_{T\bar T}$ on the covering space.  The relationship with $\X_{T\bar T}$ on the covering space can be seen explicitly as follows. We will be working at classical level, where the exact expression for $\X_{T\bar T}$ has been worked out in \cite{}. 
 %Using the same steps, one can show the deforming operator reduces to 
% 
\be
 \int d\s \, \O_{T_I\bar T_I} =  \left[H, i\! \int d\s d\tilde \s \, G(\s-\tilde \s) \H^I(\s) \mathcal{P}^I(\tilde \s)\right] + \frac{1}{R} \int d\s  d \tilde \s \left( \H^I(\s) T_{\s\s}^I(\tilde \s) -  \mathcal{P}^I(\s) \mathcal{P}^I(\tilde \s) \right) \label{strOttb}
\ee
Upon summing over the various copies, the left-hand-side of this equation, minus its diagonal piece,  corresponds precisely to $[H,\X_{T_I\bar T_I}]$. To obtain a closed-form expression for it, one may use  the $T\bar T$ trace relation in each copy 
\be
T_{\s\s}^I (\s)=\H^I(\s) - 2 \mu \O_{T_I\bar T_I}(\s) 
\ee
which holds classically as is, and quantum-mechanically  up to a total derivative. Plugging this into \eqref{strOttb} and acting with the whole equation on a twisted sector, in which the expansion of the fields takes the form \eqref{fcopytws} (we are ignoring for now the copies that do not participate in the twist), the integral over $\H^I(\s)$ that multiplies $\O_{T_I\bar T_I}$ in the second term yields a factor of $H^{(w)}/ w$, where $H^{(w)}$ is the Hamiltonian restricted to the $w$-twisted sector  and the index `$w$' is now a shorthand for the individual  $w$ copies that enter the cycle. Ultimately we find, at the classical level 

\be \label{flowcover}
\X_{T_I\bar T_I}^{( w), cls} = \frac{1}{1 + 2\mu H^{( w)}/(R  w) } \int d\s d\tilde \s \, G(\s-\tilde \s) \H^I(\s) \mathcal{P}^I(\tilde \s)
\ee
This expression corresponds to nothing but the seed $\X_{T\bar T}$ on the covering space. One way of showing this is by expanding $\H^I(\s)$ above as a function of the undeformed CFT generators $\H^{(0)}_I (\s)$ and $\mathcal{P}^I(\s)$, using the closed-form expression given in \cite{Guica:2022gts}. 
%
% is precisely that of $\X_{T\bar T}$, but on a cylinder of radius $R  w$. For this, we first expand .   
  When acting upon a twisted-sector state, which we choose again to be a single cycle of length $ w$, each of the fields can be expanded in a sum over fractional modes as in \eqref{fcopytws}.  We then  note that the integral over the sum  of $\s$, $\tilde \s$ ensures that the $I$ dependence will drop out from the integral, leading to an overall $ w$ factor. The dependence on the Fourier modes of $\H^{(0)}$ and $\mathcal{P}$ will be the same as in the double-trace expression, but with $m \r m/ w$. Since each factor of $\mu$ in the expansion is accompanied by a factor of $\H^{(0)}_I$ or $\mathcal{P}^I$, the overall factor of $w$ that multiplies the $\mu^p$ term in the expansion is $ w^{-p}\times  w \times  w$, where the last factor comes from the Fourier transform of the Green's function. The Fourier integrals also bring a factor of $R^{-(p-2)}$, by dimensional analysis. We therefore notice that the $ w$ and $R$ dependence is such that   it combines into precisely a dependence on $R  w$, which coincides, upon lifting to the  covering space, with the flow operator of the double-trace $T\bar T$  - deformed CFT. Including quantum corrections  amounts to replacing factors of $L_{m/w}$ in the expansion by factors of $n/wR$ that would result from commuting two such fractional Virasoro modes, in perfect agreement with the general counting. Thus, in the $w$-twisted sector, $\X_{T_I \bar T_I}$ acts just as  the descent of $\X_{T\bar T}$ on the covering space to the base cylinder. 
  
  % ) \emph{Understand!} %We need not worry about more complicated structure because the deforming operator is single-trace, and thus so is $\X$.

%{\color{ForestGreen} While \ref{flowcover} gives the expression for the flow operator in the cycle of length $w$ to which the copies $(1,...,w)$ participate, the total flow operator in a generic sector corresponding to a conjugacy class $[g]$ should be written as: 
%\be
%\X^{[g]} = \sum_{\s \in S_N}  \X_{T\bar T}^{(\s(1)\ldots \s( w_1))} +    \X_{T\bar T}^{(\s( w_1+1)\ldots \s( w_1+ w_2))} + \ldots 
%\ee 

More generally, if there are several twisted and untwisted sectors present - as determined by the conjugacy class, $[g]$, of the permutation group -  then \eqref{strOttb} reduces to a sum over sectors which, using the orthogonality of the various subsectors of the Hilbert space, leads  to the following expression for the total flow operator $\X_{T_I\bar T_I}$ when acting on $\H^{[g]}$
% \textbf{\emph{Normalisation?Notation ok?}}

\be
\X^{[g]}_{T_I\bar T_I} = \sum_{\s \in S_N}  \X_{T_I\bar T_I}^{(\s(1)\ldots \s( w_1))} +    \X_{T_I\bar T_I}^{(\s( w_1+1)\ldots \s( w_1+ w_2))} + \ldots  \label{genstflop}
\ee 
% where the various cycles of lengths $ w_i$ (with $\sum_i  w_i=N$)  correspond to the structure associated to $[g]$
 %
% the sum over cycles corresponds to the sum over sectors given by cycle structures $\{w_1,w_2,...\}$ 
% and the upper indices of the operators indicate the copies which participate in the given cycle.
%
Here $w_i$ are the lengths of the cycles that appear in $[g]$, each contributing with a flow operator constructed along the lines of the previous paragraph, which lifts to the flow operator of the seed $T\bar T$ - deformed CFT on the cylinder of circumference $R w_i$. The sum over permutations considers all combinations in which the copies enter the cycles of $[g]$, making the full operator well-defined on the Hilbert subspace $\H^{[g]}$.   Note that when all cycles have length one, this reduces to the expression \eqref{strflop} for $\X$ in the untwisted sector.

  Being obtained from the commutator \eqref{HcXst}, the above expression for $\X_{T_I\bar T_I}$  holds except possibly on the diagonal degenerate subspaces. However, as we have just argued, we shouldn't expect the matrix elements on $\H_D$ to be fixed; our expression \eqref{genstflop} corresponds to a particular choice.

   % We would now like to show that it can be extended to $\H_D$, too. 
 %
% While the proposed expression is very simple, the fact it is the correct answer also on the degenerate subspaces is not completely trivial 
%
%For this, we need to understand the structure of the degenerate energy subspaces in single-trace $T\bar T$ - deformed CFTs. 
%
%\medskip
%Note that it is not \emph{a priori} clear that the flow equation of the seed theory on the covering space and the flow equation \eqref{fracvirdef} on the base will agree. \emph{Move up?}

\subsection{Symmetries of single-trace $T\bar T$ and $J\bar T$-deformed CFTs\label{fractionalvirspo}}

Having discussed the flow operator, we would now like to study the extended symmetries of single-trace $T\bar T$ and $J\bar T$ - deformed CFTs, following the steps of the double-trace analysis. We therefore define a set of  Virasoro generators  as\footnote{Even though we use exactly the same notation as in the double-trace case, it should be clear from the context that these are the Virasoro generators in single-trace $T\bar T$/$J\bar T$ deformed CFTs. } 

\be
\p_\l \widetilde L_m^\l = [\X_{T_I\bar T_I}, \widetilde L_m^\l] \;, \;\;\;\;\; \widetilde L_m^{\l=0} = L_m^{\scriptscriptstyle{CFT}}
\ee

\subsubsection*{Untwisted sector analysis}

This equation is very easy to solve  in the untwisted sector: 
since $\X_{T_I\bar T_I}$ is a single-trace operator   and so is the initial $L_m^{\scriptscriptstyle{CFT}}$, it follows that the flowed $\widetilde L_m^\l$  will simply be 
\be \label{untwistedvir}
\widetilde L_m^{\l} = \sum_I \widetilde L_m^{I, \l}
\ee 
where the $\widetilde{L}_m^{I,\l}$ represent the solution to the flow equation in a single copy of a $T\bar T$ / $J\bar T$ - deformed CFT. The same definition can be extended to all the other Virasoro and Kac-Moody generators of the seed symmetric product orbifold. Their algebra consists of two-commuting copies of the Virasoro ($\ltimes$ Kac-Moody) algebra by construction. As explained in the introductory subsection, the non-trivial check that these generators represent symmetries of the theory is to prove that they are conserved. For this, we compute their commutator with the Hamiltonian 

\be
[H, \widetilde L_m^\l] = \bigg[\sum_I H_I, \sum_J\widetilde  L_m^{J,\l}\bigg] = \sum_I \a_m^I  \widetilde L_m^{I,\l} \;, \;\;\;\;\; \a_m^I \equiv \a_m (H_I, P_I)
\ee
where the $\a_m$ for the cases of interest are given in \eqref{expralm}. 
This immediately implies that $\sum_I e^{i \a_m^I t }\widetilde L_m^{I,\l}$ satisfies the conservation equation \eqref{conservation}.  While one may worry that different operators $\a_m^I$ appear in the time-dependent factors above, their expectation value in any state of the symmetric orbifold is the same, as it cannot depend on the particular copy.

\subsubsection*{Twisted sector analysis}

In the twisted sectors  on the cylinder, %where a twist operator is inserted at $t=-\infty$,} 
$\widetilde L_m^\l $ and $\X_{T_I\bar T_I}$ no longer take the form of a sum over copies. As explained in the introductory subsection, in this sector we will generically obtain fractionally-moded generators, of which the integer-moded generators are a particular case. We will  therefore discuss the preservation of  the
%
%\emph{Write nicer:} In particular, in the case of $\tilde L_0 = \sum_I \tilde L_0^I$, one would like to show that its spectrum is identical to that of the undeformed CFT. \emph{Move to previous subsection:} For this, one needs to compute the expectation value of $\sum_I H_I + \mu (H_I^2-P_I^2)$ in a twisted sector state. Using the definition of the single-trace $T\bar T$ operator, one finds that
% 
% \be
% -\p_\mu H = \sum_i H_i^2-P_i^2 + \mu \p_\mu \sum_i H_i^2 + \sum_i [H_i, \chi_i]
% \ee
% In the last sum, $H_i$ may be replaced by $H$ (as operators from different copies commute), and dropped inside an expectation value.  
%
%\subsubsection*{
fractional Virasoro and Kac-Moody symmetries generically. 
%
%\subsubsection*{Fractionally-moded generators in $T\bar T$ and $J\bar T$ symmetric orbifolds}
%
%After this somewhat lengthy introduction, we would  like
%
%\bi
%\item however, the symmetric product orbifold has a much larger symmetry group (partly given by fractional Virasoro). On the cylinder, it is
%\be
%[L_{m/w},L_{n/w}] = \frac{m-n}{w} L_{(m+n)/w} + \frac{c m (m^2-1)}{12 w^2} \delta_{m+n}\;, \;\;\;\mbox{if}\;\; L_0 |0_w\rangle = 0
%\ee
%\item one can similarly argue that the fractional Virasoro generators stay conserved by computing the $[H, L_{m/w}]$ commutator. \emph{True?}
%\item this computation seems trivial for such a single-trace twisted sector operator, since it can be explicitly written as a sum over $L_i$ with some phase factors.  \emph{Correct?}
%
%\item if true, what are the symm they generate? Is this known even in the CFT case?
%\ei
%
%\noindent Our goal is thus 
%
%to discuss fractional Virasoro modes in the single-trace $T\bar T$ and $J\bar T$ - deformed CFTs. %starting from the flowed $\widetilde L_m$.  %or, alternatively, the $Q_i$. We may try
We again define them  via the flow equation

\be \label{fracvirdef}
\p_\l \widetilde L^\l_{m/ w} = [\X_{T_I \bar T_I}, \widetilde L^\l_{m/ w}] \;, \;\;\;\;\;\;\; \widetilde L_{m/ w}^{\l=0} = \widetilde L_{m/ w}^{\scriptscriptstyle{CFT}}
\ee
This is a perfectly well-defined equation in the Hilbert space of the $T\bar T$/$J\bar T$ symmetric orbifold. Since $\X_{T_I\bar T_I}$
belongs to the untwisted sector, it infinitesimally takes a $ w$-twisted sector operator to another one. The standard Virasoro generators %(Fourier modes of \eqref{}) 
are obtained when
 $m$ is  a multiple of $ w$.

 We would now like to show that  the solution to this  flow equation corresponds precisely to the $T\bar T$ solution for the $\widetilde L_m^\l$ on the covering space - which is a cylinder of size $R w$ - reinterpreted on the base via \eqref{relfracmseed} . As discussed at the beginning of this section, this map makes no use of conformal invariance,  but only sews together the different copies. For simplicity, we assume again that the twist acts on the first $ w$ copies of the seed, and as identity on the remaining $N- w$ ones. A summation over $S_N$ will render the final result gauge-invariant.

As we have already discussed,  in the $w$-twisted sector, the flow operator on the base simply corresponds to the flow operator  $\X_{T\bar T}$ of the seed theory on the covering cylinder. Also, in the undeformed CFT, the fractional Virasoro modes correspond to integer modes on the covering. It follows that the solution to \eqref{fracvirdef} will be the same as the solution to \eqref{floweqvir} for the seed theory, but on a cylinder of radius $R  w$. Since we particularized our discussion to a given number of copies on which the single-cycle twist acts, only one of the $\X^{(w)}$ factors in \eqref{genstflop} will act non-trivially on the generator, but the final expressions will be symmetrized with respect to $S_N$. For example, in the trivial case $w=1$, we obtain $\widetilde{L}_m^{\lambda}$ in the seed for the first copy and the symmetrization will yield the untwisted sector result \eqref{untwistedvir}. 
%
%This follows from the fact that the flow operator on the base simply corresponds to $\X_{T\bar T}$ on the cover, the solution to the flow equation on the covering space will be the same as in the seed, but on a cylinder of radius $R  w$. %When $m$ is a multiple of $ w$, the solutions will need to agree with the single-trace expressions \eqref{}, but an argument similar to the above show that they will. \emph{Check!}  
%
%{\color{blue}Since we are considering single-trace twisted-sector operators, only one of the $\X$ factors in \eqref{genstflop} will act non-trivially on the generator. The final expressions will need to be symmetrized with respect to $S_N$. 
The same procedure can be applied to the right-moving Virasoro generators, as well as to the Kac-Moody currents. %{\color{ForestGreen} In the following, to simplify the notation, we will drop the $\l$ index for the generators; everything we write is valid at any value of $\l$.}
 %
 %  Since the radius enters the definition of $\X$, the solution for $\widetilde L_m^{cov}$ (\emph{Careful, this is not the rescaled cylinder!}) will depend non-trivially on $ w$ and the (Fourier modes of the) fields on the covering space.
 %
    %relations such as \eqref{relfracmseed} and \eqref{relfracmcopies}.  
 %
 %While $\X$ is an untwisted single-trace operator, there is no simple relation of the fractional Virasoros to the copies. However, since these are integrated operators
%, we may try to simply define them as flowed Virasoro modes of the seed on the covering cylinder, which has radius $ w R$. This map does not appear to require conformal invariance.
%
 %It is also possible to think about this as a purely operatorial definition, without a need for the cover.

%More precisely, note that $L_{m/ w}$ is a twisted single-trace operator, i.e. it looks like a twisted operator 

In order for the solution to  \eqref{fracvirdef} to define a set of conserved operators,  we first need to compute its commutator with the Hamiltonian. % and show an appropriate time-dependence can be added so this is the case. {\color{ForestGreen} 
This may be simply evaluated on the cover, where it yields $\a_m(H^{cov},...,  w R) \widetilde L_{m}^{cov}$, where the dots stand for the other conserved quantities that enter the definition \eqref{expralm} of $\a_m$ (i.e., momentum for $T\bar{T}$, and also $U(1)$ charge for $J\bar{T}$) and we have dropped the index `$\l$' from the generator,  to lighten the notation. Translating this to the base cylinder, we find $\a_m(H^{(w)},... ,w R) \widetilde L_{m/w}$. %where the general formula for $\a_m$ is given in \eqref{expralm}.
%
%{\color{blue}Redescending this relation to the base, we find $\a_m(H^{(w)},  w R) \widetilde L_{m/ w}$. }
%
One may also derive these expressions from the non-linear   relation \eqref{spoTTbar} between the undeformed and deformed spectrum in the twisted sector, which implies a (non-linear and $w$ - dependent) relation between $H^{( w)}$ and $\widetilde L_0^{( w)}$.  That this relation depends on the particular sector is not surprising, given that the definition of $\widetilde L_0$ is sector-dependent. 

 %, where $H^{( w)}=\sum_{i=1}^{w} H_i$ {\color{ForestGreen}(correct?)} is the unique copy of the Hamiltonian associated to that twisted sector.
 
 Adding the appropriate time-dependence to ensure conservation, and taking  into account all the possible choices of copies that can enter the cycle,
  the full answer for the  fractional Virasoro generators  then takes the form %{\color{red}\textbf{\emph{Notation? Normalisation?}}}
%{\color{ForestGreen}

\be
\widetilde{L}_{m/ w}(t) = \sum_{\s\in S_N} e^{i \a_m (H^{(\s(1)\ldots  \s(w))} ,...,  w R) t} \widetilde{L}_{m/ w}^{(\s(1)\ldots  \s(w))}
\ee
where the superscripts indicate the particular copies  entering the non-trivial cycle.
%
% which are conserved. The full time-dependent answer is best worked out on the covering spaces. 
Note this reduces to the correct answer in the untwisted sector ($ w=1$). The same holds for the right-moving Virasoro and  Kac-Moody generators.
%
%, where $m$ is a multiple of $ w$ and the irreducible cycle has length one. \emph{Is it true that whenever $m$ and $ w$ are not coprime, we are actually in the $m/ w$ - twisted sector? } 
The algebra of the fractional Virasoro generators above is the same as in the undeformed CFT, as follows from the definition \eqref{fracvirdef}.
%
%Given this expression, one can simply flow it under $T\bar T$. 
%
%{\color{blue}[The algebra is unchanged. Conservation should be implied by the commutator with the Hamiltonian, which on the cover is $\a_m(H^{cov}) L_m^{cov}$. \emph{Is this thing a product of operators on the base, with one factor $L_{m/ w}$? What is then the other factor? How does one add the time dependence?} It may be simpler and sufficcient to descend the conservation equation. 
%
%\be
%[H, L_{n/w}] = [\sum_i H_i, \sum_j L^j_{n/w} e^{2\pi i \frac{n j}{w}}] = \sum_j \a^j_{n/w} L_{n/w}^j e^{2\pi i \frac{n j}{w}}
%\ee
%
%where for the calculation we need the algebra of the fractional Virasoros, which should be identical to the CFT fractional Virasoro algebra via the flow argument.  Conservation trivially follows by attaching some time-dependent phase factors.
% We thus find the full fractional Virasoro-(Kac-Moody) symmetry is preserved. 
 In particular, for  $m$ a multiple of $ w$ we obtain the integer Virasoro modes, which are present in any sector of the theory.  
 
 %Note that the %While $H^{( w)}$ can be considered as a sum over copies, $\tilde L_0^{( w)}$ will not be so, even if it started its life as a sum over copies of the Hamiltonian of the undeformed CFT. This is not inconsistent, given that it is only defined inside the twisted sector, where different copies are allowed to mix. It appears to correspond to a multi-trace operator of zero twist in that sector. \emph{How can these interactions between the copies be ever induced by the single-trace operator $\X$?}
 
Note that in principle, these modes could also be obtained by flowing $\H_{L,R} (\s)$ and then performing the Fourier transform, as in \cite{Guica:2022gts}. Even though the operators we start from are single-trace operators, $\sum_I \H_{L,R}^I(\s)$, the solution to the flow equation 
 looks differently in the different sectors because $\X_{T_I\bar T_I}$ does. The result of the flow equation will lead to a notion of emergent field-dependent coordinates, which will also be defined in the twisted sectors; from the structure of the flow we see they will correspond precisely to the field-dependent coordinates $u,v$ on the covering space. %\emph{Can we descend non-Fourier modes back to the base?} 
 
To conclude this section, we  have explicitly shown that  single-trace $T\bar T$ and $J\bar T$ - deformed CFTs contain operators that are conserved and satisfy two commuting copies of the Virasoro ($\ltimes$ Kac-Moody) algebra, showing they possess the corresponding symmetry. %The Virasoro (-KM) algebra follows straightforwardly from the one of the copies, and the central charge picks up a factor of $N$.
The central charge and $U(1)$ level of the algebra are simply $N$ times those of the seed theories. In twisted sectors, one finds also fractional Virasoro ($\ltimes$ Kac-Moody) conserved modes.% {\color{ForestGreen} For the rest of the article, we will refer to $w$-twisted sector quantities as those for which the copies $1,...,w$ enter the cycle and assume that the final results are properly symmetrized.}

%
% Note the expectation value of this operator has a uniform time dependence, since $\langle \a_m^i \rangle$ is independent of $i$ for \emph{any} state in the symmetric product.% \emph{Check this is true also for twisted sector states!}
%Consequently,  

\subsubsection*{Other bases of generators} \label{spectrflowgen}

As discussed in section \ref{sgenerators}, the flowed Virasoro generators may not  provide 
%it is not clear whether the flowed basis is 
the most natural  basis of generators of the extended symmetries of these theories. %We may therefore prefer to use the  ``physical'' generators  in each copy. {\color{ForestGreen}
 In single-trace $J\bar T$ - deformed CFTs, the most natural basis of left-moving generators is given by the standard generators of  conformal and affine $U(1)_L$ symmetries. In the untwisted sector, they take the form 
%
%In the untwisted sector, for single-trace $J\bar T$ - deformed CFTs, these are \emph{Check conventions $\l$!} 
%
\be
L_m = \sum_I L_m^I = \sum_I \widetilde L_m^I + \frac{\l}{R} H_R^I \tilde{J}_m^I + \frac{\l^2 k}{8\pi R} H_{R,I} ^2\, \d_{m,0} \; , \;\;\;\;\;\; J_m = \sum_I J_m^I = \sum_I \widetilde J_m^I + \frac{\l k}{4\pi} H_R^I \, \d_{m,0}
\ee
and similarly for the right-moving generators, where we are now working in the convention in which the $L_m$ are dimensionful. In the $ w$-twisted sector, one can write similar relations by considering the seed on the cylinder of circumference $ Rw$, as instructed by the solution of the flow equations %\emph{I added a tilde on J}
\be \label{unflowedL}
L_{m/ w} = \widetilde L_{m/ w} + \frac{\l}{R  w} H_R^{( w)} \tilde J_{m/ w} + \frac{\l^2 k}{8\pi R w} H_{R} ^{( w)2}\, \d_{m,0} \; , \;\;\;\;\;\; J_{m/ w} = \widetilde J_{m/ w} + \frac{\l k}{4\pi} H_R^{( w)} \, \d_{m,0}
\ee
where $H_R^{( w)}$ is the (globally-defined) right-moving Hamiltonian, restricted to  the $w$ - twisted sector, with eigenvalues the $ w$-twisted sector energies \eqref{chtransf}. One may easily check that for $m=0$, these yield the correct expression \eqref{chtransf} for the deformed energies in the $w$ - twisted sector, taking into account the fact that the eigenvalue of $\widetilde L_0$ is identical  to that in the undeformed CFT.  Note that, throughout this section, the twisted generators will be built from the first $w$ copies of the seed, and symmetrization is assumed only for the final expressions.

% As a check for this proposal, we can multiply by $R$ in order to obtain dimensionless operators and consider their eigenvalues. We see that they match \eqref{conformaltwisted} (up to the central charge factor which requires mapping to the plane?) and \eqref{chtransf}.

Rewriting the relations above in terms of the effective Kac-Moody level in the $w$-twisted sector $k^{(w)}=w k$, we obtain that the relation between the two sets of generators is given by spectral flow with $\lambda H_R^{(w)}/w$. Since the relation is non-linear, the Poisson algebra spanned by the untilded generators is non-linear in these generators.

%Since the theory is still a CFT on the left-moving side, one can map the relations to the plane and  then to the covering space

%The unflowed fractional Virasoro and Kac-Moody generators is given by
%\begin{align}
%L_{n/w}=\tilde{L}_{n/w}+\frac{\lambda H_R \tilde{J}_{n/\omega}}{w}+\frac{\lambda^2 k^{(s)} H_R^2}{4w}\delta_{n,0} \hspace{1cm}J_{n/w}=\tilde{J}_{n/w}+\frac{\lambda k^{(s)}H_R}{2} \delta_{n,0}
%\end{align}  } %Note that this is consistent with our results for the left conformal dimensions and charges in the $w$-twisted sector, providing an interpretation: \eqref{conformaltwisted} are the eigenvalues of the unflowed fractional Virasoros (on the plane). \emph{Details on how factors of $ w$ appear from the covering map? Earlier?}

Note that for $m$ a multiple of $w$, we obtain the global generators of extended symmetries, which correspond to the  Fourier integrals of state (and thus sector) - independent quantities over the base cylinder. This can be established, at least classically, as follows: the solution for the generators $L_m, \bar L_m$ is the same as the double-trace solution, uplifted to the covering space. E.g., for the left-movers, we have, using \eqref{relfracmseed}  
%{\color{ForestGreen}(is it confusing to use $\s$ for the coordinates since we have permutations throughout the paper, for ex 3.52?)}

\be
L_m = L_{mw}^{cov} = \int_0^{w R} d\tilde \s \, e^{2\pi i m \tilde \s/R} \H_L (\tilde\s)
\ee 
with $\H_L (\tilde \s + w R) = \H_L (\tilde \s)$. This integral can be reduced to the integral on an interval of size $R$ of $\sum_{I=0}^{w-1} \H_L(\tilde \s+R I)$, which is periodic with period $R$. Thus, the integral above becomes a local integral of the periodic current on the base space. The same argument can also be applied to the right-movers, the only difference being that the integrand is now a  non-linear function of the fields.  We also expect it to extend to the quantum case; note it implies that  $L_m, \bar L_m$ (whose most appropriate quantum definition  may or may not exactly coincide with \eqref{physicalgen}, \eqref{rmphysgen} with $\l \r \l/w$) are the physical symmetry generators in the full theory. 

If $L_m$, $J_m$ and their right-moving counterparts are the global symmetry generators in single-trace $J\bar T$ - deformed CFTs then, given that the relation between them and $\widetilde L_m$, $\widetilde {J}_m$, etc. is that of spectral flow with parameter $\l H_R/w$, it follows that the  flowed generators are explicitly sector-dependent.  This fact is not suprising, given that the flow operator used to define these generators also  depends explicitly on the sector of the theory. It appears possible that requiring at most implicit sector dependence  of the symmetry generators in single-trace $T\bar T$/$J\bar T$ - deformed CFTs may provide a criterion for selecting the physical  basis of generators in these theories.

%At least on the left-moving side, these  should definitely not depend on $w$.  This fact can be ascertained classically from the explicit expression for $L_m$ as an integral of $\H_L$. In fact, it would be better if we could start from the integral on the cover, and then show that for the global modes this can be defined directly on the base. It also becomes clear from \eqref{unflowedL} that, even if the expression for $L_m$ has this nice property, that for $\tilde L_m$ will not, because te effective flow parameter will stay $\l/w$. 

%{\color{red}( Clearly from the flow eq it follows that flowed Virasoro depend on $w$. Can we prove that the $w$ dependence cancels in the combination above?)}

In single-trace $T\bar T$ - deformed CFTs, one may argue, at least classically, that the natural generators of ``unrescaled'' field-dependent coordinate transformations in the untwisted sector of the symmetric product orbifold should be %\textbf{\emph{Rescale by $R $?}}
% {\color{ForestGreen}(it's consistent with \eqref{relqlt})}

% to use the single-trace version of the `unrescaled' generators \emph{Naming!}
%
%which are given by $L_m = Q_m (R+ 2 \mu H_R)$ in the seed. If we study the flow equation allowing for inhomogenous terms on the RHS (roughly, $Q_m$ up to some factors of the radius) , one may be able to obtain directly the $Q_m$ from the flow equation. This indicates we should be able to define the single-trace operators

\be
Q_m = \sum_I Q_m^I = \sum_I R\, \widetilde L_m^I/R_u^I \;,\;\;\;\mbox{ with}\;\;\;\;\;\;  R_u^I = R+2\mu H_R^I \label{strQs}
\ee
where the relevant classical expression for $\widetilde L_m$ for the $T\bar T$ deformation is given in \eqref{clsttbgen}; dividing by the field-dependent radius factors yields an expression for $Q_m$ that is the integral of a quasi-local current. 
Once the appropriate quantum definition of the ``unrescaled'' generators of the field-dependent symmetries is understood in the seed theory, the result generalizes trivially in the untwisted sector as above.
%
%{\color{blue} where we have written only the relation between the classical generators (the appropriate  quantum dfinition of the generators of the field-dependent symmetries not being currently understood).} % In any case, once a relation is fixed in double-trace $T\bar T$, it single-trace generalisation is trivial. 
%In the following, we woud like to make a few simple remarks about the structure of the algebra of these generators, which in each copy are non-linearly related to the Virasoro ones. We will concentrate on the single-trace $T\bar T$ case, for simplicity. 

The algebra of these generators is % in both cases
 a sum over copies of the non-linearly-deformed  Virasoro algebras obtained in the double-trace case, given up to $\O(\hbar^2)$ by %\textbf{\emph{Factors $2\pi$?}} 
 %{\color{ForestGreen}(it's consistent with \eqref{doubletralgebra})}
 
%  {\color{blue}For example, for single-trace  $T\bar T$ - deformed CFTs, the    algebra of the ``unrescaled'' single-trace generators \eqref{strQs} is, up to $\O(\hbar^2)$}
%
%whose algebra is given by the non-linear deformation of the Virasoro algebra \emph{Need quantum version}

\be
[Q_m, Q_n] =2\pi \hbar (m-n) \sum_I \frac{Q_{m+n}^I}{R+2 \mu H_R^I} + (m-n) \sum_I \frac{8\pi \hbar\, \mu^2 H_R^I Q_m^I Q_n^I}{R\,R_u^I R_H^I} + \frac{\pi^2 c\hbar\, m^3}{3}   \sum_I \frac{1}{(R_u^I)^2} \d_{m+n} \label{strttbalg}
\ee
It is interesting to note that  \emph{new} operators appear on the right-hand-side in the single-trace case, rather than a product of operators that were already among the generators, as in the double-trace. As we already discussed,  it would be good to understand more deeply whether this basis of operators may be preferred for physical reasons and, if so, what is the significance of this non-linear algebra. 

In the $w$-twisted sector, one can introduce new fractional generators by performing the division on the covering space  of circumference $R w$, with the result %\textbf{\emph{Rescale by $R w$?}}
\begin{align}
Q_{m/ w}=\frac{ Rw\, \widetilde L_{m/ w}}{R_u^{( w)}}
\end{align}
with $R_u^{( w)}= Rw+2\mu H_R^{( w)}$. For $m$ multiple of $ w$ we obtain the integer versions on the base. These operators may be interpreted as implementing field-dependent coordinate transformations on the covering space, using  a Fourier basis. The algebra of these operators is obtained by replacing $Q^I, H_R^I$ by $Q^{(w)}, H_R^{(w)}$ in the above%{\color{red}(is it clear here that we are saying that the whole sum over I the copies in the cycle gets replaced by the $w$ quantities?)}
, $m $ by $m/w$, the sums over copies with sums over cycles.   Interestingly, if we take the expectation value of this algebra (restricted to its integer modes) in  a high-energy state, the result  precisely coincides with that of the asymptotic symmetry group analysis of the asymptotically linear dilaton black hole backgrounds performed in \cite{Georgescu:2022iyx}. % produced a result that with the expectation value of \eqref{strttbalg} in. This expectation value can be computed by assuming large $N$ factorization and then  simply  replacing $\sum_i \r N$ and $Q^i \r Q/N$ for each of the operators appearing in the sum. %\textbf{Check!} 
% {\color{red}\emph{Correct?} (I think yes)}  
 Note %that if we take the \emph{expectation value} of the RHS, we can replace the sums by appropriate factors of $N$, which will
this effectively amounts to replacing $\mu \r \mu/N$ in the double-trace algebra \eqref{doubletralgebra}, which is precisely what was found by the holographic analysis. % \cite{Georgescu:2022iyx}. 

\subsubsection*{Higher spin currents}

As discussed in the previous section, $T\bar T$ and $J\bar T$ - deformed CFTs also preserve the KdV charges associated with integrability. In their single-trace version, it is natural to introduce the single-trace analogues of the $\tilde I_s$, which in the untwisted sector are given as a sum over copies of the corresponding  expressions \eqref{kdvdef} - \eqref{kdvdef2} with $L_n^I \r \widetilde L_n^I $. Their conservation follows trivially from the conservation of the flowed KdV charges in the seed $T \bar T$/$J\bar T$ - deformed CFT, given that the Hamiltonian is a sum over the Hamiltonians in each copy.  In the twisted sectors, their conservation follows from that on the covering space.

The symmetric product orbifold of a two-dimensional CFT possesses, however, many other higher spin conserved currents, associated to its much larger symmetry algebra. 
%
% a much larger symmetry group than Virasoro; instead, one roughly expects $Virasoro^N/S_N$. \emph{More! Literature!} The Virasoro symmetries discussed above form only a small part of this algebra. 
%
%One way to discribe these additional symmetries is to note the existence of
As explained  in e.g. \cite{Apolo:2022fya},  these higher-spin primary currents  only exist thanks to multi-trace contributions to an otherwise non-primary field. One such higher-spin current that has been given as an example therein %example %in \cite{Apolo:2022fya} and, in terms of $L_m^i$, its $n^{th}$ Fourier mode reads
 is, for $R=2\pi$ and specialising to the untwisted sector
\be
(W_4)_n = \sum_{I=1}^N \sum_m  L_m^I L_{n-m}^I - \frac{3}{10} (n+2)(n+3) \sum_I L_n^I - \frac{\frac{22}{5c} +1}{N-1} \sum_{m, I\neq J} L_m^I L_{n-m}^J
\ee
%It was shown in that paper that it gets an anomalous dimension once one deforms away from the SPO point by an exactly marginal operator of non-trivial twist. \emph{Check!} The anomalous dimension is $\O(1)$ at large $N$ provided the coupling scales as \emph{What?} 
%valid in . {\color{red} (are these on the cylinder? should there be factors of $R$?)}
%
One may wonder whether this current could be transported along the single-trace $T\bar T$/$J\bar T$ flow in a similar manner to how the Virasoro and KdV currents were transported, i.e. by requiring it to be covariantly constant along the flow, which simply amonts to replacing the Virasoro modes by their flowed counterparts. Taking $n=0$ for simplicity, we would like to show that this simple procedure does not lead to a conserved charge. 

For this, we compute the commutator of $(\widetilde W_4)_0$ with the Hamiltonian. The terms that correspond to single-trace sums are conserved, following the same steps as we used to show the conservation of the KdV charges. However, the multitrace terms lead  to a different result, due to the fact that the commutator $[L_m^I, \a_n^J]$ only takes the form \eqref{comLalpha} if $I=J$, and is zero otherwise. Using this, we find, 
%
%In this paragraph, we would like to show that this current also appears to be broken under the single-trace $T\bar T$ deformation, in that the corresponding charge is not conserved, provided we simply flow it. The beaking is  due entirely to the multitrace terms, and after the summation and including the $1/N$ coefficient in front, yields a non-conservation of order \emph{What?}
%
%Using this expression,  one may straightforwardly transport $W_4$ along the flow (similarly for the KdV charges). However, it appears that this current will stop being conserved in the deformed theory. For this, we use \emph{Check!}
%
%\be
%[\widetilde{L}_m^i, H] = \a_m^i \widetilde{L}_m^i\;, \;\;\;\;\;\;\; [\widetilde{L}_m^i, \a_n^j] = \delta^{ij} (\a_{m+n}^i-\a_m^i-\a_n^i) \widetilde{L}_m^i
%\ee
%%We note that
%\be
%[L_m L_{-m},H] = L_m \a_{-m} L_{-m} + \a_m L_m L_{-m} = (\a_0 -\a_m-\a_{-m})L_m L_{-m} + (\a_{-m}+\a_m)  L_m L_{-m} =0
%\ee
%thus implying that the first term in $\int W_4$, as well as things such as KdV charges, remain conserved; however, the double-trace contribution to $\int W_4$ does not appear to be conserved
%
for $I \neq J$
%\textbf{\emph{Notation! Is this just for $I \neq J$?}} {\color{ForestGreen}(yes, otherwise $L^I$ does not commute with $\alpha^I$ to give the last equality)}
\be
[\widetilde L_m^I \widetilde L_{-m}^J, H] =  \widetilde L_m^I \a_{-m}^J \widetilde L_{-m}^J + \a_m^I \widetilde L_m^I \widetilde L_{-m}^J = (\a_m^I +\a_{-m}^J) \widetilde L_m^I \widetilde L_{-m}^J
\ee
In a CFT, $\a_m^I = m \hbar = - \a_{-m}^J$ so this term vanishes; however, in $T\bar T$ and $J\bar T$ - deformed CFTs, this is no longer the case. To leading order in $\mu$, $\a_m^I = m \hbar - 2 \mu m \hbar (H^I-P^I)$, implying that the leading term breaking the conservation is proportional to $\mu \hbar  \sum_m m (H_R^I-H_R^J) \widetilde L_m^I \widetilde L_{-m}^J$, a combination that does not appear to vanish.  Similar results hold in single-trace $J\bar{T}$ - deformed CFTs, on the right-moving side.  
 Thus, the na\"{i}ve flowed charge associated to this higher-spin symmetry is not conserved in the deformed theory, thus confirming the idea that the conservation of the flowed Virasoro generators and KdV charges is a special feature of these operators, which does not extend to arbitrary currents in the theory. This analysis points towards the  conclusion  that the associated higher spin symmetries are broken by the single-trace $T\bar T$/ $J\bar T$ deformations; one should ascertain though that it is not possible to construct corrections to the charges that would restore their conservation.

%% so it seems the sum will be non-zero. This implies $\int W_4$ will no longer be conserved in the deformed theory. It is not clear whether other multi-trace combinations will now be conserved, or the deformed theory simply has fewer symmetries than the undeformed one (which is very likely). 
%
%{\color{ForestGreen}We can expand $\alpha_m$ around $\mu=0$ (CFT):
%\begin{align}
%\alpha_m&=m\hbar-2 m \hbar(H-P)\mu+4 m\bar{h}(H+m\bar{h})(H-P)\mu^2+\mathcal{O}(\mu^3)
%\end{align}
%Hence, we obtain:
%\begin{align}
%[\int W_4,H]&=\frac{\frac{22}{5c}+1}{N-1}\sum_m \sum_{i\neq j}2m\hbar \mu (H^i-H^j-(P^i-P^j))  \tilde{L}_m^i\tilde{L}^j_{-m}+\mathcal{O}(\mu^2)
%\end{align}
%Since there are $N(N-1)/2$ choices of pairs $(i,j)$, it seems that this will scale like $N$. (is this correct?)
%}
%
%Same can be done with the other currents discussed in \cite{}. It seems that flowed currents will be conserved provided they are single-trace. \emph{True generally?} For example, the KdV charges of the SPO satisfy this requirement. 
%
%\bi
%\item what are the full symm of SPO? $Vir^N/S_N$? Are fractional Virasoros a complete basis?
%\item what is the relation between them and the higher spin primaries? 
%\item anything useful from higher spin square ($T^4$ orbifold, if I understand correctly)?
%\ei

\section{Correlation functions}\label{correlation}

In this section, we would like to discuss correlation functions of operators in symmetric product orbifolds of $T\bar T$ and, especially, $J\bar T$ - deformed CFTs. Since these theories are neither conformal, nor local, our focus and  methods will naturally be quite different from the standard discussion of correlation functions in symmetric product orbifolds of CFTs \cite{Dixon:1986qv,Lunin:2001pw,Lunin:2002fch,Dei:2019iym}, which is centered around computing correlators of
 twist operators using  non-trivial covering maps, and makes essential use of the conformal transformation properties of the operators in question.  We will instead concentrate on momentum-space operators, which are natural to consider in a non-local theory, and our main goal will be to understand how to  choose a  special basis of these operators %in single-trace $T\bar T$ and $J\bar T$ - deformed CFTs
  and compute their correlation functions in terms of the correlators of the undeformed symmetric orbifold CFT.

Our  main focus will be  $J\bar T$ - deformed CFTs, for whose double-trace version \cite{Guica:2021fkv} has proposed a concrete basis of ``primary operator analogues'' and computed their correlation functions exactly. The main goal of this section will be to adapt this prescription to the single-trace  $J\bar T $ case and use it to compute  correlation functions of both  untwisted and twisted-sector operators. It is worth mentioning that, quite recently, \cite{Aharony:2023dod} has also put forth a special basis of operators in $T\bar T$ - deformed CFTs and computed their correlation functions, obtaining  similar expressions. While a generalisation of these results to the single-trace case would be both interesting and likely possible using their formalism, we do not address this problem here, mainly because it would require a very different method than  the one we  use for the $J\bar T$ case.

 %For  the case of $T\bar T$ - deformed CFTs,  a basis of operators whose correlation functions could be computed at finite deformation has only very recently been proposed, and the generalisation to the single-trace deformation, while seemingly possible, would involve rather different methods than we will be using for $J\bar T$. The answer is expected to take the form discussed recently in \cite{Cui:2023jrb}.

% are much less understood from a field-theoretical perspective, and so we  only  make a few generic observations. %[It is likely that the $T\bar T$ case can be treated in a similar way, once the appropriate definition of the operators is understood. ]

The correlation functions in the symmetric product  orbifold of $T\bar T$ and $J\bar T$ - deformed CFTs can then be compared to the correlation functions evaluated using worldsheet  techniques in the  holographic setups that have been related to these deformations. More precisely,  they are expected to match the correlation functions of vertex operators  associated to long strings\footnote{By contrast, the short string sector  is not described by a symmetric product orbifold, and neither their spectrum, nor their correlation functions \cite{Asrat:2017tzd, Giribet:2017imm,soum} match those in $T\bar T$/$J\bar T$ - deformed CFTs.} in these backgrounds, which are well described by a symmetric product orbifold. Such worldsheet correlation functions were recently computed in \cite{Cui:2023jrb} for the case of the asymptotically linear dilaton background,  which is related to the single-trace $T\bar T$ deformation; their large-momentum behaviour in the untwisted sector agrees with that found by \cite{Aharony:2023dod}. 
%
% We comment upon the fact that the field theory computation very likely matches the worldsheet one for the long string sector, but not for the short string one, as expected. 
One may similarly compute correlation functions of long string vertex operators in warped AdS$_3$ by adapting the results of \cite{Azeyanagi:2012zd} along the lines of \cite{Cui:2023jrb}, and then compare them with  our  single-trace $J\bar T$ result; we find a slight disagreement on the non-local  side that we comment upon. 

% for the correlation function to (an adaptation of) existing results for worldsheet  correlators in warped AdS$_3$ spacetimes, and comment on the differences we find.  
%
% We show that the two results \emph{do not match}, and explain the reasons why this was  expected.
  For completeness, we  start this section with a review of the correlation functions in  double-trace $T\bar T$ and $J\bar T$ - deformed CFTs, focusing on the explicit proposal  of \cite{Guica:2021fkv} for the latter case. %We then discuss correlation functions in the symmetric product orbifold of these theories. We end with a comparison with the string-theory results. 

\subsection{Review of correlation functions in $T\bar T$ and  $J\bar T$ - deformed CFTs \label{revcorrf}}

In local QFTs, there is a special set of operators whose correlation functions are interesting to study, namely local operators.  This set is further specialised in CFTs, where much of the focus is on operators  that transform as primaries under conformal transformations, as their correlation functions are highly constrained by conformal invariance. 

Since $T\bar T$ and $J\bar T$ - deformed CFTs are non-local, it is not a priori clear which are the natural operators to consider, if a preferred basis exists at all \cite{Dubovsky:2012wk}. Indeed, due to the non-locality of these theories,  operators are best defined in momentum space. The computation of their correlation functions using e.g. conformal perturbation theory yields divergences, which need to be subtracted via counterterms. Since in a non-local QFT the structure of the allowed counterterms is in general not known, the finite part of the correlator may itself become ambiguous. The situation is under better control in $J\bar T$-deformed CFTs, whose locality and $SL(2,\mathbb{R})$ invariance on the left-moving side  single out a set of primary operators under these symmetries, for which at least the left-moving part of the correlation function is fixed \cite{Guica:2019vnb}; however, these questions remain for the right-moving, non-local piece of  the correlator. 

Despite these concerns and complications, there has been much recent progress in the computation of correlation functions of both $T\bar T$ and $J\bar T$ - deformed CFTs.
Upon making a judicious choice of the operators whose correlation functions one would like to study, it can be shown that the end result is a simple integral transform of the correlation functions of the original CFT. For the case of two-and-three point functions, this integral simply yields the CFT momentum-space correlator, with the conformal dimensions replaced by certain momentum-dependent combinations.  Thus, the deformed correlators,  though explicitly non-local, are directly and universally determined by the correlation functions in the undeformed CFT, indicating that both $T\bar T$ and $J\bar T$ - deformed CFTs have a  structure that is as rigid as that of two-dimensional CFTs.

\subsubsection*{$T\bar T$ - deformed CFTs}

The effect of the $T\bar T$ perturbation on correlation functions can be studied at small coupling using conformal perturbation theory, where one can easily note that it induces a momentum-dependent correction to the conformal dimension \cite{Kraus:2018xrn}. The first all-orders analysis of the  correlation functions of $T\bar T$-deformed CFTs  has been performed by \cite{Cardy:2019qao}, who also studied their leading UV divergences  and showed they can be absorbed into a non-local renormalization of the operators. More recently, \cite{Aharony:2023dod} approached this problem using the  path integral formulation of the $T\bar T$ deformation in terms  of coupling the undeformed CFT to JT gravity \cite{Dubovsky:2017cnj,Dubovsky:2018bmo}. As it is most clear from  its vielbein formulation, this description relates the $T\bar T$ - deformed dynamics to that of the original CFT, but seen through a set of  ``dynamical coordinates'' that are related to the  $T\bar T$ ones in a universal, but  field-dependent fashion.  These coordinates parametrize   a flat target space,  identified with the space of the undeformed CFT. The basis of operators considered in  \cite{Aharony:2023dod}  corresponds to the original CFT operators on this space or, alternatively, their Fourier transform with respect to the target space coordinates. 
%
%a non-dynamical vielbein, which maps to a target space where the undeformed theory lives. In this proposal, the operators of interest correspond to the undeformed CFT operators   assigned  to specific points in the target space. 
%
 The  correlation function of these momentum-space operators is obtained by carefully  performing the  JT  path integral and absorbing the UV divergences into a renormalization of the operators. The procedure is rather subtle and involved, and the final result takes the form %\textbf{\emph{Fix convention for $\mu$!}} 
 %{\color{ForestGreen}(we'll check again 4.1,4.2)}
 
 % taking care of fixing  the diffeomorphism and Weyl redundancies. To obtain a convergent result, certai Wick rotations are performed.  The final answer takes the form 

%e most extensively been studied in %,Aharony:2018vux}}
%. The  quoted above
%
%was attempted in {\color{ForestGreen}\cite{Kruthoff:2020hsi}}. Assuming the basis of operators does not flow, it was shown that the deformed correlation function could be expressed in terms of the lower-order correlator with some Wilson-line dressing of the operators. These operators were 
%argued that operators  have a momentum-space two-point function of the form

\be \label{TTbarcor}
\langle \O_1(p_1) \ldots  \O_n(p_n) \rangle \propto \d (\sum_{i=1}^n p_i) \int \prod_i d^2 \s_i\,  \d (\s_1)  e^{i \sum_i p_i \s_i} \mathcal{F}(p_i) \langle \O_1(\s_1) \ldots \O_n(\s_n) \rangle_{CFT} \prod_{i<j} \left( \Lambda |\s_{ij}|\right)^{\frac{\mu p_i \cdot p_j}{\pi}}
\ee
where $\Lambda$ is a renormalization scale and $\mathcal{F}(p_i) $ is a smooth function of the momenta that grows at most polynomially at large $p_i$.  If one ignores this term and considers e.g.  the two-point function, one finds it corresponds precisely to the CFT momentum-space correlator

\be
\langle \O(p,\bar p) \O(-p, - \bar p)\rangle = \frac{(2\pi)^2 }{2^{2(h+\bar{h})} \sin(\pi(h+\bar{h})) } \,\frac{p^{2h -1} \bar{p}^{2\bar{h}-1} }{\Gamma(2h)\Gamma(2\bar{h})}
\ee 
%\be
%\langle \O(p,\bar p) \O(-p, - \bar p)\rangle = \frac{(2\pi)^2 }{2^{2(h^{[\mu]}+\bar{h}^{[\mu]})} \Gamma(2h^{[\mu]})\Gamma(2\bar{h}^{[\mu]})\sin(\pi(h^{[\mu]}+\bar{h}^{[\mu]})) } p^{2h^{[\mu]} -1} \bar{p}^{2\bar{h}^{[\mu]}-1} 
%\ee  
but with the dimensions replaced by the momentum-dependent combinations (in our conventions)

\be
h= h(p,\bar p) = h_{\scriptscriptstyle{CFT}} + \frac{\mu}{\pi} p \bar{p}\;, \;\;\;\;\;\;\; \bar{h}=\bar h(p,\bar p) =\bar{h}_{\scriptscriptstyle{CFT}} +\frac{\mu}{\pi} p \bar{p}
\ee
This is precisely the answer obtained in \cite{Cui:2023jrb} by computing the correlation functions of long string worldsheet vertex operators, as we review in section \ref{holography}. It is interesting to ask whether a different renormalization of the operators in \cite{Aharony:2023dod} could reproduce this exact formula.

% Using Stirling's approximation for the $\Gamma$ function, the answers agree in the large momentum limit. Whether Lorentzian momentum-space correlators can also be defined and whether the result can be Fourier-transformed back to position space are interesting questions that are likely to be investigated in the near future. 

\subsubsection*{$J\bar T$ - deformed CFTs}

As already stated, the main goal of this section  is to compute correlation functions of both untwisted and twisted-sector operators in single-trace   $J\bar T$ - deformed CFTs.
 For this, we will adapt the prescription of \cite{Guica:2021fkv} for defining appropriate analogues of primary operators and computing  their correlation functions. In order to facilitate the generalisation of these results to the single-trace case, 
 we  present  the construction of \cite{Guica:2021fkv}  in a slightly different fashion, which we hope also makes its physical interpretation   more transparent.

%The study of correlation functions in $J\bar T$ - deformed CFTs could be put on firmer grounds, thanks to the lesser degree of non-locality of these theories.

This construction relies on the interplay of physical and flowed Virasoro generators in $J\bar T$ - deformed CFTs. Let us start by discussing the left-moving sector, where the meaning of the various operators is very clear.  As we already explained, the standard $SL(2,\mathbb{R})_L$ conformal symmetry of the theory unambigously identifies a set of left primary states whose conformal dimensions  become   momentum-dependent  due to the irrelevant deformation %, as expected 
\cite{Guica:2010sw} 
%\emph{Factors $2\pi$!}
\be \label{momentumdepcf}
h(\bar p) =  h^{[0]} + \l  q^{[0]} \bar p + \frac{ \l^2 k \bar p^2}{4} \;, \;\;\;\;\;\; q(\bar p) =  q^{[0]} + \frac{ \l k \bar p}{2}
\ee
where $ h^{[0]},  q^{[0]}$ are the undeformed left conformal dimension and charge.  Note this relation takes precisely the form of a spectral flow by $\l \bar p$. To simplify the expressions in this section,  we have rescaled the deformation parameter $\l\to 2\pi\l$ with respect to the previous ones. We also set $R=2\pi$. % for most of this section. 
% . 
%{\color{blue}[The left conformal dimensions of the operators can be obtained from the finite-size spectrum via an infinite boost together with a limiting procedure on the radius.]}

The state-operator correspondence, which is still valid  \cite{Nishida:2007pj}, 
%
%The left conformal invariance 
then predicts the existence of a set of primary operators with these conformal dimensions, which obey the standard Ward identities with respect to the true conformal generators $L_m, J_m$, and for which the left-moving part of the correlation function is fixed by $SL(2,\mathbb{R})_L$ in the standard way. % Of course, the much more interesting question is whether there exists a preferred basis of operators, for which the right-moving part of the correlation function is also fixed. %An affirmative answer was given in \cite{Guica:2021fkv}, who provided a concrete construction of these operators and computed their correlation functions in terms of those in the undeformed CFT. 
To construct them, one starts by formally defining the `flowed' operators  \cite{Kruthoff:2020hsi}

\be \label{floweqoppr}
\p_\l \widetilde \O^\l(\zeta,\bar \zeta) = [\X_{J\bar T}, \widetilde \O^\l(\zeta,\bar \zeta)]
\ee 
where $\zeta, \bar \zeta$ are the  coordinates on the cylinder and $\X_{J\bar T}$ is the operator that drives the flow of the energy eigenstates in the $J\bar T$ - deformed CFT\footnote{The flow is defined on a fixed time slice, say $\tau=0$. The operator at $\tau \neq 0$  is simply defined as $\O_{\scriptscriptstyle{CFT}} (\tau, \s) = e^{i H_{\scriptscriptstyle{CFT}} \tau} \O_{\scriptscriptstyle{CFT}} (0,\sigma) e^{-i H_{\scriptscriptstyle{CFT}} \tau}$, and the whole expression is then flowed. }. By construction, their correlation functions in the flowed vacuum state and  their commutation relations with  $\widetilde L_n, \widetilde J_n$ are identical to those in the undeformed CFT
\be
[\widetilde L_n , \widetilde \O(\zeta,\bar{\zeta})] = e^{ n \zeta} \bigg(  n\,  h^{[0]} \, \widetilde{\O}(\zeta,\bar{\zeta}) + \p_\zeta \widetilde{\O}(\zeta,\bar{\zeta})\bigg) \;, \;\;\;\;\;\;[\widetilde{J}_n, \widetilde{\O} (\zeta,\bar \zeta) ] = e^{ n \zeta}  q^{[0]} \, \widetilde{\O} (\zeta,\bar \zeta)
\ee
and similarly on the right-moving side.  Despite this fact, the $\tilde{\O} (\zeta,\bar \zeta)$ do not correspond to physical operators in the theory and are non-local even on the left-moving side; thus, $\zeta,\bar \zeta$ should be simply viewed as labels inherited from the undeformed CFT.

The operators we are interested in should instead be primary with respect 
%Let us start by discussing the left-moving sector, where the meaning of the various generators is very clear. The physical operators that should be considered are primary operators with respect
to the untilded Virasoro generators \eqref{physicalgen}, which in our new conventions read

\be
L_m = \widetilde L_m + \l H_R \widetilde J_m + \frac{\l^2 k}{4} \, H_R^2\, \d_{m,0} \;, \;\;\;\;\;\; J_m = \widetilde J_m + \frac{\l k}{2}\, H_R \, \d_{m,0} \label{lmgenph}
\ee
with left dimensions given by \eqref{momentumdepcf}. In order for these dimensions to make sense, they should also be 
 eigenoperators of $H_R$, with eigenvalue $\bar p$. 
  We  will find it convenient to work in a mixed basis, $(\zeta, \bar p)$. The operators should satisfy 
 
 \be
 [H_R,\O(\zeta,\bar{p})]=\bar{p}\,\O(\zeta,\bar{p})
 \ee
 Since the relation between the undeformed and deformed dimension is given by spectral flow, the relation between the physical and tilded operator should involve dressing by an appropriate vertex operator. %{\color{ForestGreen} For spectral flow in a CFT, such vertex operator can be written as $e^{i\eta (\varphi(\zeta)+\bar{\varphi}(\bar{\zeta}))}$, with
%  \begin{align}
%  \varphi(\zeta)=\phi_0-i \tilde{J}_0\zeta +i\sum_{n\neq 0}\frac{\tilde{J}_{n}}{n}e^{-n\zeta} \hspace{1cm}   \bar{\varphi}(\zeta)=\bar{\phi}_0-i \tilde{\bar{J}}_0 \bar{\zeta}+i\sum_{n\neq 0}\frac{\tilde{\bar{J}}_{n}}{n}e^{-n\bar{\zeta}}
%  \end{align}
%  
%  }
%  
Note, however, that $\widetilde \O$ already carries the correct charges with respect to $J_0$ and $L_0$,  implying that we should remove by hand the zero mode of the vertex operator involved. Our Ansatz is, therefore

%Based on \cite{Guica:2021fkv}, we look at:
\be\label{leftmovop}
\O(\zeta,\bar{p})=\int d\bar{\zeta\, }e^{-\bar{\zeta}\bar{p}}:\widetilde{\V} \widetilde{\O} (\zeta,\bar \zeta):  e^{ \mathcal{Y}_\O \zeta + \bar{\mathcal{Y}}_\O \bar \zeta } \ee
where the normal-ordered dressed operator is given by 

\be
 :\widetilde{\V}\widetilde{\O}(\zeta,\bar{\zeta})\!: \; \equiv\; \widetilde{\V}_{\eta}^{+}(\zeta)\widetilde{\bar{\V}}_{\eta}^{+}(\bar{\zeta})\, \widetilde{\O}(\zeta,\bar{\zeta})\, \widetilde{\bar{\V}}^{-}_{\eta}(\bar{\zeta}) \widetilde{\V}^{-}_{\eta}(\zeta)
\ee
Here, $\widetilde{\V}^\pm$ represent the positive and, respectively,  negative-frequency parts  of the dressing operator and the $\, \widetilde{}\,$ stands for the fact that they are simply the $J\bar T$ flow of an identical expression in the undeformed CFT.  With foresight, we have included vertex operator dressings for both the left and right-movers, and have left the spectral flow parameter arbitrary in order to be able to reuse this computation in the single-trace case.  
The argument above yields the following expression for the left dressing operators 

\be \label{vertexopleft}
\widetilde{\V}_{\eta}^{+}(\zeta)=e^{\eta\sum_{n=1}^{\infty}\frac{1}{n}\widetilde{J}_{-n} e^{ n\zeta} }\;, \;\;\;\;\;\;\; \widetilde{\V}_{\eta}^{-}(\zeta)=e^{\eta  \left( \widetilde{J}_0 \, \zeta - \sum_{n=1}^{\infty}\frac{1}{n}\widetilde{J}_{n} e^{- n\zeta} \right) } 
\ee
Note in \eqref{leftmovop} we have also allowed for a correction,  $\mathcal{Y}_{\O}$, to the zero mode of the current, which % is linear in the left-moving coordinate. The operator  $\mathcal{Y}_{\O}$  
commutes with all the left generators and will be fixed by the commutation relations of $\O (\zeta, \bar p)$ with the physical Virasoro generators.

 In order to prepare the ground for generalizing these results to the single-trace case, it will be useful to compute of the Ward identities obeyed by $\O(\zeta,\bar{p})$  %this simple computation 
in two steps:
%so as to not rely much on these explicit expressions 
%and try to understand these operators for arbitrary $\eta$, from the point of view of the undeformed CFT. The
 first, we compute the commutation relations of $:\!\widetilde{\V} \widetilde{\O}\!\:$ with the left-moving flowed generators using the explicit expressions \eqref{vertexopleft} for the dressings; then, we assemble them into commutation relations of the non-linear combinations \eqref{lmgenph} with $\O(\zeta, \bar p)$. The advantage of performing the first step separately is that the commutation relations of $:\!\widetilde{\V} \widetilde{\O} \! :$ with the flowed generators are \emph{identical} to the corresponding expressions in the undeformed CFT, given that the flow acts by  simple conjugation. The result is %\emph{\textbf{Widetilde everywhere!}}

\be
[\widetilde{J}_m,\,  :\hspace{-0.1cm}\widetilde{\V} \widetilde{\O}(\zeta,\bar{\zeta})\hspace{-0.1cm}: ]=e^{ m\zeta}\bigg(q^{[0]}+\frac{\eta k}{2}(1-\delta_{m,0})\bigg):\hspace{-0.1cm}\widetilde{\V} \widetilde{\O}(\zeta,\bar{\zeta})\hspace{-0.1cm}: \label{wardL}
\ee
\be
[\widetilde{L}_m,\, :\hspace{-0.1cm}\widetilde{\V} \widetilde{\O}(\zeta,\bar{\zeta})\hspace{-0.1cm}:]=e^{ m \zeta}\left( m(h^{[0]}+\eta q^{[0]}+\frac{k \eta^2}{4}) -\frac{\eta^2 k}{4}(1-\delta_{m,0})+\partial_{\zeta}  \right) :\hspace{-0.1cm}\widetilde{\V} \widetilde{\O}(\zeta,\bar{\zeta})\hspace{-0.1cm}:- \eta :\hspace{-0.1cm}\widetilde{\V} \widetilde{\O}(\zeta,\bar{\zeta})\hspace{-0.1cm}: \tilde{J}_m  \nonumber
\ee
and corresponds, as it is clear from  \eqref{vertexopleft}, % and the fact that $\tilde \O$ carries charge $\tilde q$, these correspond precisely
 to the CFT Ward identities for a normal-ordered left vertex operator of charge $ q^{[0]} + \eta k/2$, but with part of its zero mode left out (so that the charge carried with respect to $\widetilde J_0$ is just $q^{[0]}$). Note the $\widetilde{\bar{\V}}_{\eta}$ (constructed from right-moving current modes only) do not affect the left Ward identities. %, we discuss them later.

% Note that - barring the factor of $\mathcal{Y}_\O$ -  this corresponds precisely to the Ward identities, in the \emph{undeformed CFT}, for a (normal-ordered) left vertex operator of charge $\tilde q + \eta k/2$, but with part of its zero mode left out. This is of course explicit from the expression \eqref{vertexopleft} and the fact that $\tilde \O$ carries charge $\tilde q_0$.  

  %\emph{Check! $\mathcal{Y}_O$ is now probably on the other side}

%
%{\color{blue}
%\begin{align}
%[\tilde{J}_m,\O(\zeta,\bar{p})]&=e^{ m\zeta}\bigg(\tilde{q}+\frac{\eta k}{2}(1-\delta_{m,0})\bigg)\O(\zeta,\bar{p}) \\
%[\tilde{L}_m,\O(\zeta,\bar{p})]&=e^{ m \zeta}\bigg( m(\tilde{h}+\eta\tilde{q}+\frac{k \eta^2}{4})\O(\zeta,\bar{p}) +\partial_{\zeta}\O(\zeta,\bar{p})- \eta\O(\zeta,\bar{p})\tilde{J}_m -\frac{\eta^2 k}{4}(1-\delta_{m,0})\O(\zeta,\bar{p})-\mathcal{Y}_{\O}\O(\zeta,\bar{p}) \bigg)
%\end{align}
%}
It is then not hard to check that the commutation relations of $\O(\zeta, \bar p)$ with the  generators \eqref{lmgenph} are precisely those of a primary of the expected dimension and charge, provided we set $\eta=\lambda\bar{p}$ and 

\be
\mathcal{Y}_{\O}=\lambda q H_R + \l \bar p q^{[0]}+ \frac{k \l^2 \bar p^2}{4} \label{exprYO}
\ee
%yields the usual momentum space Ward identities for $\O$ with respect to the unflowed generators. 
This can also be seen from the fact that the left-moving piece of $\O(\zeta, \bar p)$ organises into a left vertex operator of charge $q$ in the deformed theory, i.e. constructed from the current modes $J_m$: the dressings \eqref{vertexopleft} shift the charge from $q^{[0]}$ to $q$ for the non-zero modes, whereas the the first term in the expression for $\mathcal{Y}_\O$ shifts the $\widetilde J_0$ term in \eqref{leftmovop} (whose total coefficient is $2 q/k$) to $J_0$. 
%
%Writing $\widetilde \O$ in terms of a vertex operator (with parameter $2 q^{[0]}/k$) that carries all the charge, and a piece that commutes with $\widetilde J_0$, we find that  at the level of left vertex operators, , whereas
The last two terms correspond to the difference  $h(\bar{p})-h^{[0]}$ in the operator dimensions, which becomes important when translating the operators defined on the cylinder to those on the plane via the standard map  $\O_{(cyl)} (\zeta) = e^{h \zeta } \O_{(pl)} (z)$. Finally, the total charge of the operator, as measured by $J_0$, is $q$. %, which makes up the new operator of dimension $h$ on the cylinder. %\emph{Argue correlation function is precisely that of vertex operators!}
One may check that, upon performing a conformal transformation to the plane, the left-moving piece of \eqref{leftmovop} corresponds precisely to a local left vertex operator in the deformed theory with charge $q$, with no mismatch between the coefficients of the zero and non-zero modes.  

Having understood in detail the construction of the left-moving piece of the operator - which is entirely fixed by conformal symmetry -  via the dressing discussed above, the proposal of \cite{Guica:2021fkv} consists in choosing the dressing operators for the right-movers to have an identical form, but with all quantities barred. Requiring that the resulting operator be an eigenoperator of $H_R$\footnote{In practice, this is obtained by writing the commutator of $\O$ with $\widetilde{\bar{L}}_0$ in two different ways, see \cite{Guica:2021fkv} for details.}, we obtain the following expression for $\bar{\mathcal{Y}}_\O$ 

\be
\bar{\mathcal{Y}}_{\O}=\lambda\bar{p}(\widetilde{J}_0-\widetilde{\! \bar J}_0)+\lambda q H_R +\lambda\bar{p} q^{[0]}+\frac{k\lambda^2\bar{p}^2}{4}
\ee
The right-moving piece of $:\widetilde{\V} \widetilde{\O}:$ can again be interpreted as a vertex operator in the undeformed CFT, of charge $\bar q^{[0]} + k \eta/2$, but with a discrepancy between the coefficient of the zero and non-zero modes, which is then trivially flowed to the deformed theory.  
A factor of $\l \bar q H_R$ in the expression for $\bar{\mathcal{Y}}_\O$ could again be understood as a correction to the zero mode of the right-moving current. However, since this term does not commute with the right current generators, we no longer have a useful interpretation for the full operator as a vertex operator in the deformed theory. In addition, there are a number of discrepancies involving winding between the operator we obtain and a na\"{i}ve spectrally-flowed operator on the right-moving side, at least at finite $R$. These discrepancies can presumably be traced back to the fact that $\bar L_0$ and $H_R$ differ by winding terms. In any case, given the explicit form \eqref{leftmovop} of the operator, the 
%
% The winding factor is though more mysterious, and it breaks the simple spectral flow interpretation at finite $R$. However, as we will show, it disappears in the large $R$ limit. With this expressions for the operators it can be showed, see \cite{Guica:2021fkv},  that
  the commutation relations with the  generators \eqref{rmphysgen} can be shown to take the form of CFT Ward identities in the limit $R\rightarrow\infty$, with  %{\color{ForestGreen}(I didn't redo these computations from the paper)}
\be 
\bar h (\bar p) = \bar h^{[0]} + \l \bar q^{[0]} \bar p + \frac{k \l^2 \bar p^2}{4} \;, \;\;\;\;\;\; \bar q(\bar p) = \bar q^{[0]} + \frac{k \l \bar p}{2} \label{rmspecfl}
\ee
which resembles a  spectral flow transformation by $\l \bar p$ of the right-moving conformal dimensions and charges.% In this sense, we will call these $\bar{h}$ ``conformal dimensions", although the theory does not have conformal symmetry on the right side.

Let us now compute correlation functions of these operators. Since  $\mathcal{Y}_\O,\bar{\mathcal{Y}}_\O$ only involve $H_R$ and the winding and the $\O(\zeta,\bar p)$ are eigenoperators of both, the correlation function can be simplified to  
%\emph{\textbf{Widetildes!}}
\begin{align}
\langle \O_1 (\zeta_1,\bar p_1) \ldots \O_n (\zeta_n,\bar p_n) \rangle &= \int \prod_{i=1}^n  d\bar \zeta_i \, e^{- \sum_i  \bar p_i \bar \zeta_i} \, e^{\sum_i(\l \bar p_i q_i^{[0]} + \frac{k \l^2 \bar p_i^2}{4} ) (\zeta_i+\bar \zeta_i)}\, e^{\l H_R^{vac} q_i (\zeta_i + \bar \zeta_i)}  \times \nonumber \\[2pt]
&  \hspace{-3 cm} \times \; e^{\sum_{i<j}(\l \bar{p}_i(q_j^{[0]}-\bar{q}_j^{[0]})\bar{\zeta}_i+\l q_i \bar{p}_j(\zeta_i+\bar{\zeta}_i))} \,  \langle  :\! \widetilde{\V}_1  \widetilde{\O}_1 (\zeta_1,\bar \zeta_1)\!: \ldots :\!\widetilde{\V}_n \widetilde{\O}_n (\zeta_n,\bar \zeta_n)\!:  \rangle \label{corfunctstep}
\end{align}
%
%\begin{align}
%\langle \O_1 (\zeta_1,\bar p_1) \ldots \O_n (\zeta_n,\bar p_n) \rangle &= \int \prod_{i=1}^n \big( d\bar \zeta_i \, e^{-   \bar p_i \bar \zeta_i +  \l \bar p_i (\tilde q_i + \frac{\l k \bar p_i}{4} ) (\zeta_i+\bar \zeta_i)}\, e^{\l H_R^{vac} q_i (\zeta_i + \bar \zeta_i)} \big) \times \nonumber \\
%&  \hspace{-3 cm} \times \; e^{\sum_{i=1}^{n-1}(\l \bar p_i \bar \zeta_i \sum_{j=i+1}^n (q_j-\bar q_j) + \l q_i (\zeta_i + \bar \zeta_i) \sum_{j=i+1}^n \bar p_j )}  \langle  : \V_1 \bar \V_1 \tilde \O_1 (\zeta_1,\bar \zeta_1): \ldots :\V_n \bar \V_n \tilde \O_n (\zeta_n,\bar \zeta_n):  \rangle 
%\end{align}
%
%\bea
%\langle \O_1 (\zeta_1,\bar p_1) \ldots \O_n (\zeta_n,\bar p_n) \rangle & = & \int \prod_{i=1}^n d\bar \zeta_i \, e^{-   \bar p_i \bar \zeta_i +  \l \bar p_i (\tilde q_i + \frac{\l k \bar p_i}{4} ) (\zeta_i+\bar \zeta_i)} \, e^{\l \bar p_i \bar \zeta_i \sum_{i+1}^n (q_j-\bar q_j) } \times \nonumber \\
%&  & \hspace{-3 cm} \times \;  e^{ \l q_i (\zeta_i + \bar \zeta_i) (H_R^{vac} +\sum_{i+1}^n \bar p_j) }  \langle  :\V_1 \bar \V_1 \tilde \O_1 (\zeta_1,\bar \zeta_1): \ldots :\V_n \bar \V_n \tilde \O_n (\zeta_n,\bar \zeta_n):  \rangle
%\eea
where $H_R^{vac}$ represents the eigenvalue of $H_R$ in the flowed vacuum.
The last term in the integrand is almost a correlation function of vertex operators of charges $q_i$ in the undeformed CFT, up to the missing zero modes. Using the explicit expressions for the vertex operators to evaluate the zero mode contribution, we obtain
%\emph{Compute prefactor!}
\begin{align} \label{cftresultcor}
&\langle  : \! \widetilde{\V}_1 \widetilde{\O}_1 (\zeta_1,\bar \zeta_1)\!: \ldots :\!\widetilde{\V}_n  \widetilde{\O}_n (\zeta_n,\bar \zeta_n)\!:\rangle \,  =\,  e^{-\sum_{i<j}\l\bar{p}_{j}(q_i\zeta_i+\bar{q}_i\bar{\zeta}_i) } e^{-\sum_i (\l\bar{p}_{i}(q_i^{[0]}\zeta_i+\bar{q}_i^{[0]}\bar{\zeta}_i)+\frac{k\l^2\bar{p}_i^2}{4}(\zeta_i+\bar{\zeta}_i))} \, \times \nonumber\\[2pt]
&\hspace{0.05cm}\times \prod_{i<j} \left( e^{\frac{\zeta_{ij}}{2}}-e^{-\frac{\zeta_{ij}}{2}} \right)^{\frac{2}{k} (q_i q_j- q_i^{[0]} q^{[0]}_j)} \left( e^{\frac{\bar{\zeta}_{ij}}{2}}-e^{-\frac{\bar{\zeta}_{ij}}{2}} \right)^{\frac{2}{k} (\bar q_i \bar q_j- \bar{q}^{[0]}_i \bar{q}^{[0]}_j)}  \langle \widetilde{\O}_1(\zeta_1, \bar \zeta_1) \ldots \widetilde{\O}_n(\zeta_n, \bar \zeta_n) \rangle
\end{align}
where $\zeta_{ij}=\zeta_i-\zeta_j$, $\bar{\zeta}_{ij}=\bar{\zeta}_i-\bar{\zeta}_j$. The second line simply represents the spectrally-flowed correlation function on the cylinder.  The first exponential factor on the first line corresponds to  the correction due to  the missing zero modes, while the second one accounts for the   change in the definition of the cylinder operators due to the shift in their dimensions.  The same result can be obtained by commuting the current modes until they annihilate the (flowed) vacuum. Plugging in this expression into \eqref{corfunctstep} and reinstating the factors of the radius,  the final result for the $n$-point function simplifies to
\begin{align}
\langle \O_1 (\zeta_1,\bar p_1) \ldots \O_n (\zeta_n,\bar p_n) \rangle &= \int \prod_{i=1}^n d\bar{\zeta}_i \, e^{- \sum_i \bar p_i \bar \zeta_i}\, e^{\frac{2\pi}{R}\mathcal{A}} \prod_{i<j} \left(e^{\frac{\pi\zeta_{ij}}{R}}-e^{-\frac{\pi\zeta_{ij}}{R}}\right)^{\frac{2}{k} (q_i q_j- q^{[0]}_i  q_j^{[0]})}\,\times \nonumber \\
&\hspace{-1cm}\times\, \left(e^{\frac{\pi \bar \zeta_{ij}}{R}}-e^{-\frac{\pi\bar \zeta_{ij}}{R}}\right)^{\frac{2}{k} (\bar q_i \bar q_j-\bar q^{[0]}_i \bar q^{[0]}_j)}  \langle \O_1^{\scriptscriptstyle{CFT}}(\zeta_1, \bar \zeta_1) \ldots \O^{\scriptscriptstyle{CFT}}_n(\zeta_n, \bar \zeta_n)  \rangle
\end{align}
%where we put back the radius, anticipating the $R\to\infty$ limit. H
where we plugged in the fact that the correlation function of the $\widetilde \O$ is identical to the corresponding  undeformed CFT correlator 
%
%$\O^{[0]}$ are the operators in the undeformed CFT that flow to $\widetilde{\O}$ in the deformed theory 
and the exponent $\mathcal{A}$ can be written, using charge conservation, %, namely $\sum_i \bar{p}_i=0, \sum_i \tilde{q}_i=0,\sum_i \tilde{\bar{q}}_i=0$, 
as
\begin{align}
\mathcal{A}& =\sum_{i=1}^n \l H_R^{vac} q_i(\zeta_i+\bar{\zeta}_i)+\sum_{i<j} \l \bar{p}_j(q_i-\bar{q}_i) \bar{\zeta}_{ij}
\end{align}
If we ignore this exponential factor, then the result of the integral is simply the spectrally-flowed CFT correlation function in the mixed $(\zeta_i, \bar p_i)$ basis. Note that the spectral flow only affects the operator dimensions, but not the OPE coefficients of appropriate Virasoro-Kac-Moody primaries. In particular,  for the case of two-and three-point function, the spectral flow will simply shift the operator dimensions to the momentum-dependent combinations \eqref{momentumdepcf}, \eqref{rmspecfl} in the corresponding mixed  position/momentum-space correlator. Including the exponential factor will simply shift the coefficients of the $\bar \zeta_i$ in the Fourier integral by a  factor that, importantly, is proportional to $1/R$. The result will be the spectrally-flowed CFT momentum-space correlator evaluated at momenta that differ from $\bar p_i$ by terms that vanish in the decompactification limit. 

%Except for this exponential factor, the rest is just a CFT momentum space $n$-point function on the cylinder, for operators with conformal dimensions spectrally flowed by $\l \bar{p}$ on both sides. However, it is clear from the form of the extra  factors that they cannot modify the ``conformal dimensions", but only shift the values of the momentum. In the limit $R\rightarrow\infty$, in which the operators satisfy the CFT-like Ward identities for the RM also, the shifts vanish and the effect of the deformation is fully captured by the momentum-dependent spectral flow of the conformal dimensions. 

Note that on the left-moving side, the only term that could potentially upset the conformal invariance of the spectrally-flowed correlator is the shift proportional to $H_R^{vac} \zeta_i$ in the expression for $\mathcal{A}$. As discussed in \cite{Guica:2021fkv}, this discrepancy is related to the lack of standard $SL(2,\mathbb{R})_L$ invariance of the flowed vacuum in finite size, which also disappears in the $R \r \infty$ limit.

{\color{blue}

}

\subsection{Correlation functions  in single-trace  $J\bar T$ - deformed CFTs  \label{strcorrf}}
%{\color{ForestGreen} Let us now consider single-trace $J\bar{T}$-deformed CFTs, for which we want to define a set of ``physical" operators in the corresponding SPOs and compute their correlation functions.}
Let us, for completeness, start with some generalities. 
There are several classes of operators one may discuss in a symmetric product orbifold QFT: untwisted or twisted, single-trace or multitrace.  Operators that are untwisted take the form %(in position space \emph{Correct?})

\be
\sum_{\s \in S^N} \O_{\s(1)}^1 \otimes \O^2_{\s(2)} \otimes \ldots \otimes \O^N_{\s(N)}
\ee
where all insertions are at the same position $(\zeta, \bar \zeta)$. In the particular case where all but one operator are the identity, this reduces to $\sum_I \O_I (\zeta, \bar \zeta)$, which is a single-trace untwisted operator. When more than one insertions are non-trivial, the operator is called  multitrace. 
The Fourier transform of these operators can be written as a sum over copies, e.g. $\sum_I \O_I (p,\bar{p})$ in the single-trace case, provided one acts on a state from the untwisted sector. The multi-trace untwisted operators take a similar form, but with several sums over copies and an integral over all the possible ways to distribute the momenta among them: e.g., in the double-trace case, the relevant operators take the form $\sum_{I,J} \int d^d k \, \O^1_I (k) \O^2_J(p-k)$. 
 %These operators  take states in a given twisted sector to states in the same twisted sector. 
The untwisted-sector correlation functions of such operators can be straightforwardly computed using those of the seed: in the single-trace case, they are simply proportional to seed correlators; for multi-trace operators, one obtains a momentum-space convolution  of seed correlation functions, which may be evaluated if the latter are known explicitly. 

Thus, all the new correlators with respect to the seed lie in the twisted sector.
%
%The twisted operators, by definition, contain twist operators inserted at their location. 
One may again classify the twisted-sector operators into single-trace and multi-trace.  
For the single-trace operators, the twist  involves only one non-trivial cycle; for the multitrace ones, several cycles should be considered. The end result should of course be symmetrized with respect to all permutations of the various  copies. % In the following, we would like to discuss  untwisted and twisted sector operators in a symmetric product orbifold of $T\bar T$/$J\bar T$ - deformed CFTs. 

The construction of twisted-sector correlation functions in  single-trace $T\bar T$ and $J\bar T$ - deformed CFTs should depend on the particular method used to build the corresponding  double-trace correlators. In the $T\bar T$ case, the method of \cite{Aharony:2023dod} uses the JT formulation of the deformation, which involves maps from the space where the theory is defined to a flat target space. Given the similarity between  this formalism and one involving the string worldsheet, for example in what regards the computation of  the finite-size spectrum
%For the case of the finite-size spectrum, the contributions of the twisted sectors were included in the very similar formalism of
 \cite{Hashimoto:2019wct,Hashimoto:2019hqo}, it appears reasonable to assume that the correlation functions in the twisted sectors will be captured by 
  considering ``worldsheets'' in the JT formulation that wind around the target space. %; this can presumably be adapted to the computation of the correlation functions in the JT formulation. % {\color{ForestGreen}A similar technique, involving maps from the string worldsheet to the target space, was used in \cite{Hashimoto:2019wct,Hashimoto:2019hqo} for the computation of the finite-size spectrum in the single-trace $T\bar{T}$ case. Different windings of the worldsheet around the target space correspond to the different twisted sectors; this can presumably be adapted to the computation of the correlation functions, generalizing the computation of \cite{Asrat:2017tzd} for the $w=1$ case.}
 The string worldsheet computation of \cite{Cui:2023jrb} suggests the effect of the winding on the correlation functions is extremely straightforward, as it simply amounts to replacing the $T\bar T$ coupling $\mu$ by $\mu/w$ in the twisted sectors. 

The main goal of this section is to put forth an appropriate basis of twisted-sector operators  in single-trace $J\bar T$ - deformed CFTs and to  compute  their correlation functions. We will be following step-by-step  the procedure used in  the double-trace case. 
%Of course, the most interesting case is that of the twisted sector operators. We focus on single-trace $J\bar T$ - deformed CFTs, where we can simply follow the procedure from the double-trace case. 
%{\color{ForestGreen} We will start again by discussing the LM, for which we can use conformal symmetry.
 We will concentrate on single-trace operators, which contain a twist  corresponding to a single non-trivial cycle of length $w$, inserted at their location. %\footnote{In the discussion of section \ref{fractionalvirspo}, this twist corresponded to a twist on the cylinder, viewed as the insertion of a twist operator at $- \infty$. We would now like to interpret it as the twist of the operator inserted at a finite distance, which is possible provided  no other insertions of twist operators are present on the cylinder. }
 The operators of interest are primary  on the left-moving side, with the following momentum-dependent $SL(2,\mathbb{R})_L$ dimensions and $U(1)$ charges  %(notation, should we put $w$ indices?)
%The infinite boost on the twisted-sector deformed energies yields the following $SL(2,\mathbb{R})_L$ dimensions

\be \label{confdinfrac}
h^{(w)} (\bar p) =  h^{(w)}_{\scriptscriptstyle{CFT}} + \frac{\l \bar p}{w}\, q^{[0]} + \frac{k \l^2 \bar p^2}{4 w} \;, \;\;\;\;\; q= q^{[0]} + \frac{\l k \bar p}{2}
% \frac{ h(\bar p)}{w} + \frac{c}{24} (w-\frac{1}{w}) \;, \;\;\;\;\;   h(\bar p) =  h^{[0]} + \l  q^{[0]} \bar p + \frac{\l^2 k \bar p^2}{4}
\ee
%
%
%\be \label{confdinfrac}
%h = \sum_{i \in cycles} \frac{ h_i(\bar p_i)}{w_i} + \frac{c}{24} (w_i-\frac{1}{w_i}) \;, \;\;\;\;\;   h(\bar p_i) =  h_i^{[0]} + \l  q_i^{[0]} \bar p_i + \frac{\l^2 k \bar p_i^2}{4}\;, \;\;\;\;\;
%\sum_i \bar p_i = \bar p
%\ee 
which were computed in section \ref{ttjtspectrum} via an infinite boost of the twisted-sector deformed energies.  $h^{(w)}_{\scriptscriptstyle{CFT}}$ is the dimension of the twisted-sector operator in the undeformed symmetric product orbifold CFT, which is related to a  seed dimension by the holomorphic counterpart of \eqref{conformaldim}. 
 Note $\l$ has been 
 rescaled by a factor of $2\pi$ with respect to \eqref{conformaltwisted}. % and {\color{ForestGreen} we restricted to the case of a single cycle of length $w$ to which the first $w$ copies participate}., whose conformal dimensions and charges were written in \ref{conformaltwisted}. 
 The extension to several non-trivial cycles is straightforward, as the dimensions are additive.

%{\color{ForestGreen} In order to introduce the analogue of \eqref{primaryJTbar}, we use the fact that the $w$-twisted sector is equivalent to the seed on a $w$ covering of the base cylinder. Thus, we will lift all the computations to this covering, where we can use the results from the double-trace case, first to define the flowed operators and then the unflowed ones which have the conformal dimensions \ref{conformaldinfrac}.
%}\emph{Actually, this is not what we do.}

The relation between the deformed and undeformed twisted-sector dimensions corresponds to a transformation known as \emph{fractional} spectral flow \cite{Martinec:2001cf,Chakrabarty:2015foa}, defined as the base space transformation that lifts to the standard spectral flow on the covering space. It acts as

\be  \label{fracspecfl}
h^{(w)} \r h^{(w)} + \eta\,  q^{(w)} + \frac{k w \, \eta^2}{4} \;,\;\;\; \;\;\;\;\; q^{(w)} \r q^{(w)} + \frac{k w \, \eta}{2}
\ee
on the base. In the standard discussions, $\eta$ is taken to be fractional, hence the name. In our case,   $\eta=\l \bar p/w$  is a continuous parameter, so  its fractionality is not very important; what is important is that the  level that enters the spectral flow transformation is the effective level in the $w$-twisted sector, $k^{(w)} = k w$.

%where we remind that  is $k w$ and $h^{[0]},q^{[0]}$ are now the conformal dimensions and charges in the CFT SPO. In the language of the undeformed CFT SPO, 

%We immediately note that the relation between the deformed and undeformed twisted-sector operator is given by fractional spectral flow, with parameter $\l \bar p/w$. \emph{References!} This is a notion of spectral flow in symmetric product orbifold CFTs, which corresponds to standard spectral flow from the perspective of the covering space \cite{}. On the base space, the dimensions and charges are shifted as \emph{Fix factors!}
%
%\be
%h \r h + \eta q + k w \eta^2 \;, \;\;\;\;\; q \r q + k w \eta
%\ee
%In our case, since $\eta$ is a continuous parameter, its fractionality is not very important. What is important is the effective level $k w$ in the twisted sector, which affects the overall scaling with $w$. 

As argued in section \ref{fractionalvirspo}, the physical generators of conformal symmetries in the $w$ - twisted sectors are the integer modes\footnote{In principle, one could also consider fractional Virasoro modes around the location of the twist operator. The cylinder fractional modes appearing in  \eqref{unflowedL} are associated to a twist operator inserted at $-\infty$, whereas in this section we will be considering twists inserted at finite distance on the cylinder. The two sets of fractional generators can be related provided no other operator insertion is present - an overly restrictive requirement. On the other hand, the integer-moded Virasoro generators 
 are everywhere well-defined.} of \eqref{unflowedL}, which are related to the flowed generators in that sector by a transformation resembling fractional spectral flow  with parameter  $\l H_R/w$ 

\be \label{unflowgen}
L_m = \widetilde L_m + \frac{\l }{w} \, H_R \widetilde J_m + \frac{\l^2 k}{4 w} H_R^2 \,  \d_{m,0} \;, \;\;\;\;\;\;\; J_m = \widetilde J_m + \frac{\l k}{2} H_R \,\d_{m,0}
\ee
where we have again set $R= 2 \pi$ , rescaled $\l$ by a $2\pi$ factor and dropped the index indicating the twist sector from $H_R$. Throughout this section, we will work with operators corresponding to the first $w$ copies of the seed QFT,  and symmetrization is assumed only for the final expressions. Our first task is  to construct a basis of operators that are primary with respect to the generators above, with conformal dimensions  and charges given in \eqref{confdinfrac}.

 We may again proceed by defining a flowed operator $\widetilde{\O}^{(w)}(\zeta,\bar \zeta)$ via an equation of the form \eqref{floweqoppr}, with the initial condition that at $\lambda=0$, %{\color{ForestGreen} the operator is the part of a $w$-twisted operator in the symmetric product orbifold CFT which corresponds to the copies $1,...,w$ entering the $w$-cycle, in this order} %
 the operator is a $w$-twisted operator in the undeformed  CFT\footnote{Strictly speaking, since $\O^{(w)}$ is only associated with the first $w$ copies of the QFT, it should be flown with the flow operator associated with these copies - the quantum analogue of \eqref{flowcover}.   While  this analogue of \eqref{floweqoppr} is a well-defined operator equation, it is difficult to lift it to the covering space due to the presence of the twist operator at the location of $\widetilde{\O}^{(w)}$. In particular, the mapping to the larger cylinder used in section \ref{fractionalvirspo}, which did not rely on  conformal invariance, cannot be used anymore. }. 
% {\color{ForestGreen}We will refer in the following to such operator as an operator in the undeformed symmetric product, even if it is not properly symmetrized with respect to $S_N$.}
This operator will have the same conformal dimension and will satisfy the same Ward identities with respect to the flowed  Virasoro and Kac-Moody currents as in the undeformed theory.
% \eqref{Wardidcft1}, \eqref{Wardidcft2}, where now  $\zeta,\bar \zeta$ should be simply thought as labels. 
%
As in the previous subsection,  one should add an appropriate dressing to this operator, so it becomes primary with respect to the standard conformal generators, with the expected dimension \eqref{conformaltwisted}. However, since the relation between the deformed and undeformed dimension is now given by \emph{fractional} spectral flow, we can no longer write a simple, explicit expression for the dressing vertex as in \eqref{vertexopleft}; instead, any explicit expression would involve fractional current modes. %{\color{blue}, which are only defined around the location of the given twist operator.}
%
%Note, importantly, that the relation between flowed and unflowed  generators on the cylinder depends on the twist, as it corresponds to a spectral flow by $\l H_R/w$. As discussed in section \ref{fractionalvirspo}, this is consistent with the fact that the flowed generators are not globally defined, which leads to the main complication in our construction, with respect to the double-trace case: we cannot write a simple expression for the dressing vertex operators in terms of global current modes as in \eqref{vertexopleft}. (One can actually write the dressing factors just like in the double-trace case but with the $\tilde{J}$ replaced by their fractional versions and with an overall $1/w$ factor which can be understood as coming from the effective level $k\to kw$. 
%
%However, these explicit 
%
%Such formulae cannot be used to compute correlation functions of deformed twisted-sector operators, because
Since the fractional modes are defined only locally around operators of generically  different twists,  it does not make sense to commute them as we did in the double-trace case, and thus we need a different way to evaluate the correlation functions of interest.

The approach that we will instead take will be to reformulate, to the largest extent possible,  the computations in terms of undeformed CFT ones - where the lift to  the appropriate covering space is standard - and a $J\bar T$ flow, which does not affect the end result for correlators and commutators. Concretely, we will  write, in analogy with \eqref{leftmovop}

%The main complication that arises is that the flowed generators are not globally defined, 

%To obtain an operator that is primary with respect with the unflowed generators \eqref{}, we dress the flowed operator $\tilde \O^{(w)}$ as in \eqref{}.  

%The main difference with respect to the double-trace case is that now the dressing vertex operators will be implementing fractional spectral flow and, as  such, will not have a simple expression in terms of global current modes as in \eqref{vertexopleft}.  

\be\label{leftmovopstr}
\O^{(w)}(\zeta,\bar{p})=\int d\bar{\zeta\, }e^{-\bar{\zeta}\bar{p}}:\!\widetilde{\V}_\eta \widetilde{\O}^{(w)} (\zeta,\bar \zeta)\!:  e^{ \mathcal{Y}^{(w)}_\O \zeta + \bar{\mathcal{Y}}^{(w)}_\O \bar \zeta } \ee
where $:\!\widetilde{\V}_\eta \widetilde \O^{(w)}\!:$ now represents the single-trace $J\bar T$ flow of an operator in the undeformed theory that is almost given by the fractional spectral flow, with parameter $\eta = \l \bar p/w$, of the original $w$-twisted CFT operator%{\color{ForestGreen} $:\!(\V_\eta \O^{(w)})_{CFT}\!:$, corresponding to the copies $1,...,w$ involved in the $w$-cycle}
, up to some missing zero modes and some rescaling factors. %{\color{ForestGreen} Note that the flow equation whose solution is $:\!\widetilde{\V}_\eta \widetilde \O^{(w)}\!:$ is given by the quantum version of the flow operator \ref{flowcover} involving the same copies. The proper symmetrization with respect to $S_N$ is done only at the end of the construction, for \ref{leftmovopstr}. }
The fractional spectral flow in the undeformed CFT may be implemented by lifting to the covering space, where it becomes usual spectral flow. 
%
%Nevertheless, as explained in the double-trace case, we can trace back the construction to the undeformed theory, where covering space techniques are available. More precisely, we can dress $\widetilde{\O}^{(w)}$, seen as an operator in the CFT SPO, by CFT vertex operators which implement fractional spectral flow, which lifts to the usual spectral flow on the covering space. The operator dressed as such is a local operator in the undeformed CFT with charge $q^{[0]}+ k w \eta/2$. 
Guided by the double-trace case, one should subtract  by hand\footnote{Note that this procedure is not  as innocuous as it may sound. In the CFT, the transformation under the covering map is understood for  local operators, but once we remove some zero modes e.g. from the operator on the covering space, it is no longer clear how to map it back to the base. } from the result  the zero mode of the dressing vertex operator, so that its commutation relations with the undeformed CFT generators - which, upon $J\bar T$ flow, will be the same as those of the  flowed generators with $:\!\widetilde{\V} \widetilde{\O}\!:$ - take precisely the form
%
%which modifies its commutation relations with $wide\tilde{J}_0,\widetilde{L}_0$. The resulting operator, which we denote by $\O^{(w)} (\zeta,\bar p)$ is non-local in the undeformed theory and its commutation relations with the flowed generators are just given by 
\eqref{wardL}, but with $k \r k w$. Note that, in order to reach this conclusion, we do not absolutely need the explicit form of $\widetilde{\V}$.

Using the fact that, by assumption, $\O^{(w)} (\zeta, \bar p)$
 is  an eigenoperator of the  right Hamiltonian  $H_R$, with eigenvalue $\bar{p}$, and  requiring that the commutation relations with the generators \eqref{unflowgen}  are CFT  Ward identities for operators with conformal dimension \eqref{conformaltwisted}, we find  $\eta = \l \bar p/w$, and% \emph{\textbf{Note $H_R$ is a globally defined operator.}} 
%\textbf{\emph{Check and work out!}}

\be
\mathcal{Y}^{(w)}_\O= \frac{1}{w}\bigg(\l q H_R  +\l \bar{p} q^{[0]}+\frac{k\l^2\bar{p}^2}{4} \bigg) %\;, \;\;\;\;\; \bar{\mathcal{Y}}^{(w)}_\O = \frac{1}{w}\bigg(\bigg)
\ee
Assuming that, as in the double-trace case, the commutation relation of $\widetilde{\bar{L}}_0$ with $:\! \widetilde{\V} \widetilde{\O}\! (\zeta, \bar \zeta):$ still yields a simple $\bar \zeta$ derivative of the operator, 
%obtained in the double-trace case for vertex operators that implement spectral flow, but with the zero mode removed, 
we  also deduce that 
\begin{align}
\bar{\mathcal{Y}}^{(w)}_\O = \frac{1}{w}\bigg( \l q H_R + \l \bar{p} q^{[0]} + \frac{k\l^2 \bar{p}^2}{4}+\l\bar{p}(\tilde{J}_0-\tilde{\bar{J}}_0)\bigg)
\end{align}
Thus, choosing $:\!\widetilde{\V} \widetilde{\O}\!:$ to be a fractionally spectrally flowed operator on the left (with some  zero modes removed) reproduces the correct left Ward identities,  even in  absence of explicit expressions for the dressing factors. Our proposal for the right-moving piece of this operator - which will fix the basis \eqref{leftmovopstr} of operators we consider in single-trace $J\bar T$ - deformed CFTs - is that it behaves exactly the same in the undeformed CFT, i.e. it is a fractional spectral flow on the right with parameter $\l \bar p/w$. The commutation relations with the flowed right-moving generators then follow from the CFT ones, and are given by \eqref{wardL} with all quantities barred and $k \r k w$.  The commutation relations of $\O^{(w)} (\zeta, \bar p)$ with $\bar L_m, \bar J_m $, whose relationship to the flowed right-moving generators is given by \eqref{unflowgen} with all quantities barred, then follows straightforwardly from their definitions. It is not hard to show that
%
 %Finally, the analogous computations in the double-trace case, together with the equivalence between the fractional spectral flow on the base and the usual spectral flow on the covering space in the undeformed theory, indicate that the commutation relations with the right unflowed Virasoro generators 
 in the $R\rightarrow\infty$ limit , they correspond to standard CFT right-moving Ward identities with 
 %\textbf{\emph{Did you check this?}}
\be  \label{fracspecflright}
\bar h^{(w)} (\bar p) = \bar h^{(w)}_{\scriptscriptstyle{CFT}} + \frac{\l \bar q^{[0]} \bar p}{w} + \frac{k \l^2 \bar p^2}{4 w} \;, \;\;\;\;\;\; \bar q(\bar p) = \bar q^{[0]} + \frac{k \l \bar p}{2} 
\ee
which resembles a right-moving spectral flow with parameter  $\l \bar p/w$.
%Since these quantities are given by spectral flow withof the right-moving conformal dimensions and charges, we will call them right ``conformal dimensions" and charges in the $w$-twisted sector of the deformed theory.

%To summarize,, we have sketched the construction of operators $\O^{(w)}$ in the $w$-twisted sector of the deformed theory that have CFT-like Ward identities with respect to the left-moving unflowed Virasoro and Kac-Moody generators and with respect to the right-moving unflowed generators in the $R\rightarrow\infty$ limit only. Our argument was based on reformulating the double-trace construction in terms of the undeformed CFT only, where there exist vertex operators that implement fractional spectral flow. \emph{Shorten?}

%Since these are almost the commutation relations of local operators, they can be obtained via covering space techniques. 

%{\color{blue}
%\be
%[L_m^{[0]}, \O(\zeta,\bar \zeta)] = e^{2\pi m\zeta/R} ( \frac{2\pi}{R} m( h^{(w)}_{CFT} + \eta q^{[0]} + \frac{k w \eta^2}{4}) +\p_\zeta   \O(\zeta,\bar \zeta) + \frac{2\pi}{R} [\eta(\tilde J_0-\tilde J_m)-A_\O]\O
%\ee 
%When we compute the commutation relations with the unflowed generators, which have $\eta = \l \bar p/w$ in the $w$-twisted sector, we obtain the required Ward identities, up to some extra terms. By taking $\O\mapsto e^{2\pi A_{\O}\zeta/R}\O$, one can choose $A_{\O}$ in order to cancel those terms. Note these unflowed Virasoros are cylinder generators, but can be made to act on $\O$ by assuming no other insertions are present. }

Let us now discuss the correlations functions of these operators, which %we claim can be computed in the CFT using standard techniques. More precisely, we are interested in computing
take the form
\begin{align} \label{corrtwist}
\langle \O_1^{(w_1)} (\zeta_1,\bar p_1) \ldots \O_n^{(w_n)} (\zeta_n,\bar p_n) \rangle
\end{align}
where each $\O_i^{(w_i)}$ is  now a gauge-invariant operator in the $w_i$-twisted sector of the theory, and we assume that the twists are such that the correlation function is non-vanishing. %, namely it is the Fourier transform of an operator which has a twist operator $\sigma_{w_i}$ inserted at its position. 
%The gauge invariant twist operators can be written as sums over twist operators corresponding to certain elements of $S_N$. In order for the $n$-point function to be nonzero, the product of the single-cycle permutations that give these twist operators should give identity. For the rest of the section, we will assume this is the case. 
For simplicity, we will consider only correlation functions of single-trace twist operators. 

As in the previous section, we can evaluate the correlation function in two steps: first, we use the fact that  \eqref{leftmovopstr} are eigenoperators of  $\mathcal{Y}^{(w_i)}_{\O},\bar{\mathcal{Y}}^{(w_i)}_{\O}$ to reduce the computation  to the evaluation of a correlation function of ``almost CFT vertex operators", as in \eqref{corfunctstep} %\emph{\textbf{Fix factors!}}{\color{ForestGreen}(I put the factors)}
\bea \label{twistedcorrelator}
\langle \O_1^{(w_1)} (\zeta_1,\bar p_1) \ldots \O_n^{(w_n)} (\zeta_n,\bar p_n) \rangle &= &\int \prod_{i=1}^n  d\bar \zeta_i \, e^{- \sum_i  \bar p_i \bar \zeta_i} \, e^{\sum_i(\l \bar p_i q_i^{[0]} + \frac{k \l^2 \bar p_i^2}{4} ) \frac{(\zeta_i+\bar \zeta_i)}{w_i}}\, e^{\l H_R^{vac} q_i \frac{(\zeta_i + \bar \zeta_i)}{w_i}}  \times \nonumber \\[2pt]
& & \hspace{-4 cm} \times \; e^{\sum_{i<j}\frac{1}{w_i}(\l \bar{p}_i(q_j^{[0]}-\bar{q}_j^{[0]})\bar{\zeta}_i+\l q_i \bar{p}_j(\zeta_i+\bar{\zeta}_i))} \,  \langle  :\! \widetilde{\V}_1  \widetilde{\O}_1^{(w_1)} (\zeta_1,\bar \zeta_1)\!: \ldots :\!\widetilde{\V}_n \widetilde{\O}_n^{(w_n)} (\zeta_n,\bar \zeta_n)\!:  \rangle 
\eea
The computation of the correlator of $:\widetilde{\V}_i \widetilde \O_i:$ can then be traced back (by inverting the $J\bar T$ flow) to the cylinder correlation function of the original CFT operators, fractionally  spectrally flowed by an amount $\l \bar p_i/w_i$, up to some  missing zero modes. %[In the double-trace case, one could use the explicit expression for the vertex operators to compute the contribution of the zero modes, which yielded $\exp\{-\sum_{i<j}\l \bar{p}_j (q_i \zeta_i+\bar{q}_i \bar{\zeta}_i) \}$. As discussed, such a computation is no longer possible in the single-trace case, as the fractional current modes are only defined locally since the zero mode and the $\tilde{J}_0,\tilde{\bar{J}}_0$ are only defined locally (they belong to different operators and it does not make sense to commute them since they are not globally defined) \emph{\textbf{True?}}. However, for the LM one can still interpret the operators as being uncharged operators dressed with vertex operators constructed from the unflowed current modes, with charge $q^{[0]}+\frac{k\l\bar{p}}{2}$, independent of the particular sector. Hence, for the LM part of the correlation function, the contribution from the missing zero modes needs to cancel precisely the $e^{\mathcal{Y}^{(w_i)}_{\O}\zeta}$ contribution. Thus, one can easily see that the missing zero modes should come with a factor of $\exp\big\{-\sum_{i<j} \frac{\l\bar{p}_j}{w_i}q_i\zeta_i\big\}$. While there is no such argument available for the RM, we can naively guess that in the single-trace case, for \eqref{corrtwist}, the contribution is given by:]
%\begin{align}
%\exp\bigg\{-\sum_{i<j} \frac{\l\bar{p}_j}{w_i}(q_i\zeta_i+\bar{q}_i\bar{\zeta}_i) \bigg\}
%\end{align}
%This assumption is based on the fact that in each $w_j$-twisted sector the spectral flow parameter is $\l p_j/w_j$, the only other difference from the double-trace case being the effective level $k w_j$, which does not enter in the expression (this is wrong, doesn't give the correct result for LM). 
Note that, unlike for the standard spectral flow, whose action on a general correlation function is known in closed form \eqref{cftresultcor}, for fractional spectral flow involving operators from different twisted sectors no such formula appears to exist. % (as one can explicitly check for the three-point function),  the map to the covering space not being known in general.
 On the other hand, there does exist a well-defined prescription for computing such a  correlation function by mapping the CFT correlators to the covering space, so we may consider this part of the problem as being in principle solved.  

 The additional issue of subtracting the zero modes is complicated by the fact that: i) unlike in the double-trace case, we no longer have  a globally defined explicit expression for the dressing vertex operators and ii) the maps to the covering space are for local operators, but locality is spoiled by the removal of the zero modes. Note one constraint on the corrections is that they should be consistent with translation invariance on the cylinder, which the prefactor in \eqref{twistedcorrelator} currently violates, as did the analogous prefactor in \eqref{corfunctstep}. We will not attempt to estimate the zero-mode contribution herein. Instead, we argue, guided by the explicit expressions in the double-trace case, that the corrections from such zero modes will take the form of exponential factors, which are all suppressed in the $R\r \infty$ limit, and will thus not contribute to the final correlation function on the plane. 
%
%After accounting for the missing zero modes, we are left with an $n$-point function in the undeformed CFT symmetric product orbifold for operators with charges $q_i,\bar{q}_i$, which we denote below by $\hat{\O}_i^{(w_i)}$. Overall, we can write 
%\begin{align}
%\langle \O_1^{(w_1)} (p_1,\bar p_1) \ldots \O_n^{(w_n)} (p_n,\bar p_n) \rangle &= \int \prod_{i=1}^n d^2\zeta_i \, e^{-(p_i \zeta_i+ \bar p_i \bar \zeta_i)}\,\, e^{\frac{2\pi}{R}\mathcal{A}} \langle 0| \hat{\O}_1^{(w_1)}(\zeta_1, \bar \zeta_1) \ldots \hat{\O}^{(w_n)}_n(\zeta_n, \bar \zeta_n) |0 \rangle_{CFT} 
%\end{align}
%where we denoted again by $\mathcal{A}$ the exponent
%\begin{align}
%\mathcal{A}&=\sum_{i=1}^n \frac{1}{w_i} (\l \bar{p}_i q^{[0]}_i+k \l^2 \bar{p}_i^2/4+\l q_i H_R^{vac} )(\zeta_i+\bar{\zeta}_i)+\sum_{i<j}\big[\l\bar{p}_i(q_j-\bar{q}_j)+\l\bar{p}_j(q_i-\bar{q}_i)\big]\frac{\bar{\zeta}_i}{w_i} 
%\end{align}
%Just like in the double-trace case, the structure of the CFT correlation function cannot be modified by the $e^{2\pi\mathcal{A}/R}$ factor, which only shifts the momenta at which the operators are evaluated. In the plane limit $R\rightarrow\infty$, this shift vanishes and
Thus, in this limit we are left with usual CFT $n$-point function in mixed position/ momentum space, for twisted-sector operators with fractionally spectrally flowed conformal dimensions given by \eqref{confdinfrac}, \eqref{fracspecflright} %. For example, for the two-point function, where the twists of the two operators involved are necessarily equal,  we find \emph{Notation? Do we want this explicit?} {\color{ForestGreen}(maybe, instead of this, it's more useful to write schematically our conclusion in a box, like "n-pt function in single tr $J\bar{T}$ = CFT SPO n-pt function with spectrally flowed cf dim")}
%\emph{\textbf{Remove widetildes!}}
\bea
\langle \O_1^{(w_1)} (\zeta_1,\bar p_1) \ldots \O_n^{(w_n)} (\zeta_n,\bar p_n) \rangle_{plane} & = &\\
&&
\hspace{-5cm} =\; \int \prod_{i=1}^n  d\bar \zeta_i \, e^{- \sum_i  \bar p_i \bar \zeta_i} \,  \langle  :\! (\V_1 \O_1^{(w_1)})_{\scriptscriptstyle{CFT}} (\zeta_1,\bar \zeta_1)\!: \ldots :\!(\V_n \O_n^{(w_n)})_{\scriptscriptstyle{CFT}} (\zeta_n,\bar \zeta_n)\!:  \rangle _{ plane}  \nonumber
\eea

Thus, we have succeeded in evaluating correlation functions of a particular basis of  of twisted-sector operators in single-trace $J\bar T$ - deformed CFTs in terms of correlation functions of the undeformed theory. In order to compare with the holographic computation of the next section, it is useful to fully transform the result to momentum space and directly consider the Fourier transform of the euclidean correlator. For the specific case of the two-point function, the result is  %\emph{\textbf{Fix factors and where is the minus!}}{\color{ForestGreen}(we'll check one more time)}

% As far as the two-point functions are concerned,  they both take the form of CFT momentum-space correlators with momentum-dependent conformal dimensions that are quadratic in the momenta. 

%{\color{ForestGreen}
%\be \label{twist2pt}
%\hspace{-0.24cm}\langle \O^{(w) \dag} (p, \bar p) \, \O^{(w)} (p, \bar p)\rangle = \frac{(2\pi)^2 p^{2 h^{(w)}(\bar p)-1} \bar p^{2 \bar h^{(w)}(\bar p ) -1}}{2^{2(h^{(w)}(\bar{p})+\bar{h}^{(w)}(\bar{p}))}\Gamma(2h^{(w)}(\bar{p}))\Gamma(2\bar{h}^{(w)}(\bar{p}))\sin(\pi(h^{(w)}(\bar{p})+\bar{h}^{(w)}(\bar{p})))}  
%\ee

\be \label{twist2pt}
\langle \O^{(w) \dag}(p,\bar p) \O^{(w)} (-p, - \bar p)\rangle = \frac{(2\pi)^2 }{2^{2(h^{(w)} +\bar{h}^{(w)} )} \sin(\pi(h^{(w)} +\bar{h}^{(w)} )) } \,\frac{p^{2h^{(w)}  -1} \bar{p}^{2\bar{h}^{(w)} -1} }{\Gamma(2h^{(w)} )\Gamma(2\bar{h}^{(w)} )}
\ee 
where $h^{(w)}, \bar h^{(w)}$ stand for the momentum-dependent combinations \eqref{confdinfrac},  \eqref{fracspecflright}.

\subsection{Comparison with holographic results  \label{holography}}
 
The single-trace $T\bar T$ and $J\bar T$ deformations have been linked to holography for certain non-AdS backgrounds, namely an asymptotically linear dilaton spacetime for the case of $T\bar T$ \cite{kutasov}, and warped AdS$_3$ for $J\bar T$ \cite{Apolo:2018qpq,Chakraborty:2018vja}. Both of these backgrounds are supported by pure NS-NS flux and flow to AdS$_3$ in the interior; the full spacetimes can  be viewed as non-normalizable deformations thereof. %Consequently the holographic duals are expected to correspond to  irrelevant deformations of the IR CFT$_2$.
Perturbative worldsheet string theory in these backgrounds is given by the $SL(2,\mathbb{R})$ WZW model (for level $N_5>1$), deformed by a certain class of exactly marginal current-current operators. Such deformations are exactly solvable. %The resulting background interpolates between $AdS_3$ in the IR to flat spacetime with a liner dilaton in the UV
 In certain examples, the deformed string background can be thought of as the  near-horizon  geometry of a stack of $N_5$ NS5 branes and  $N_1\gg1$ F1 strings  \cite{kutasov,Chakraborty:2020swe}.   

 %{\color{blue}{The worldsheet sigma model for strings propagating in these backgrounds - at least when the number, $N_5$, of NS5 branes supporting the solution is sufficiently large - and corresponds to an exactly marginal current-current deformation of the $SL(2,\mathbb{R})$ WZW model that describes the IR AdS$_3$ \cite{}. }}

For $N_5>1$, the full CFT dual to the infrared AdS$_3$ is known to \emph{not} be described by a symmetric product orbifold. It is a somewhat singular theory \cite{Seiberg:1999xz}, due to the presence of states with a continuous spectrum known as long strings, which wind around the asymptotic AdS$_3$ boundary. Its long string subsector  has, on the other hand,  been argued  to be described by a symmetric  orbifold \cite{Seiberg:1999xz,Giveon:2005mi}; note, however, that it captures only a small fraction of the system, at least in the regime of interest \cite{Giveon:2005mi}.  Given this structure, it should be  clear  that the symmetric product orbifold of $T\bar T$/$J\bar T$-deformed CFTs cannot be exactly dual to these non-AdS backgrounds for $N_5>1$, because the remaining  sectors do not possess a symmetric  orbifold structure\footnote{Nevertheless, there exists an apparently independent link between asymptotically linear dilaton backgrounds and single-trace $T\bar T$, suggested by the fact that black hole entropy \cite{kutasov,Chakraborty:2020swe,Chakraborty:2023mzc} and the infinite asymptotic symmetries of the linear dilaton spacetime \cite{Georgescu:2022iyx} precisely agree with those of single-trace $T\bar T$. Since only universal quantities match, this suggests the dual could be part of a  ``universality class'' of $T\bar T$-deformed CFTs. 
%In $J\bar T$, there are claims the entropy matches, but this has so far only been shown for very special, rather than generic, charges.  
}; at the same time, the long string  subsector survives in the deformed backgrounds and, moreover, its dynamics are well-described by single-trace $T\bar T$ and, respectively, $J\bar T$ - deformed CFTs. Concretely, the spectrum of long strings has been shown \cite{kutasov,Giveon:2017myj,Apolo:2018qpq,Chakraborty:2018vja} to perfectly match that of single-trace $T\bar T$/ $J\bar T$ (as derived in this article, or also in \cite{Apolo:2023aho}); more recently, correlation functions of long string vertex operators in the asymptotically linear dilaton background have been computed \cite{Apolo:2023aho} and appear to match well with the recent $T\bar T$ results of \cite{Aharony:2023dod}. 

The aim of this section is to  compare the correlation functions of our proposed set of ``primary operator analogues'' in single-trace $J\bar T$ - deformed CFTs with the those of long string vertex operators  in warped AdS$_3$. The latter will be estimated by adapting the short-string computation of \cite{Azeyanagi:2012zd} for the same background to long strings, along the lines of \cite{Cui:2023jrb}. The two results turn out to not exactly match. To explain this, we first review some relevant prior work.

%%and can thus. The IR AdS$_3$ is supported by pure NS-NS flux created by $k$ NS5 and $p$ F1 strings. In order for supergravity to be valid, one needs $k>>1$. In this case, the CFT dual is not an SPO, implying that the deformed backgrounds cannot be dual to an exact SPO of $T\bar T$/ $J\bar T$ (this is only possible for $k=1$). On the other hand, the dual CFT contains a subsector, denoted as the ``long string'' subsector,   which is described  by an SPO. Most of the matching obtained for these spacetimes only refer to this subsector, and do not extend to the entire theory
%\footnote{However, in ALD, there exist indications that the rel to $T\bar T$ extends beyond this subsector, such as entropy and ASG. }. 

Let us start by considering worldsheet vertex operators of type II superstring theory in $AdS_3\times\mathcal{N}$ in the presence of pure NS-NS flux. Here $\mathcal{N}$ is a 7-dimensional compact manifold. % with a left moving $U(1)$ at level $k'$. \emph{Why primed?}
 Long strings correspond to worldsheet vertex operators  that belong to the continuous series representation of $SL(2,\mathbb{R})$. %Such vertex operators are dual via AdS/CFT to operators of the boundary field theory with left and right moving dimensions $(h,\bar{h})$. 
Their worldsheet dimension
%s of such physical (on-shell) worldsheet vertex operators are 
is given by 
\begin{equation}
\begin{aligned}\label{vir1}
&\Delta=-\frac{j(j+1)}{N_5}-w\left(h+\frac{N_5 w}{4}\right)+\Delta_{\N} +N \\
&\bar{\Delta}=-\frac{j(j+1)}{N_5}-w\left(\bar{h}+\frac{N_5 w}{4}\right)+\bar{\Delta}_{\N} +\bar{N}
\end{aligned}
\end{equation}
where
% the rightmost equality is the physical on-shell condition for the vertex operator,  
$j\in -1/2+i\mathbb{R}$ labels the Casimir of the global $SL(2,\mathbb{R})$ algebra, $w\geq 1$ denotes the integer spectral flow in $AdS_3$ - identified with the winding of the long string around the AdS$_3$ boundary, $(\Delta_{\N},\bar{\Delta}_{\N})$ are left/right  vertex operator dimensions of the worldsheet CFT in $\N$, $(N,\bar{N})$ are the left/right oscillator numbers in $AdS_3$ and, finally, $(h,\bar{h})$  represent the eigenvalues of the $J^3, \bar J^3$ zero modes of the worldsheet  $SL(2,\mathbb{R})$. Note that for continuous series representations, these are unrelated to the eigenvalue of the Casimir. In global AdS$_3$, $(h, \bar h)$ are identified with the left/right energies of the state on the cylinder, and thus to the dual operator dimensions  via the standard state-operator map. The physical on-shell condition for these superstring vertex operators is $\Delta=\bar \Delta = 1/2$.
 They can be constructed explicitly \cite{Kutasov:1999xu}, and their correlation function takes the form\footnote{In this article, we normalize the worldsheet operators such that the two-point function of the dual CFT operators takes the form $x_{12}^{-2h}\bar{x}_{12}^{-2\bar{h}}$.  Note this normalization  is different from the standard convention in string theory in $AdS_3$.  }
\be\label{vv2pt}
\langle V (z_1, x_1) V(z_2, x_2) \rangle = \frac{1}{z_{12}^{2\Delta} \bar z_{12}^{2\Delta} x_{12}^{2 h} \bar x_{12}^{2 \bar h}}
\ee
where $x, \bar x$ are auxiliary coordinates that become identified with the space where the dual CFT lives. Integrating over the worldsheet coordinates, one obtains the standard correlation function of CFT operators on the boundary. One may also perform a Fourier transform of the latter, to obtain the momentum-space boundary correlator.    %{\color{ForestGreen}{ For the computation of correlation using worldsheet techniques we would like to restrict ourselves to the case $h=\bar{h}$.
 For example, the momentum space two-point function takes the form %\emph{\textbf{Check!}}

\be\label{VVws2pt}
\langle V (z_1,p) V(z_2,-p) \rangle =\frac{(2\pi)^2 p^{2h -1} \bar p ^{2\bar h-1} }{2^{2(h+\bar h)} \sin (\pi (h+\bar h)) \Gamma(2h) \Gamma(2\bar h)}   \frac{1}{z_{12}^{2\Delta} \bar z_{12}^{2\bar \Delta}} 
\ee
which will be useful in this section.  
 %{\color{ForestGreen}The relation \ref{vir1} holds also for short strings, that correspond to worldsheet vertex operators that belong to the discrete series of representations of $SL(2,\mathbb{R})$, but in this case $j$ is real and positive and, different from the long strings, $j$ and $h,\bar{h}$ are not independent variables, namely $h=j+m,\bar{h}=j+\bar{m}$ where $m,\bar{m}$ are positive integers. In what follows, we will restrict ourselves to long strings, for comparison with our symmetric product orbifold results.}

%Let us consider worldsheet vertex operators of string theory in $AdS_3\times\mathcal{N}$ in the presence of pure NS-NS flux. Here $\mathcal{N}$ is a 7-dimensional compact manifold with a left moving $U(1)$ at level $k'$. In the discussion that follows, we will restrict ourselves to long strings. 

%These form the continuous series states of the  boundary theory. Note that for the case of continuous states (long strings), the $j$ and $h,\bar{h}$ are independent variables. This is not the case for the discrete states (short strings). 

The asymptotically linear dilaton and warped AdS$_3$ backgrounds can both be obtained from AdS$_3$ $\times \; \mathcal{\N}$ via a transformation known as TsT: T-duality, shift, T-duality. In both cases, the effect of the TsT transformation can be encoded in a non-local coordinate transformation, which is equivalent to twisting the  boundary conditions of the fields on the  AdS$_3$ worldsheet theory in a charge-dependent way \cite{Alday:2005ww}. This mildly affects the relationship between $\Delta$ and $h$, by adding %{\color{ForestGreen}(we need to check again $J\bar{T}$)}

%\textbf{\emph{Factors $\pi$! $J\bar T$ looks wrong!!!!}}
%\cite{Apolo:2019zai,Cui:2023jrb}

\be\label{deltah}
\d_{T\bar T} \Delta =   \d_{T\bar T}\bar  \Delta  = \frac{\mu}{\pi}\, p\bar{p} \;, \;\;\;\;\;\; \d_{J\bar T} \Delta =   \d_{J\bar T}\bar  \Delta = \l\bar{p}\left(q^{[0]}+\frac{\lambda k \bar{p}}{4}\right)
\ee
on the left-hand side of \eqref{vir1}. If $(h, \bar h)$ are interpreted as the left/right global energy, then these shifts yield the correct deformed energy formulae to match to single-trace $T\bar T$/$J\bar T$.

The vertex operators in the deformed theory may be expressed in terms of the vertex operators in the undeformed theory, with some appropriate dressing. This has been worked out in  \cite{Azeyanagi:2012zd} for the  warped $AdS_3$ background  and in  \cite{Cui:2023jrb} for the asymptotically linear dilaton one; see also \cite{demisethesis} for relevant prior results. % From the OPE of the deformed energy-momentum tensor with the deformed vertex operators one can deduce that the dressing factors induce a shift in conformal dimensions, which is referred to as spectral flow \footnote{This spectral flow should not be confused with the usual spectral flow automorphism on the worldsheet.}.  Very recently, \cite{Cui:2023jrb} has been able to compute correlation function of the associated vertex operators in ALD. 
It turns out that the correlation function of the appropriate vertex operators still takes the form \eqref{VVws2pt}, %{\color{red}{(Comment: I think the worldsheet vertex operator correlation functions in the deformed theory can only be written in momentum space. I don't know of a neat way of writing worldsheet correlation function in x-space.) }} 
but the relationship between $h$ and $\Delta$ is modified by the shifts \eqref{deltah}.  Since $\D$ must equal $1/2$, this translates into a shift of $h, \bar h$ that takes the form: 
%\textbf{\emph{Check factors!}}  
%{\color{ForestGreen}(we need to recheck factors)}

\bea
&T\bar T &: h \r h + \frac{\mu}{w\pi} p \bar p \;, \;\;\;\; \hspace{1.8cm}\bar h \r \bar h + \frac{\mu}{w\pi} p \bar p \nonumber \\[2pt]
&J\bar T & :  h \r h+\frac{\lambda q^{[0]}\bar{p}}{w}+\frac{\lambda^2 k \bar{p}^2 }{4 w}\;, \;\;\;\;\;\; \;\; \bar h \r  \bar h+\frac{\lambda q^{[0]}\bar{p}}{w}+\frac{\lambda^2 k \bar{p}^2 }{4 w}
\eea
and can be obtained by combining \eqref{vir1} with the appropriate shift in \eqref{deltah}. %{\color{blue}{[ Since the shifts are momentum-dependent, one should consider \eqref{} in the case of $T\bar T$, and  a mix of both for $J\bar T$. ]}} 
Thus, we obtain a two-point function that is identical to the momentum-space two-point function in a CFT, but with shifted dimensions\footnote{Note that this argument appears to imply that, if the normalization of the undeformed AdS$_3$ vertex operators is rescaled by some function of the dimensions, then all these factors would become momentum-dependent. This is \emph{not} what happens if we perform a similar rescaling in the field-theory analysis of section \ref{revcorrf}, where the only functions of $h$ that acquire a momentum dependence are those that result from the Fourier transform. This suggests that the argument of \cite{Cui:2023jrb} may be more subtle than it na\"{i}vely appears.  }.  The $T\bar T$ shift of $h, \bar h$  nicely agrees with \cite{Aharony:2023dod} for the case $w =1$. More generally, one obtaines the shifts in the twisted sectors to be the same, but controlled by $\mu/w$.
%
%%heir main result is that the 2-point function of such operators takes precisely the form of the 2-point function of vertex operators in $AdS_3$, namely
%
%where $\Delta$ and $h$ are related via \eqref{vir1}, but with shifted $\D, \bar \D$ and spectrally flowed $h,\bar{h}$. 
%

Note that the shift in the left-moving single-trace $J\bar T$ dimension precisely agrees with our previous analysis \eqref{conformaltwisted}, including the $w$ dependence. 
However,    the right-moving piece of the correlator \eqref{VVws2pt}  does not agree with \eqref{twist2pt}, which involves $\bar{q}^{[0]}$ instead of $q^{[0]}$ in the right-moving dimension $\bar h^{(w)} (\bar p)$. It would be interesting to understand the origin of this mismatch. Remember, in particular, that in our  field-theory analysis we did encounter shifts of the right-movers that involved $J_0$ instead of $\bar J_0$; however, the discrepancies produced by these terms disappeared in the $R \r \infty$ limit\footnote{The reason was that these discrepancies only appeared in a  single exponential factor. Had they appeared in a $\sinh$ factor,   they would have contributed to the shift in dimension, which would have then agreed with the long string result.}. It would be interesting to perform a more careful worldsheet analysis, possibly on the cylinder,  in order to track this discrepancy. Of course, it could also be that  the  string theory vertex operators  simply are a different set of  operators from those for which we computed correlation functions in field theory. Our criterion for fixing the right-moving piece of the operators in field  theory was based on symmetries, i.e. by requiring that they satisfy CFT-like Ward identities with respect to the right-moving generators \eqref{rmphysgen}, which were related to the flowed right-moving Virasoro generators by an operator-dependent spectral flow.
%which single out a basis of operators and allow for a notion of ``conformal dimensions" for them, even in the absence of standard conformal invariance for the right-movers. {\color{ForestGreen} 
Since these Virasoro symmetries are not yet understood from the worldsheet perspective, the analogous way to fix the operator basis is not yet available on the string theory side. % and there is no reason apriori that the two sets of operators correspond.
%Another possibility is that there exist additiona consistency criteria that
Conversely, it could be that consistency of the worldsheet vertex operators (e.g., mutual locality) would single out a different set of constraints on the operators that would be natural to impose.  
In any case, it would be interesting to further explore the properties of the two sets of operators and 
% relation between them and 
 check whether one  may be preferred to the  other.

%Such approach is not available for now from the worldsheet \emph{how about constructing Virasorors from worldsheet?}, so there is no reason apriori that the two sets of operators correspond and it would be interesting to explore the relation between them. In addition, checking whether one set of operators is more consistent than the other is worth. 

%One can in principle do the same computation in the warped backgrounds related to $J\bar T$. One such computation was performed in \cite{}. Importantly, it ed to shifts in $\Delta, \bar \Delta$ that were \emph{identical} and likely of the form \emph{Check!}
%
%
%{\color{blue}
%\be
%\D \D = \l q \bar p + \frac{\l^2 \bar p^2}{4}
%\ee
%Pluggin this into the dimension formula, we find that for long-string states, $h, \bar h$ shift by $1/w$ times this amount. This agrees with the shift we find for $h$, but  disagrees with  the one for $\bar h$, which involves $\bar q$. It would be interesting to understand whether the vertex operators in string theory are different, and why. 
%
%
%The same calculation can be applied to short string states. There, the shift in $\D$ will result into a shift of the Casimir operator, and thus}

% Thus, we have succeeded in constructing correlation functions of both untwisted and twisted sector operators in single-trace $J\bar T$ - deformed CFTs, as well as untwisted ones in single-trace $T\bar T$ - deformed CFTs. 

Finally, let us mention that exactly the same method may be used to compute correlation functions of the short string vertex operators \cite{Asrat:2017tzd,Giribet:2017imm,soum}; one simply needs to reinterpret the relation between $\D$ and $h$ for the discrete $SL(2,\mathbb{R})$ representations on the worldsheet. In this case, one most commonly  consisders $w=0$ short strings.  Since now the representation is  lowest-weight, the spacetime dimension is related to the worldsheet Casimir as $h=j+1$. Taking into account the shift in the relation between $\D$ and the worldsheet dimension, one finds 
\be
\hspace{-0.01cm}h_{ALD} = \frac{1}{2} + \sqrt{\left(h-\frac{1}{2}\right)^2+\frac{\mu N_5}{2\pi}\,p\bar{p}}\;, \;\;\;\;\; h_{wAdS} =\frac{1}{2}  + \sqrt{\left(h-\frac{1}{2}\right)^2 + \lambda N_5 q^{[0]}\, \bar{p}+\frac{\lambda^2N_5 k}{4}\, \bar{p}^2} 
\ee
%
 %Two-point functions for the $w=0$ short strings in single-trace $T\bar{T}$, namely for operator built upon discrete $SL(2,\mathbb{R})$ representations, have been computed from worldsheet and supergravity approaches in \cite{Asrat:2017tzd,Giribet:2017imm,soum}. The momentum dependence of the conformal dimensions yields a square root of a quadratic 
%
%\be
%h_{ALD} =  1 + \sqrt{1+m^2 - \mu p \bar p}\;, \;\;\;\;\;\;h_{wAdS}(p) = 1 + \sqrt{1+m^2 + \l^2 \bar p^2}
%\ee
%{\color{Red}{(Comment: In the case of wAdS, $h_{wAdS}(\bar p)$ is the dimension of operators  with $q^{[0]}=0$. For $q^{[0]}\neq0$ we need to add the term $4\lambda N_5 \bar{p}q^{[0]}$ inside the square-root.)}}
Since  the information about the $T\bar{T}$ deformed symmetric product is only captured  by the long string sector %\footnote{This can be seen already at the level of the spectrum: the deformation of the short string spectrum does not  match the spectrum of symmetric product of $T\bar{T}$ deformed CFT, as expected \cite{demisethesis}.}
, it is not surprising that the momentum dependence of the conformal dimensions is different from \cite{Aharony:2023dod}. The same comment applies to $J\bar{T}$, where the short string sector conformal dimensions do not match with our symmetric product orbifold results.

 \section{Conclusions \label{discussion}}
 
In this article, we have studied various  properties of symmetric product orbifolds of $T\bar T$ and $J\bar T$ - deformed CFTs - namely the spectrum, the symmetries and the correlation functions -  from a purely field-theoretical perspective.   Our derivations relied mostly on Hilbert space techniques and made no use of conformal invariance, which is not present in these models. 
 
The first observable we discussed was the torus partition function. We showed that  the group-theoretical techniques \cite{Bantay:1998fy,Bantay:1999us,Bantay:2000eq} that were previously developed to determine the partition function of a symmetric product orbifold of  \emph{C}FTs in terms of that of the seed can be easily generalised to two-dimensional \emph{Q}FTs (not necessarily Lorentz-invariant) by appropriately taking into account the  dependence of the partition function on the size of the circle on which the theory is defined.  We also showed that the modular properties of the symmetric product orbifold followed from those of the seed theory. %: for example  in the $T\bar{T}$ case, the torus partition function is modular invariant, while in the $J\bar{T}$ case, it is a Jacobi form.
We then applied these results to the symmetric product orbifold of $T\bar{T}$ and $J\bar{T}$ - deformed CFTs, reproducing the finite-size spectra % and deformed charges (in the $J\bar{T}$ case) of these theories which
that were previously computed using worldsheet techniques. It would be interesting to find other classes of UV-complete QFTs with dimensionful couplings  whose symmetric product orbifold can be studied with this method. %; note in particular that, while the partition function of $T\bar T$/$J\bar T$ is analytic near zero \emph{True?} and admits a Taylor expansion, the general behaviour is expected to be non-analytic. 

%, we studied the torus partition function of arbitrary SPO QFTs assuming the well-definiteness of the torus partition function of the seed QFT. Following the analysis of \cite{}, we computed the left conformal dimension of the $w$-twisted sector in the $J\bar{T}$ SPO and interpreted it as a fractional spectral flow with right-moving energy.

Our second result was a proof that the full Virasoro $\times $ Virasoro ($\ltimes$ Kac-Moody$^2$) symmetries of the symmetric product orbifold CFT, including their fractional counterparts, survive the single-trace $T\bar T$/ $J\bar T$ deformation.
As in the double-trace case  \cite{Guica:2021pzy}, the argument was based on transporting the extended symmetry generators of the undeformed CFT along the irrelevant flow, and then showing that they remain conserved. The ``physical'' symmetry generators may now be singled out by the fact that they correspond to integrals of quasi-local current densities descended from  the covering space, whereas the flowed ones explicitly depend on the twist sector. %\emph{Comment higher spin?} 
We further exploited these symmetries -  following the steps of the double-trace  analysis  \cite{Guica:2021fkv} -  to single out a special basis of operators in single-trace $J\bar T$ - deformed CFTs, both from the untwisted and the twisted sector, and compute arbitrary correlation functions thereof. 
It seems reasonable to hope that similar techniques may be used in the future to construct the correlation functions of single-trace $T\bar T$ - deformed CFTs, provided the construction of correlation functions in the double-trace $T\bar T$-deformed CFTs can be recast in the same language as in the $J\bar T$ case, namely using an interplay of the symmetries and the flow equation.

Our results show that the `QFT data' of single-trace $T\bar T$ and $J\bar T$ - deformed CFTs are rigidly determined by the corresponding observables in the seed double-trace-deformed theories, which in turn are universally determined by those of the undeformed CFT. Thus,  from the point of view of the program set forth by \cite{Smirnov:2016lqw} - of understanding the space of UV-complete\footnote{We replaced the integrability requirement of \cite{Smirnov:2016lqw} by UV-completeness, because for just the standard $T\bar T$/ $J\bar T$ deformations, the universal modification of the S-matrix does not require integrability  of the underlying theory.} two-dimensional quantum field theories - these theories are not  significantly more general than their double-trace counterparts, which  themselves can be understood as mostly kinematical deformations of the underlying CFT \cite{videoDubovsky}.
%, whose dynamical data are as constrained as that of the CFT. 
 On the other hand,  symmetric  product orbifold  CFTs do sometimes  allow  for non-universal exactly marginal deformations that break the symmetric orbifold structure; if such  deformations could also be applied to a $T\bar T$/$J\bar T$ symmetric product orbifold in such a way that its UV completeness  is preserved, then this would  have the potential of significantly enlarging the space of known UV-complete, yet non-local QFTs.

%The data we have derived in this article - spectrum, symmetries, correlation functions - characterizes to a large extent the behaviour of single-trace $J\bar T$ and $T\bar T$ - deformed CFTs and is universally determined by that of the seed $T\bar T$/$J\bar T$ - deformed CFTs, which in turn is determined by that of the undeformed CFT.  It is certainly interesting to have such data from the point of view of understanding the space of possible QFTs. Note these theories are as constraind as CFTs. While the leap was certainly taken with the SZ proposal, the symmetric product orbifolds of these theories, while rigidly determined by the, are significantly more interesting from the holographic perspective; also, from the field-theory one, they may be allowing for deformations that are not obviously present in the double-trace case.  

Another important motivation for understanding the detailed properties of single-trace $T\bar T$/$J\bar T$ - deformed CFTs is their application to non-AdS holography. 
%
% 
%\bi
%\item showed how to compute arbitrary SPO QFT partition function, reproducing results in the lierature
%\item showed Virasoro, including fractional and KdV  survive, but not sure about higher-spin currents
%\item maybe then the symmetry algebra of $T\bar T$ SPO is as much as that of CFT in the moduli space, but keeping SPO structure; gaps may not be large
%\item showed how to compute correlators without using conformal symmetry; maybe also $T\bar T$?
%\item starting point for further extending the space of integable QFTs
%\item weakly-coupled dual to srtingy ALD/ warped AdS
%\ei 
% 
%The structure of the SPO is rather rigid and so, in a certain sense, the properties of the single-trace $T\bar T$ theories we found is as universal as that of its double-trace counterpart. However, this structure is different and one can imagine various ways of deforming it, which may yield less universal results. 
%
%One of our main motivations for this work has been to understand non-AdS holography. 
The exact symmetric orbifolds  we studied should be holographically dual to a highly stringy spacetime, corresponding to  $N_5=1$ in our analysis of section \ref{holography}. The worldsheet theory for a string propagating in such a background is no longer described by the RNS formalism, but could in principle be studied with the methods of \cite{Gaberdiel:2018rqv,Eberhardt:2018ouy}, which would be very interesting to adapt to this non-AdS setting. 
%
% most likely the non-AdS analogue of the spacetime studied in \cite{}.
Alternatively, one could concentrate on the 
 %
%  Of course, in order to understand the
   dual description of the weakly-curved spacetimes that are usually  of most interest in holography, which involve  deformations of these theories that break the symmetric orbifold structure. % But even regardless of that, there are certain features of the deformed theories that may be universal, e.g. the symmetries, which we found to match.  
   It would be very interesting to understand % how to define such deformations away from the symmetric orbifold point, and
    which features of the single-trace $T\bar T$/$J\bar T$ - deformed CFTs we studied -  the entropy, the symmetries, the structure of the correlators - remain universal  once one moves off the symmetric orbifold point. The results of \cite{kutasov,Georgescu:2022iyx} suggest that at least the entropy and the extended symmetries should be part of the list.

\subsubsection*{Acknowledgements}

 We are grateful to Luis Apolo, Brando Bellazzini, Alex Maloney, Sylvain Ribault and especially  Alex Belin  for insightful conversations. The work of SC received funding under the ``Horizon 2020" Framework Program for Research and innovation under the Marie Sklodowska-Curie grant agreement number 945298. The work of SG is supported by the PhD track fellowship of the \'Ecole Polytechnique. The work of SG and MG was supported in part by the ERC starting grant 679278 Emergent-BH.

\end{document}